\documentclass{aa} 

\usepackage{graphicx}
\usepackage{txfonts}
\usepackage{amsmath}
\usepackage{physics}    
\usepackage{color}              
\usepackage{verbatim}   
\usepackage{threeparttable} 
\usepackage{ulem}
\usepackage{xcolor}
\usepackage[colorlinks,linkcolor=blue,urlcolor=blue,citecolor=blue]{hyperref} 
\usepackage{placeins}

\def\photoz{\mbox{photo-\textit{z}}}
\def\specz{\mbox{spec-\textit{z}}}
\def\Msol{{\rm M}_\odot}
\def\zPDF{\textit{z}PDF}
\def\classic{\textsc{Classic}}
\def\lephare{\texttt{LePhare}}
\def\eazy{\texttt{EAZY}}
\def\tractor{\texttt{The Tractor}}
\def\farmer{\texttt{The Farmer}}

\usepackage{caption} 
\definecolor{Blue}{rgb}{0,0.25,0.9}
\definecolor{Red}{rgb}{0.85,0.08,0.05}
\definecolor{Green}{rgb}{0.35,0.45,0.25}
\definecolor{Orange}{rgb}{1.0,0.5,0.15}
\definecolor{Brown}{rgb}{0.7,0.25,0.0}

\renewcommand\arcsec{\mbox{$^{\prime\prime}$}}

\begin{document}

\title{COSMOS2020: UV-selected galaxies at $z\geq7.5$}

\author{
O.~B.~Kauffmann\inst{1}
\and O.~Ilbert\inst{1,2,3}
\and J.~R.~Weaver\inst{4,5}
\and H.~J.~McCracken\inst{6}
\and B.~Milvang-Jensen\inst{4,5}
\and G.~Brammer\inst{4,5}
\and I.~Davidzon\inst{4,5}
\and O.~Le~F\`evre\inst{1,14}
\and D.~Liu\inst{7}
\and B.~Mobasher\inst{8}
\and A.~Moneti\inst{6}
\and M.~Shuntov\inst{6}
\and S.~Toft\inst{4,5}
\and C.~M.~Casey\inst{9}
\and J.~S. Dunlop\inst{10}
\and J.~S.~Kartaltepe\inst{11}
\and A.~M. Koekemoer\inst{12}
\and D.~B. Sanders\inst{13}
\and L.~Tresse\inst{1}
}

\institute{
Aix Marseille Univ, CNRS, CNES, LAM, Marseille, France 
\and
California Institute of Technology, Pasadena, CA 91125, USA 
\and
Jet Propulsion Laboratory, California Institute of Technology, Pasadena, CA 91109, USA 
\and
Cosmic Dawn Center (DAWN)
\and
Niels Bohr Institute, University of Copenhagen, Jagtvej 128, 2200 Copenhagen, Denmark
\and
Sorbonne Universit{\'e}s, CNRS, UMR 7095, Institut d'Astrophysique de Paris, 98 bis bd Arago, 75014 Paris, France 
\and
Max-Planck-Institut f\"ur Extraterrestrische Physik (MPE), Giessenbachstr. 1, D-85748 Garching, Germany
\and
Department of Physics and Astronomy, University of California, Riverside, 900 University Avenue, Riverside, CA 92521, USA 
\and
The University of Texas at Austin 2515 Speedway Blvd Stop C1400 Austin, TX 78712 USA
\and
Institute for Astronomy, University of Edinburgh, Royal Observatory, Edinburgh, EH9 3HJ, UK
\and
School of Physics and Astronomy, Rochester Institute of Technology, 84 Lomb Memorial Drive, Rochester NY 14623, USA
\and
Space Telescope Science Institute, 3700 San Martin Drive, Baltimore, MD 21218, USA
\and
Institute for Astronomy, University of Hawaii, 2680 Woodlawn Drive, Honolulu, HI 96822, USA
\and
Deceased
}

\date{Received ??? / Accepted ???}

\abstract{
This paper presents a new search for $z\geq7.5$ galaxies using the COSMOS2020 photometric catalogues. Finding galaxies at the reionisation epoch through deep imaging surveys remains observationally challenging. The larger area covered by ground-based surveys such as COSMOS enables the discovery of the brightest galaxies at these high redshifts. Covering $1.4$\,deg$^2$, our COSMOS catalogues were constructed from the latest UltraVISTA data release (DR4) combined with the final \textit{Spitzer}/IRAC COSMOS images and the Hyper-Suprime-Cam Subaru Strategic Program DR2 release. We identified $17$ new $7.5<z<10$ candidate sources, and confirm $15$ previously published candidates. Using deblended photometry extracted by fitting surface brightness models on multi-band images, we selected four candidates which would be rejected using fixed aperture photometry. We tested the robustness of all our candidates by comparing six different photometric redshift estimates. Finally, we computed the galaxy UV luminosity function in three redshift bins centred at $z=8,9,10$. We find no clear evolution of the number density of the brightest galaxies $M_\text{UV}<-21.5$, in agreement with previous works. Rapid changes in the quenching efficiency or attenuation by dust could explain such a lack of evolution between $z\sim 8$ and $z\sim 9$. 
A spectroscopic confirmation of the redshifts, already planned with JWST and the Keck telescopes, will be essential to confirm our results.

}


\maketitle
%

\section{Introduction}

\defcitealias{stefanon_brightest_2019}{S19}
\defcitealias{bowler_lack_2020}{B20}
\defcitealias{bruzual_stellar_2003}{BC03}

Star-forming galaxies are expected to be major contributors to the reionisation process in the primordial Universe \citep[e.g.][]{2018PhR...780....1D,finkelstein_conditions_2019}. During this time, when the Universe was less than 1\,Gyr old, Lyman continuum photons emitted by newborn massive stars ionised the intergalactic medium around them. The efficiency of this process is expected to depend on the density of galaxies, their star formation rate (SFR), and the fraction of ionising photons escaping from their interstellar medium \citep[e.g.][]{bouwens_reionization_2015}. The latest results from the {\it Planck} survey \citep{Planck20} give the reionisation redshift mid-point $z_\text{re}=7.67\pm0.73$, while the reionisation probably ended around redshift $z=6$ (e.g. \citealt{fan_constraining_2006}). However, an accurate timeline showing the interplay between the formation of the first galaxies and reionisation has not been established yet. At high redshifts, the galaxy UV luminosity function (UVLF) is the main observable to establish the cosmic star formation rate density (SFRD). The number of ionising photons generated by galaxies can be derived from the SFRD, assuming a number of Lyman continuum photons per second produced per unit SFR, and a fraction of them escaping from galaxies \citep[e.g.][]{Robertson15, Vanzella18}. 
Furthermore, because of the fluctuation of the inter-galactic medium (IGM) opacity on large scales \citep[e.g.][]{Kulkarni19}, the observed redshift of the end of the reionisation may depend on sight lines. 
Therefore, a census of star-forming galaxies at $z>6$ in deep fields is essential to understand the origin and physics of reionisation, as well as the formation of the first galaxies.

Neutral hydrogen in the IGM, the circumgalactic medium, and the interstellar medium absorbs the flux blueward of the Lyman limit at $912$\,\AA{}, with a flux expected to be consistent with zero below this limit \citep[e.g.][]{bouwens_uv_2015}. Moreover, the IGM absorbs the light bluer than the Lyman alpha emission line at $1215.67$\,\AA{}. These features have been used to select high-redshift galaxies in deep imaging surveys \citep[][]{steidel_spectroscopic_1996}. The search for these Lyman-break galaxies (LBGs) is carried out by using fluxes straddling the Lyman emission wavelength at any given redshift and detecting the amplitude of this feature (the height of the Lyman break). 
The colour redward of the break constrains the rest-frame UV slope, which is sensitive to star formation, dust, and metallicity. Photometric redshifts are also commonly used to select high-redshift candidates, through spectral energy distribution (SED) fitting \citep[e.g.][]{mclure_new_2013,finkelstein_evolution_2015,bowler_galaxy_2015}. This method presents the advantage of using more than three bands if available, as well as providing a redshift probability distribution function (hereafter \zPDF{}) for each candidate.

Recent deep and wide-area surveys have provided the necessary multi-wavelength data to select high-redshift galaxies, in particular at near-infrared wavelengths ($1-2$\,$\mu$m). The Wide Field Camera 3 (WFC3) onboard the \textit{Hubble} space telescope (HST), with its unmatched sensitivity at near-infrared wavebands, has revolutionised our knowledge of the distant Universe. One of the major surveys in this context is the Cosmic Assembly Near-infrared Deep Extragalactic Legacy Survey \citep[CANDELS;][]{grogin_candels:_2011,koekemoer_candels:_2011}, covering a total of $750$\,arcmin$^2$ with deep HST imaging. The CANDELS fields include the Great Observatories Origins Deep Survey \citep[GOODS;][]{giavalisco_great_2004} with two fields (GOODS-South and GOODS-North), COSMOS \citep{scoville_cosmic_2007}, the Extended Groth Strip \citep[EGS;][]{davis_all-wavelength_2007}, and the UKIDSS Ultra-deep Survey field \citep[UDS;][]{cirasuolo_evolution_2007}. Furthermore, the \textit{Hubble} Ultra-Deep Field (HUDF) covering $4.7$\,arcmin$^2$ \citep{illingworth_thehstextreme_2013} has allowed for the deepest ever astronomical images to be taken, reaching depths of $30$\,mag in the optical and near-infrared. Alternatively, the \textit{Hubble} Frontier Fields \cite[HFFs;][]{lotz_frontier_2017} include deep imaging of six massive galaxy clusters, which are used to study extremely faint background galaxies magnified through gravitational lensing.
Using these multi-wavelength surveys, over a thousand galaxies with $z>6$ have been identified, reaching $z\sim10$ \citep[e.g.][]{mcleod_z_2016,oesch_most_2014,oesch_dearth_2018,bouwens_uv_2015,bouwens_newly_2019}.

Based on galaxy candidates detected in these HST deep fields, there is evidence that the Schechter function \citep[][]{schechter_analytic_1976} still provides a reasonable fit to the UVLF at high redshift \citep[e.g.][]{schmidt14,bouwens_uv_2015}. The shape of the UVLF appears to evolve with redshift, with a steepening of the faint-end slope $\alpha$ from $-1.6$ at $z=4$ to $-2.3$ at $z=10$ \citep[e.g.][]{bouwens_uv_2015,finkelstein_evolution_2015,oesch_dearth_2018}. The observed steep slope for the UVLF at high redshifts predicts a high abundance of faint galaxies in the Universe during the reionisation epoch. However, there is no consensus whether the observed evolution of the UVLF at $z>6$ is driven by changes in the characteristic absolute magnitude or in the normalisation \citep[e.g.][]{mclure_new_2013,finkelstein_evolution_2015}. This may be due to an evolving shape of the LF towards a non-Schechter form. When converted in SFRD, most of the studies are consistent with a steady decrease in the SFRD from its peak at $z=2-3$ to $z=10$ \citep[e.g.][]{madau14_review, finkelstein16_review}. \citet[][]{oesch_dearth_2018} find that this decrease is even more rapid between $z=8$ and $z=10$ by more than a factor ten, possibly explained by the fast buildup of the underlying dark matter mass function in the primordial Universe. 

While an extremely powerful technique to study the faint galaxy populations, the current deep pencil-beam surveys do not cover sufficient cosmological volumes to capture the rare and bright sources which would constrain the bright end of the UVLF. For instance, \citet[][]{oesch_dearth_2018} analysed $800$\,arcmin$^2$ of archival data, but did not constrain the LF brighter than $M_\text{UV}<-21.2$. \citet[][]{bridge19_borg} found eight galaxies brighter than $M_\text{UV}<-21.5$ at $7<z<8$ in the Brightest of Reionising Galaxies (BoRG) survey, a parallel HST survey specially designed to observe the brightest galaxies. 
In this regime, deep ground-based imaging surveys are invaluable. The combination of deep optical ($27-28$\,mag) and near-infrared ($25-26$\,mag) imaging over a degree-scale area has made it possible to isolate the brightest galaxies in the $z>7-8$ Universe \citep[e.g.][]{bowler_bright_2014}. Sufficiently deep \textit{Spitzer} coverage with the InfraRed Array Camera (IRAC) at $3.6$ and $4.5$\,$\mu$m complements these observations, allowing for the efficient removal of the intermediate-redshift contaminants \citep[e.g.][]{oesch12}, as well as the direct detection of high-redshift sources. 


Using the available data from the UltraVISTA survey \citep[][]{mccracken_ultravista_2012} in the COSMOS field, multiple samples of galaxy candidates have already been identified at $z>6$ \citep{bowler_bright_2014,bowler_galaxy_2015} and $z\geq7.5$ \citep[][hereafter S19]{stefanon_hst_2017,stefanon_brightest_2019}, imposing strong constraints on the bright end of the galaxy UVLF. \citet[][hereafter B20]{bowler_lack_2020} found 27 LBGs over a $6$\,deg$^2$ area combining data from the COSMOS and XMM-LSS fields. The bright end ($-23<M_\text{UV}<-21$) of the resulting galaxy number density is in excess compared to the exponential decline predicted from a Schechter parametrisation, suggesting a double power law may be more appropriate. 
Early models have linked the UVLF evolution with the one of the dark matter halo mass functions \citep[][]{Cooray06_CLF,Bouwens08,Tacchella13}.
The change of shape at the bright end of the UVLF compared to the halo mass function is interpreted by the combination of several physical processes, including quenching and dust extinction \citep[e.g.][]{Harikane22_goldrush}. Indeed, for a given massive galaxy, its host dark matter halo or its own intrinsic stellar content may not be massive enough to trigger internal star-formation quenching \citep[][]{peng10}. Moreover, dust extinction is known to bend the bright end of the UVLF at intermediate redshift \citep[][]{Reddy10}. The formation of dust in the interstellar medium (ISM) of galaxies at $z>6$ may still be immature, thus high SFR galaxies at this epoch may experience less attenuation \citep{finkelstein_evolution_2015}. This would result in a UVLF that is well populated at the bright end, describing a double power law \citep{bowler_bright_2014,bowler_galaxy_2015}.


With the COSMOS2020 photometric catalogues \citep{2021arXiv211013923W}, we have the opportunity to improve the search and identify the brightest star-forming galaxies at the epoch of reionisation. Candidates are identified using deep near-infrared imaging from UltraVISTA DR4, complemented with IRAC images from the Cosmic Dawn Survey \citep{2021arXiv211013928M} to detect the galaxy rest-frame optical emission. While \citetalias{bowler_lack_2020} used the latest UltraVISTA release (DR4), we used, in addition, the latest deep optical images from the public DR2 of the Hyper Suprime-Cam (HSC) Subaru Strategic Program (HSC-SSP; \citealt{aihara_hyper_2018}) and the final \textit{Spitzer} images in the mid-infrared. We also validated our results using the recently released DR3 of the HSC-SSP \citep{Aihara21}. These data are essential to improve the purity of the high-redshift sample and to extend the area under investigation. Furthermore, COSMOS2020 photometric extractions have been made with two different techniques: a traditional approach using aperture photometry and a surface brightness profile-fitting technique using multi-band images. In this paper, we combine both photometric catalogues, and multiple photometric redshift codes, leading to a more robust final sample of candidates. 


For this study, we searched for galaxies at $z\geq7.5$ in the COSMOS field using the COSMOS2020 photometric catalogue, and we estimated the galaxy rest-frame UVLF at $8\leq z\leq10$. The paper is structured as follows. Section~\ref{sec:data} presents the imaging data used in this work, as well as the estimated source photometry. Section~\ref{sec:selection} describes the high-redshift galaxy selection. The sample of candidates is presented in Sect.~\ref{sec:sample} and compared with previous studies. The updated constraints on the high-redshift UVLF resulting from the selected galaxies are presented in Sect.~\ref{sec:UVLF}. We discuss these results in Sect.~\ref{sec:discussion} and summarise our conclusions in Sect.~\ref{sec:conclu}. 

We adopted the standard $\Lambda$ Cold Dark Matter cosmology with $\Omega_\text{m}=0.3$, $\Omega_\Lambda=0.7$, and $H_0=70$~km s$^{-1}$ Mpc$^{-1}$. We used the initial mass function (IMF) from \citet{2003PASP..115..763C}. Magnitudes are given in the AB system \citep{oke_absolute_1974}.

\section{Data}
\label{sec:data}

This work is based on the COSMOS2020 photometric catalogues \citep{2021arXiv211013923W}, which includes the currently deepest optical to mid-infrared imaging in the COSMOS field. The imaging data, as well as the extracted photometry are briefly described here. Full details can be found in \citet{2021arXiv211013923W}.

\subsection{Imaging}

\begin{table}
\small
\centering
\renewcommand{\arraystretch}{1.2}
\setlength{\tabcolsep}{4pt}
\begin{threeparttable}
\caption{Depth of the deepest optical and infrared broad bands used in this work.}
\begin{tabular}{cccc}
\hline\hline
Band & $m_{3\sigma,\text{ AB}}$\tnote{a} & $m_{3\sigma,\text{ AB}}$ & Source \\
 & ultra-deep\tnote{b} & deep & \\
\hline
$u$ & 27.8 & ... & CFHT/MegaCam \\
$u^{*}$ & 27.7 & ... & ... \\
$g$ & 28.1 & 27.8 &  Subaru/HSC \\
$r$ & 27.8 & 27.4 & ... \\
$i$ & 27.7 & 27.2 & ... \\
$z$ & 27.2 & 26.7 & ... \\
$y$ & 26.5 & 25.8 & ... \\
$Y$ & 26.6 & 25.3 &  UltraVISTA \\
$J$ & 26.4 & 25.2 &  ... \\
$H$ & 26.1 & 24.9 &  ... \\
$K_s$ & 25.7 & 25.3 &  ... \\
$[3.6]$ & 26.5 & 26.4 &  \textit{Spitzer}/IRAC \\
$[4.5]$ & 26.5 & 26.2 & ... \\
\hline
\end{tabular}
\begin{tablenotes}
\item[a] Depth at $3\sigma$ computed on PSF-homogenised images (except for IRAC images) in empty $2\arcsec$ diameter apertures. 
\item[b] The deep and ultra-deep regions are distinct for the HSC and UltraVISTA data.
\end{tablenotes}
\label{tab:bands}
\end{threeparttable}
\end{table}

The COSMOS field is covered by several multi-wavelength deep imaging surveys. While the photometric catalogue consists of observations in more than $35$ bands, here we only discuss the deepest and reddest broad-band imaging, which are the most relevant datasets to search for high-redshift galaxies. The photometric depths of these broad-band images are given in Table~\ref{tab:bands}. 
We stress that the depths are not homogeneous over the full field and differ by wavelength. We discuss the impact on the selection of the high-redshift candidates in Sect.~\ref{sec:sample}.

\begin{figure}
  \includegraphics[width=\hsize]{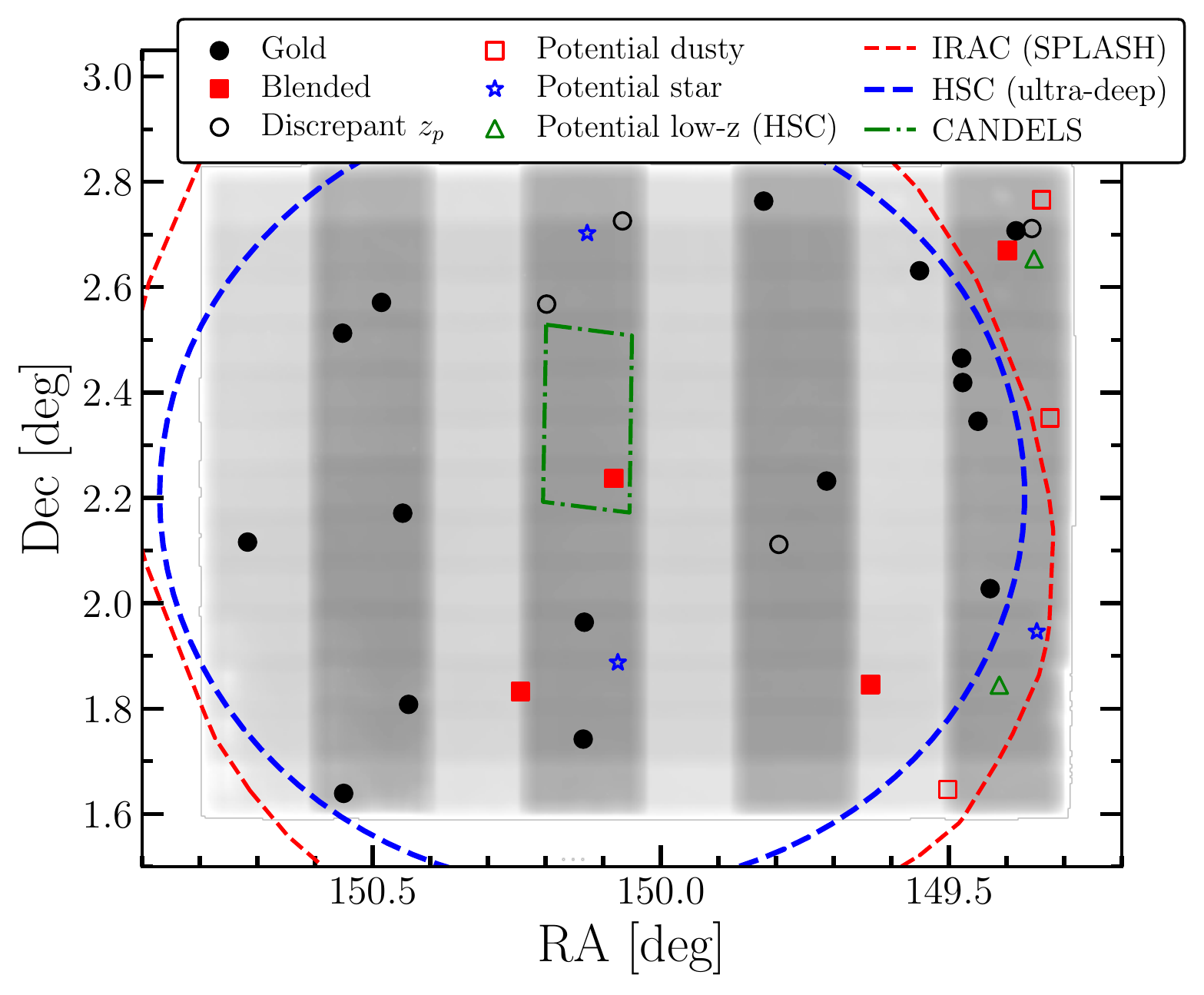}
  \caption{Imaging data in COSMOS. The background image shows the UltraVISTA $H$-band weight map, where the vertical dark grey stripes represent the UltraVISTA ultra-deep stripes. The red dashed line indicates the \textit{Spitzer}/IRAC $[3.6]$ coverage of the SPLASH survey, the blue dashed line represents the ultra-deep region of the HSC images, and the green dot-dashed line corresponds to CANDELS. The filled circles indicate our gold sample (see Sect.~\ref{sec:robustness}). Open symbols indicate sources for which we cannot exclude a low redshift or star solution based on template fitting, or because of a low S/N detection in one HSC band.
  }
  \label{fig:cand_coords}
\end{figure}

The UltraVISTA survey \citep{mccracken_ultravista_2012} provides deep near-infrared imaging over $1.5$\,deg$^2$ of the COSMOS field, as shown in Fig.~\ref{fig:cand_coords}. The exposure time is not homogeneous over the entire field. Four ultra-deep stripes across the field covering $0.62$\,deg$^2$ have deeper exposure times (vertical dark grey stripes in Fig.~\ref{fig:cand_coords}). We use the UltraVISTA DR4 data, with a near-infrared coverage in four broad bands, $YJHK_s$. The additional narrow-band \textit{NB}$118$ covers the ultra-deep stripes. 

The mid-infrared data comes from the Cosmic Dawn Survey \citep{2021arXiv211013928M}, in which all the \textit{Spitzer} observations in the IRAC/$[3.6]$, $[4.5]$ bands are processed (the full list of included programs is given in their Table~C.1). This notably includes the COSMOS \textit{Spitzer} survey (S-COSMOS; \citealt{sanders_s-cosmos_2007}) over $2$\,deg$^2$, the \textit{Spitzer} Large Area Survey with Hyper Suprime-Cam (SPLASH; \citealt{steinhardt_star_2014}) over $1.8$\,deg$^2$, the \textit{Spitzer} Extended Deep Survey (SEDS; \citealt{ashby_seds_2013}), the deep \textit{Spitzer} Matching Survey of the UltraVISTA ultra-deep Stripes survey (SMUVS; \citealt{ashby_spitzer_2018}) and Completing the Legacy of \textit{Spitzer}/IRAC over COSMOS (COMPLETE, COMPLETE2; \citealt{labbe_spitzer_2016,stefanon_spitzer_2018}). The IRAC depths reported in Table~\ref{tab:bands} are separately computed inside the deep and ultra-deep UltraVISTA stripes. The $3\sigma$ depths of S-COSMOS data outside the SPLASH field (shown in Fig.~\ref{fig:cand_coords}) reach $25.3$\,mag in $[3.6]$ and $25$\,mag in $[4.5]$. 

The optical data provided by the Hyper Suprime-Cam (HSC) Subaru Strategic Program (HSC-SSP; \citealt{aihara_hyper_2018}), include deep imaging in the $g,r,i,z,y$ broad bands over $2.2$\,deg$^2$. We use the HSC-SSP DR2 images \citep{aihara_second_2019}, stacked for the COSMOS2020 catalogue. The central region of the COSMOS field includes ultra-deep HSC imaging within a $0.75$\,deg radius circle (blue circle in Fig.~\ref{fig:cand_coords}), and deep imaging in the extended COSMOS survey, about $0.5$\,mag shallower than in the centre. Since the COSMOS2020 catalogue was created, a third release of the HSC-SSP became available \citep{Aihara21}. While these data are not included in our analysis, we used these images to insure the robustness of our candidates. 
In addition, we include the reprocessed Subaru Suprime-Cam images with 12 medium and two narrow bands in optical \citep{taniguchi_cosmic_2007,taniguchi_subaru_2015}, the $u$ and $u^*$ bands from the CFHT Large Area $U$-band Deep Survey (CLAUDS; \citealt{sawicki_cfht_2019}), and the UV photometry from GALEX \citep{zamojski_deep_2007}. 

\subsection{Photometry}
\label{photometry}

Source detection is performed in the combined \texttt{CHI\_MEAN} image constructed with \texttt{SWarp} \citep{bertin_terapix_2002} from the HSC/$i,z$ and UltraVISTA/$Y,J,H,K_s$ bands. This stacked detection image provides an advantage for faint, high-redshift sources compared to single images.
Photometry is extracted following two independent approaches, leading to two separate photometric catalogues, called the \classic{} catalogue and \farmer{} catalogue.
In the \classic{} approach, the photometry of the high-resolution images is performed with \texttt{SExtractor} \citep{bertin_sextractor:_1996} in dual-image mode. 
The photometry is extracted in fixed $2\arcsec$ diameter apertures. To ensure that the apertures include the same features at all wavelengths, the point-spread functions (PSF) of the science images are homogenised. Multiple corrections are applied to the measured magnitudes, including the magnitude error scaling and the aperture-to-total magnitude corrections. The photometry of the low-resolution IRAC images is performed with the software \texttt{IRACLEAN} \citep{hsieh_taiwan_2012}, using the high-resolution images as prior to extract the photometry of the confused sources. In this case, the IRAC PSF is iteratively subtracted from the IRAC images centred within the source boundary, as defined by the high-resolution prior.
In the second approach, all bands are extracted with \farmer{} (Weaver et al., in prep.), which uses \tractor{} software \citep{lang_tractor_2016} to obtain more accurate photometry by fitting galaxy profiles with parametric models. The morphology of the sources is determined through a decision tree, separating point and extended sources. In contrast with the \classic{} approach, this method directly provides total magnitudes, performs an improved deblending in the high-resolution images, and extracts all the images consistently.

\section{Galaxy selection criteria}
\label{sec:selection}

Here we describe different steps taken to select a complete sample of galaxies at high redshifts. The resulting candidates are discussed in Sect.~\ref{sec:sample}.

We exclusively search for candidates over the UltraVISTA area, by rejecting sources located in the masked regions near bright stars in the HSC images. This corresponds to an area of $1.404$\,deg$^2$. 
We require the candidates to be detected in three bands among the $H$, $K_s$, [3.6] and [4.5] bands. We require at minimum a $5\sigma$ detection in one of these bands, one at $3\sigma$ in a second band, and at $1\sigma$ in a third band. This ensures that at least two colours are reliable to estimate \photoz{}. Sources only detected in IRAC images are not included. We apply these selection criteria separately on both the \classic{} and \farmer{} catalogues. 

We reject candidates detected in any broad band blueward of the Lyman alpha break based on visual inspection, rather than explicit magnitude cuts. The measured flux may be inconsistent with zero in the case of a noise local maximum at the source coordinates. Moreover, nearby sources likely contaminate aperture fluxes for blended candidates, and generate a significant flux in the photometric apertures. In these cases, we estimate whether the observed flux originates from the detected source or from nearby sources. In the latter case, we flag these sources as blended. Artefacts are also rejected through the visual inspection of the science images, as discussed in Appendix~\ref{sec:appendix_cross-talks}.

\subsection{Photometric redshift selection}
\label{sec:photoz_selection}

Our initial candidate lists are selected taking into account both the photometric redshifts and the posterior probability redshift distributions (\zPDF{}) computed for all sources in COSMOS2020 using \lephare{} \citep{arnouts_measuring_2002,ilbert_accurate_2006} with both the \classic{} and \farmer{} catalogues. 
We first use the default COSMOS2020 configuration of \lephare{}, as done in \citet[][]{ilbert_mass_2013}, \citet[][]{laigle_cosmos2015_2016} and \citet{2021arXiv211013923W}, with the templates, dust attenuation curves and the recipe to add emission lines from \citet[][]{ilbert_photoz_2009}. The galaxy library includes $33$ templates, covering various SED types, from quenched to starbursting galaxies. The most relevant ones for this work are $12$ templates generated with the 
\citet[][]{bruzual_stellar_2003} (hereafter BC03) stellar population synthesis models, with ages ranging from $30$\,Myr to $3$\,Gyr, and including sub-solar metallicities ($Z=0.004$, $0.008$, $0.02$). We add the dust attenuation as a free parameter, allowing $E(B-V)$ to vary between $0$ and $0.5$ for two different attenuation curves: \citet{calzetti_dust_2000} and \citet{prevot_typical_1984}. Several emission lines are added using an empirical relation between the UV luminosity corrected for dust attenuation and H$_\alpha$ emission line flux. Physically constrained ratios are considered between the intrinsic emission lines. We added [OII], H$_\beta$, [OIII], and H$_\alpha$ emission lines, as well as Lyman$_\alpha$, despite the large uncertainties that potentially affect the modelling of this line. Dust attenuation is applied to the emission line fluxes using the same dust model as for the stellar continuum. The normalisation of the emission line fluxes are allowed to vary by a factor of two (using the same ratio for all lines) during the fitting procedure. IGM absorption is implemented following the analytical correction of \citet{madau95}.

As in \citet[][]{laigle_cosmos2015_2016}, we perform the fit using fluxes (and not magnitudes), even when a source is extremely faint or non-detected in one band. This type of approach is suitable as long as uncertainties are measured consistently. Therefore, we do not have to include upper-limits in the fitting procedure. 

\lephare{} provides the redshift likelihood distribution for each source, after a marginalisation over the galaxy templates and the dust attenuation. We use it as the posterior redshift probability density function (\zPDF{}), assuming a flat prior. The \photoz{} point estimate, $z_\text{phot}$, is defined as the median of the \zPDF{}. We also consider the \photoz{} which minimises the $\chi^2$ over the full template library as an alternate \photoz{} point estimate, to strengthen our selection of high-redshift sources. 
We select any source with a solution at $z\geq7.5$ with \lephare{}, in either the \classic{} or \farmer{} catalogue. This is defined by imposing that both the median of the \zPDF{} and the redshift which minimises the $\chi^2$ are at $z\geq7.5$. 

\subsection{Complementary template-fitting procedures}
\label{sec:photoz_confirmation}

We produce photometric redshifts with other template-fitting procedures for the selected candidates. These estimates are not directly used to select the candidates, but are used in Sect.~\ref{sec:robustness} to assess their robustness.

The photometric redshifts computed for all sources using \eazy{} \citep{brammer_eazy_2008} are available in COSMOS2020. The adopted strategy is equivalent to \lephare{} and corresponds to the technique described in \citet{2021arXiv211013923W}. The fitted galaxy library consists of $17$ templates derived from the Flexible Stellar Population Synthesis models \citep{conroy_relevance_2009,conroy_model_2010}. Moreover, \eazy{} allows the combination of the templates in the fitting procedure. 

Furthermore, we compute new photometric redshifts for the selected candidates using a different \lephare{} configuration, hereafter noted \lephare{} BC03. Such a configuration is optimised to model sources with extreme colours. Since we select a really specific population with extreme properties, a larger coverage of the parameter space (in terms of dust, ages, star formation histories) should be explored to reject potential intermediate redshift contaminants. In this case, we include a set of \citetalias{bruzual_stellar_2003} templates assuming different star formation histories (exponentially declining and delayed), as described in \citet{ilbert_evolution_2015}. Each of these $12$ templates is generated at $43$ ages from $50$\,Myr to $13.5$\,Gyr. During the fitting procedure, no template with an age older than the age of the Universe is considered. We assume two attenuation curves \citep{calzetti_dust_2000,arnouts_encoding_2013} with $E(B-V)$ varying from $0$ to $2$. The dust attenuation reaches $A_{V}\sim8$, enabling potentially extremely dusty sources at lower redshift to be rejected \citep{Dunlop07}. We add the emission line fluxes with a recipe described in \citet{saito_synthetic_2020}. For each template, we derive the number of ionising photons by integrating the SED blueward of the Lyman break, and convert it into H$_\beta$ luminosity following \citet{schaerer_impact_2009}. Then, the ratios between H$_\beta$ and other emission lines are given by \citet{osterbrock_astrophysics_2006, anders_spectral_2003}. Emission line fluxes are allowed to vary by a factor of two during the fitting procedure, to reproduce potential variations around the expected value. 
We do not use this configuration to compute photometric redshifts for the full COSMOS2020 catalogue, since the large parameter space covered by the templates increases the risk of degeneracy in the colour-redshift space and creates a larger fraction of catastrophic failures for the general population\footnote{Reducing the parameter space could be seen as a prior for the general population.}.

\subsection{Brown dwarf contamination}
\label{sec:brown_dwarfs}

 The selected high-redshift candidates may be contaminated by cool Milky Way brown dwarfs, because of their similar near-infrared colours \citep[e.g.][]{Wilkins14}. Given the predictions by \citet{Ryan16}, we could expect 277/deg$^2$ brown dwarfs at $J<25$.
 
 To isolate these contaminants, we fit brown dwarf templates to COSMOS2020 photometry. We include the simulated high-resolution brown dwarf templates from \citeauthor{2015A&A...577A..42B} (\citeyear{2015A&A...577A..42B}, BT-Settl/CIFIST2011\_2015) and \citet{2012ApJ...756..172M,2014ApJ...787...78M} in \lephare{}. The modelled emission extends to at least 10\,$\mu$m in the mid-infrared. In addition, the nonphysical templates are rejected following constraints from \citet{2008ApJ...689.1327S} in the effective temperature versus surface gravity space, based on the predicted evolution of cool brown dwarfs.
Then, we compare the minimum $\chi^2$ obtained with the galaxy templates (we use the standard \lephare{} configuration described in Sect.~\ref{sec:photoz_selection}) and the one obtained with the stellar library including brown dwarf templates. We compute $\Delta \chi^2 = \chi_\text{star}^2-\chi_\text{gal}^2$ with $\chi_\text{gal}^2$ being the minimum $\chi^2$ obtained with the galaxy templates, and $\chi_\text{star}^2$ the one obtained with the stellar templates. We select candidates with $\Delta \chi^2 > 0$.

\subsection{The case of ID720309}
\label{sec:ID720309}

The source ID720309 has the highest \photoz{} among the candidates extracted from the COSMOS2020 catalogue. However, we identified a bright star at the coordinates R.A.=09h59m10.81s, Dec.=+2d11m04.29s which could potentially generate a cross-talk signal (see Appendix~\ref{sec:appendix_cross-talks}) at a position 09h59m10.81s +2d06m42.40s (assuming a native pixel scale of $0.341$\arcsec{} expected for the detector~7 of the VIRCAM camera). Each ultra-deep stripe is the combination of VIRCAM observations taken at three different declinations, each called ``paw'' (see \citealt{mccracken_ultravista_2012}). This source presents two components in one paw (paw~1), and just one component in the other (paw~2). It suggests that ID720309 in COSMOS2020 is the superposition of a cross-talk and a real source separated by $0.7$\arcsec{}. It contributes to the signal in the $2$\arcsec{} aperture of the \classic{} catalogue. As \farmer{} photometry is centred on the real source, the cross-talk has less impact on the profile-fitting photometry. We also notice that the $[3.6]$ and $[4.5]$ positions are well aligned with the $K_s$ position of the source in paw~2.

\begin{figure}
  \centering
  \includegraphics[width=\hsize]{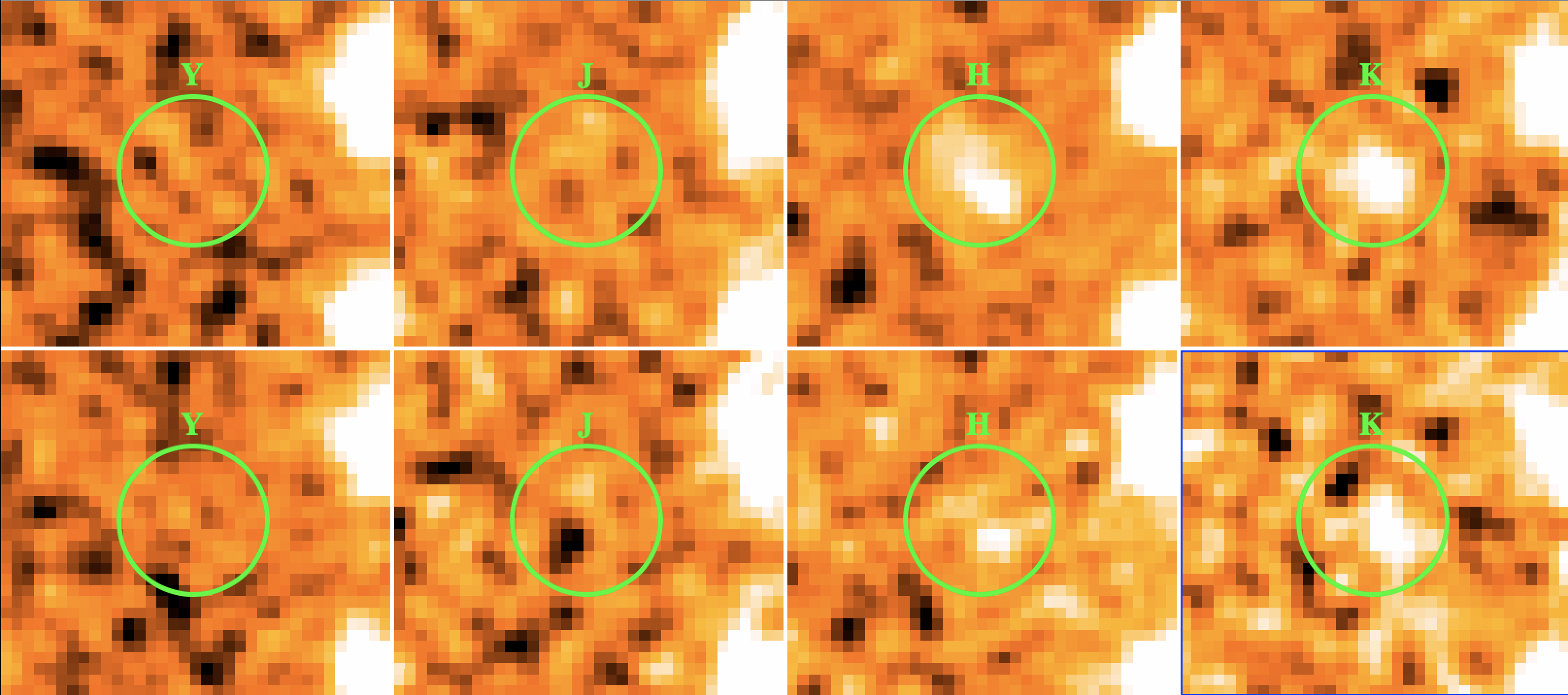}
  \caption{Images of the source ID720309 in the $Y$, $J$, $H$, $K_s$ UltraVISTA broad bands (from left to right), in the full stack (top) and in the stack including only the images not contaminated by the potential cross-talk artefact (bottom). The circles represent the $2$\arcsec{} diameter aperture used in the \classic{} photometry. }
  \label{fig:ID720309}
\end{figure}

The star responsible for the artefact is not included in 80\% of the images taken in the paw~2. Therefore, we created new stacks in $Y$, $J$, $H$, $K_s$ using only these observations. This impacts the sensitivity of our dataset, but it ensures that the source is not affected by the cross-talk. Figure~\ref{fig:ID720309} shows the difference between the images on the full stack (top panels) and the stack not affected by the cross-talk (bottom). The main difference appears in $H$ with a change in the shape of the source. We recomputed the flux using \farmer{} on this new stack. The resulting magnitudes are given in Table~\ref{tab:candidates_mag}, showing fainter magnitude in $H$ and $K_s$, and a non-detection in the $Y$-band.

\section{Resulting galaxy sample}
\label{sec:sample}

Our final galaxy sample comprises $32$ candidates at $z\geq7.5$. Based on the \lephare{}/\farmer{} photometric redshifts, the sample includes $15$ candidates in the range $7.5<z<8.5$, $11$ candidates at $8.5<z<9.5$ and $1$ candidates at $z>9.5$  (some \photoz{} are below $z<7.5$ in the \lephare{}/\farmer{} configuration). 
Tables~\ref{tab:candidates_mag} and \ref{tab:candidates_photoz} present the coordinates, photometry, photometric redshifts and absolute magnitudes of each object. The identifiers are from the \classic{} catalogue, and are always indicated starting with the letters ``ID'' in the following discussion. We report the magnitudes from \farmer{}, corrected for the Milky Way extinction and systematic zero-point offsets. 
Appendix~\ref{sec:appendix_seds_stamps} describes the photometry, the best-fitting templates and the \zPDF{} estimated with \lephare{} together with a detailed discussion for each candidate. 
Figure~\ref{fig:cand_coords} illustrates the coordinates of the identified candidates over the COSMOS field. We emphasise an important aspect of our selection: four candidates are located in the UltraVISTA ``deep stripes''. These areas are one magnitude shallower than the ultra-deep stripes in $YJH$. This makes it more difficult to meet selection criteria requiring an unambiguous three-band detection.

We identify $17$ new unpublished candidates at $z\geq7.5$, many of which are located within the westernmost UltraVISTA ultra-deep stripe (at low right ascension), as shown in Fig.~\ref{fig:cand_coords}. This region is fully covered with the ``deep'' HSC-SSP DR2 survey, about $0.5$\,mag shallower on the outer part of the field compared to the centre. In contrast, the Suprime-Cam data did not cover this part of the field, which made impossible the search of $z>7$ galaxies in this region in previous works, because of the lack of a sufficiently deep optical photometry. 

We find some differences between the total apparent magnitudes between the \classic{} and \farmer{} photometric COSMOS2020 catalogues, as shown in Appendix~\ref{sec:appendix_seds_stamps}. \farmer{} catalogue has the advantage of photometric measurements performed consistently over the full wavelength range. Moreover, \farmer{} measures fluxes more accurately in crowded fields. Therefore, we adopt this photometry by default for our UVLF analysis.

\begin{table*}[t!]
\footnotesize\centering
\renewcommand{\arraystretch}{1.35}
\setlength{\tabcolsep}{6pt} 
\begin{threeparttable}
\caption{Coordinates and observed photometry of the selected $z\geq7.5$ candidates. The first columns indicate the ID and coordinates from the \classic{} catalogue. The other columns give the photometry from \farmer{} catalogue, corrected for Milky Way extinction and systematic zeropoint offsets. The upper-limits correspond to the 3$\sigma$ depth of the image at the source coordinates (10$\sigma$ for IRAC), for sources with $\text{S/N}<1$.
}
\begin{tabular}{ccccccccccc}
 \hline \hline
ID & R.A. & Dec. & ID & $Y$ & $J$ & $H$ & $K_s$ & [3.6] & [4.5] & Flag\tnote{a} \\
\classic{} & [J2000] & [J2000] & \texttt{Farmer} & [mag] & [mag] & [mag] & [mag] & [mag] & [mag] &  \\
 \hline
234500  & 10:02:12.08 & 01:38:20.22 & 153872 & $>24.8$ & $25.76_{-0.12}^{+0.13}$ & $25.60_{-0.09}^{+0.10}$ & $25.27_{-0.09}^{+0.10}$ & $24.67_{-0.03}^{+0.03}$ & $24.22_{-0.02}^{+0.02}$ & ... \\
241443  & 09:58:00.45 & 01:38:46.82 & 228669 & $27.12_{-0.30}^{+0.41}$ & $25.31_{-0.07}^{+0.08}$ & $24.87_{-0.07}^{+0.07}$ & $24.41_{-0.06}^{+0.06}$ & $23.97_{-0.03}^{+0.03}$ & $23.83_{-0.02}^{+0.02}$ & 1 \\
336101  & 10:00:32.32 & 01:44:31.22 & 499535 & $26.83_{-0.13}^{+0.15}$ & $25.68_{-0.05}^{+0.06}$ & $25.28_{-0.05}^{+0.05}$ & $25.59_{-0.09}^{+0.10}$ & $24.74_{-0.02}^{+0.02}$ & $24.44_{-0.02}^{+0.02}$ & ... \\
403992  & 10:01:45.04 & 01:48:28.42 & 214940 & $27.87_{-0.35}^{+0.52}$ & $25.96_{-0.08}^{+0.08}$ & $26.11_{-0.11}^{+0.13}$ & $25.99_{-0.13}^{+0.15}$ & $25.85_{-0.10}^{+0.11}$ & $25.00_{-0.05}^{+0.05}$ & ... \\
441697  & 09:57:39.01 & 01:50:40.05 & 598190 & $>26.7$ & $26.02_{-0.09}^{+0.09}$ & $24.96_{-0.04}^{+0.05}$ & $24.89_{-0.06}^{+0.06}$ & $26.16_{-0.19}^{+0.24}$ & $24.30_{-0.04}^{+0.05}$ & 1 \\
428351  & 10:00:58.48 & 01:49:55.97 & 615402 & $26.79_{-0.21}^{+0.26}$ & $25.27_{-0.07}^{+0.07}$ & $25.51_{-0.11}^{+0.12}$ & $25.26_{-0.09}^{+0.10}$ & $25.01_{-0.03}^{+0.04}$ & $25.07_{-0.04}^{+0.04}$ & DB \\
442053  & 09:58:32.63 & 01:50:43.59 & 711973 & $>25.5$ & $25.95_{-0.20}^{+0.24}$ & $24.96_{-0.11}^{+0.13}$ & $25.08_{-0.09}^{+0.10}$ & $24.82_{-0.04}^{+0.04}$ & $24.54_{-0.03}^{+0.03}$ & DB \\
485056  & 10:00:17.89 & 01:53:14.39 & 184348 & $26.99_{-0.14}^{+0.16}$ & $25.99_{-0.07}^{+0.07}$ & $26.29_{-0.12}^{+0.13}$ & $27.24_{-0.35}^{+0.52}$ & $>25.1$ & $27.03_{-0.18}^{+0.21}$ & ... \\
545752  & 09:57:23.39 & 01:56:45.93 & 758856 & $26.20_{-0.09}^{+0.10}$ & $24.92_{-0.03}^{+0.03}$ & $24.89_{-0.04}^{+0.04}$ & $25.09_{-0.07}^{+0.07}$ & $25.60_{-0.14}^{+0.17}$ & $25.52_{-0.08}^{+0.09}$ & 1 \\
564423  & 10:00:31.87 & 01:57:50.12 & 78629  & $>26.6$ & $25.95_{-0.07}^{+0.07}$ & $25.40_{-0.05}^{+0.06}$ & $25.43_{-0.08}^{+0.08}$ & $26.18_{-0.07}^{+0.08}$ & $24.83_{-0.02}^{+0.02}$ & ... \\
631862  & 09:57:42.84 & 02:01:39.64 & 747154 & $27.27_{-0.22}^{+0.28}$ & $25.89_{-0.08}^{+0.08}$ & $25.59_{-0.08}^{+0.08}$ & $25.42_{-0.09}^{+0.10}$ & $25.27_{-0.04}^{+0.04}$ & $24.80_{-0.03}^{+0.03}$ & ... \\
720309  & 09:59:10.82 & 02:06:41.96 & 859236 & $27.76_{-0.30}^{+0.42}$ & $>26.4$ & $25.12_{-0.05}^{+0.05}$ & $24.88_{-0.05}^{+0.05}$ & $24.36_{-0.02}^{+0.02}$ & $24.06_{-0.02}^{+0.02}$ & ... \\
720309\tnote{b}  & ... & ... & ... & $>26.9$ & $27.89_{-0.59}^{+0.59}$ & $25.53_{-0.1}^{+0.1}$ & $25.1_{-0.1}^{+0.1}$ & ... & ... & ... \\
724872  & 10:02:52.10 & 02:06:57.91 & 729770 & $26.56_{-0.27}^{+0.36}$ & $25.07_{-0.08}^{+0.09}$ & $24.74_{-0.08}^{+0.09}$ & $25.45_{-0.11}^{+0.12}$ & $25.04_{-0.04}^{+0.04}$ & $24.71_{-0.04}^{+0.04}$ & D \\
784810  & 10:01:47.48 & 02:10:15.43 & 749805 & $>26.5$ & $25.72_{-0.06}^{+0.06}$ & $25.69_{-0.07}^{+0.08}$ & $25.82_{-0.11}^{+0.12}$ & $26.02_{-0.09}^{+0.09}$ & $25.16_{-0.04}^{+0.04}$ & ... \\
852845  & 09:58:50.94 & 02:13:55.09 & 371044 & $27.13_{-0.19}^{+0.23}$ & $26.01_{-0.08}^{+0.09}$ & $25.25_{-0.05}^{+0.05}$ & $25.32_{-0.07}^{+0.08}$ & $25.03_{-0.03}^{+0.03}$ & $24.30_{-0.02}^{+0.02}$ & ... \\
859061  & 10:00:19.59 & 02:14:13.28 & 443686 & $>26.6$ & $26.29_{-0.10}^{+0.11}$ & $25.95_{-0.10}^{+0.11}$ & $26.14_{-0.16}^{+0.19}$ & $24.84_{-0.03}^{+0.03}$ & $24.88_{-0.03}^{+0.03}$ & B \\
978062  & 09:57:47.90 & 02:20:43.55 & 764263 & $>26.7$ & $25.07_{-0.04}^{+0.04}$ & $24.71_{-0.03}^{+0.03}$ & $24.72_{-0.05}^{+0.05}$ & $24.60_{-0.03}^{+0.03}$ & $24.20_{-0.02}^{+0.02}$ & ... \\
984164  & 09:57:18.00 & 02:21:05.90 & 774509 & $>26.4$ & $26.01_{-0.08}^{+0.09}$ & $25.53_{-0.07}^{+0.08}$ & $25.46_{-0.09}^{+0.10}$ & $24.53_{-0.20}^{+0.24}$ & $23.48_{-0.10}^{+0.10}$ & 2 \\
1055131 & 09:57:54.25 & 02:25:08.42 & 518572 & $>26.3$ & $25.42_{-0.04}^{+0.04}$ & $25.57_{-0.07}^{+0.07}$ & $25.48_{-0.09}^{+0.10}$ & $25.10_{-0.03}^{+0.04}$ & $24.52_{-0.03}^{+0.03}$ & ... \\
1103149 & 09:57:54.69 & 02:27:54.95 & 71035  & $28.39_{-0.44}^{+0.75}$ & $25.81_{-0.06}^{+0.06}$ & $25.91_{-0.09}^{+0.09}$ & $25.86_{-0.12}^{+0.14}$ & $25.05_{-0.03}^{+0.03}$ & $24.92_{-0.05}^{+0.05}$ & ... \\
1151531 & 10:02:12.54 & 02:30:45.84 & 776124 & $27.99_{-0.38}^{+0.59}$ & $25.03_{-0.03}^{+0.03}$ & $25.15_{-0.05}^{+0.05}$ & $25.82_{-0.13}^{+0.14}$ & $24.91_{-0.03}^{+0.03}$ & $24.13_{-0.02}^{+0.02}$ & ... \\
1209618 & 10:00:47.53 & 02:34:04.50 & 270250 & $>26.8$ & $25.85_{-0.07}^{+0.07}$ & $25.66_{-0.08}^{+0.08}$ & $26.48_{-0.21}^{+0.26}$ & $25.65_{-0.06}^{+0.07}$ & $25.19_{-0.04}^{+0.04}$ & ... \\
1212944 & 10:01:56.33 & 02:34:16.22 & 391218 & $28.09_{-0.40}^{+0.65}$ & $25.94_{-0.08}^{+0.08}$ & $26.22_{-0.13}^{+0.15}$ & $25.98_{-0.14}^{+0.16}$ & $26.13_{-0.10}^{+0.11}$ & $25.17_{-0.04}^{+0.04}$ & ... \\
1274544 & 09:58:12.23 & 02:37:52.34 & 50411  & $25.69_{-0.13}^{+0.14}$ & $25.09_{-0.08}^{+0.09}$ & $25.01_{-0.11}^{+0.12}$ & $24.87_{-0.06}^{+0.07}$ & $23.86_{-0.01}^{+0.01}$ & $23.50_{-0.01}^{+0.01}$ & D \\
1297232 & 09:57:24.53 & 02:39:13.18 & 292281 & $26.77_{-0.13}^{+0.15}$ & $25.16_{-0.04}^{+0.04}$ & $25.08_{-0.05}^{+0.05}$ & $25.38_{-0.08}^{+0.09}$ & $25.03_{-0.10}^{+0.11}$ & $24.59_{-0.09}^{+0.10}$ & 2 \\
1313521 & 09:57:35.64 & 02:40:12.09 & 454321 & $>26.4$ & $25.86_{-0.07}^{+0.08}$ & $25.21_{-0.06}^{+0.06}$ & $25.21_{-0.07}^{+0.08}$ & $24.03_{-0.03}^{+0.03}$ & $23.66_{-0.05}^{+0.05}$ & 2B \\
1346929 & 10:00:30.65 & 02:42:09.10 & 764277 & $>26.7$ & $25.71_{-0.07}^{+0.07}$ & $25.36_{-0.06}^{+0.07}$ & $26.11_{-0.16}^{+0.18}$ & $25.26_{-0.05}^{+0.05}$ & $26.76_{-0.18}^{+0.22}$ & ... \\
1352064 & 09:57:32.07 & 02:42:25.56 & 740295 & $27.11_{-0.19}^{+0.23}$ & $25.67_{-0.06}^{+0.06}$ & $25.41_{-0.06}^{+0.07}$ & $25.39_{-0.08}^{+0.09}$ & $25.88_{-0.18}^{+0.22}$ & $24.42_{-0.07}^{+0.07}$ & 2 \\
1356755 & 09:57:25.45 & 02:42:41.22 & 9881 & $>26.4$ & $26.67_{-0.15}^{+0.15}$ & $24.65_{-0.32}^{+0.32}$ & $24.53_{-0.40}^{+0.40}$ & $23.68_{-0.03}^{+0.03}$ & $23.87_{-0.05}^{+0.05}$ & ... \\
1371152 & 10:00:15.97 & 02:43:32.91 & 247885 & $>26.7$ & $26.24_{-0.10}^{+0.11}$ & $25.50_{-0.07}^{+0.08}$ & $25.50_{-0.10}^{+0.11}$ & $26.41_{-0.16}^{+0.19}$ & $25.74_{-0.09}^{+0.10}$ & ... \\
1409328 & 09:59:17.15 & 02:45:48.22 & 666370 & $26.60_{-0.15}^{+0.18}$ & $25.90_{-0.10}^{+0.11}$ & $25.50_{-0.09}^{+0.10}$ & $25.31_{-0.10}^{+0.11}$ & $25.30_{-0.06}^{+0.06}$ & $24.38_{-0.03}^{+0.03}$ & ... \\
1412106 & 09:57:21.36 & 02:45:57.47 & 331814 & $>26.2$ & $25.30_{-0.06}^{+0.06}$ & $25.47_{-0.08}^{+0.09}$ & $26.12_{-0.20}^{+0.25}$ & $25.21_{-0.11}^{+0.12}$ & $25.29_{-0.17}^{+0.20}$ & 2 \\
\hline
\end{tabular}
\begin{tablenotes}
\item[a] The flags are the following. 1: outside the HSC ultra-deep region, 2: outside the IRAC SPLASH region, B: blended, D: inside the UltraVISTA deep stripe
\item[b] UltraVISTA photometry measured on a new stack not affected by the contribution of a potential cross-talk artefact (see Sect.~\ref{sec:ID720309}). 
\end{tablenotes}
\label{tab:candidates_mag}
\end{threeparttable}
\end{table*}

\begin{table*}[t!]
\footnotesize\centering
\renewcommand{\arraystretch}{1.35}
\setlength{\tabcolsep}{6pt} 
\begin{threeparttable}
\caption{Photometric redshifts estimated with \lephare{} and \eazy{} as in \citet{2021arXiv211013923W}, and with \lephare{} using BC03 templates, for the $z\geq7.5$ candidates. The UV absolute magnitudes are derived from a Monte Carlo sampling of the galaxy \zPDF{}, using \farmer{} and \lephare{}. The three last columns list identifiers from \citetalias{stefanon_brightest_2019} and \citetalias{bowler_lack_2020}, flags\tnote{a} and \specz{} from \citet{Bouwens21_REBELS}.}
\begin{tabular}{ccccccccccc}
 \hline \hline
ID & \multicolumn{6}{c}{$z_\text{phot}$} & $M_\text{UV} $ & B20/S19 & Flag & $z_\text{spec}$ \\
& \multicolumn{2}{c}{\lephare{}} & \multicolumn{2}{c}{\eazy{}} & \multicolumn{2}{c}{\lephare{} BC03} & & & & \\
& \classic{} & \texttt{Farmer} & \classic{} & \texttt{Farmer} & \classic{} & \texttt{Farmer} & & & & \\

\hline
\multicolumn{9}{l}{All criteria satisfied (gold sample)}\\
\hline
234500          &  $9.15_{-1.10}^{+0.83}$        &  $8.20_{-0.35}^{+0.43}$                 &  $9.40_{-0.39}^{+0.36}$     &  $8.14_{-0.22}^{+0.23}$                  &  $7.53_{-4.10}^{+2.20}$       &  $8.48_{-0.48}^{+0.78}$        & $-21.56_{-0.17}^{+0.50}$      & ...     & ...  & ...  \\
336101          &  $7.45_{-0.29}^{+0.42}$        &  $7.51_{-0.09}^{+0.09}$         &  $7.30_{-0.16}^{+0.16}$        &  $7.46_{-0.11}^{+0.07}$               &  $7.69_{-0.37}^{+0.45}$       &  $7.77_{-0.12}^{+0.07}$   & $-21.40_{-0.10}^{+0.11}$         & 213/Y3  & ...  & 7.306/R25  \\
403992          &  $8.98_{-0.87}^{+0.88}$        &  $8.52_{-0.51}^{+0.31}$         &  $8.98_{-0.12}^{+0.31}$        &  $8.59_{-0.58}^{+0.19}$               &  $8.72_{-3.70}^{+1.10}$       &  $8.77_{-0.60}^{+0.74}$        & $-21.20_{-0.19}^{+0.25}$      & 266     & ...  & ...  \\
564423          &  $8.82_{-0.61}^{+0.62}$        &  $9.17_{-0.17}^{+0.15}$         &  $8.92_{-0.17}^{+0.21}$        &  $8.98_{-0.03}^{+0.05}$               &  $8.71_{-0.73}^{+1.00}$       &  $9.83_{-0.19}^{+0.12}$        & $-22.00_{-0.11}^{+0.12}$      & 237/Y5  & ...  & $-$/R24 NIRSpec  \\
631862          &  $8.62_{-0.80}^{+0.67}$        &  $8.51_{-0.85}^{+0.46}$         &  $8.81_{-0.27}^{+0.26}$        &  $7.55_{-0.08}^{+0.10}$               &  $8.68_{-0.93}^{+0.96}$       &  $8.38_{-0.54}^{+1.00}$        & $-21.52_{-0.30}^{+0.31}$      & ...     & ...  & ...  \\
724872          &  $8.03_{-2.20}^{+1.40}$        &  $8.15_{-0.53}^{+0.51}$         &  $8.57_{-1.40}^{+0.53}$        &  $7.64_{-0.07}^{+0.07}$               &  $7.53_{-5.30}^{+1.80}$       &  $8.04_{-0.24}^{+0.66}$        & $-22.12_{-0.38}^{+0.18}$      & ...     & D    & ...  \\
784810          &  $8.37_{-0.55}^{+0.52}$        &  $8.68_{-0.33}^{+0.23}$         &  $8.36_{-0.25}^{+0.29}$        &  $8.79_{-0.13}^{+0.10}$               &  $8.58_{-0.62}^{+0.77}$       &  $9.05_{-0.65}^{+0.72}$        & $-21.63_{-0.16}^{+0.17}$      & 598/Y10 & ...  & ...  \\
852845          &  $8.50_{-1.20}^{+0.82}$        &  $9.16_{-0.21}^{+0.17}$         &  $8.74_{-0.42}^{+0.31}$        &  $8.96_{-0.11}^{+0.10}$                       &  $7.83_{-4.50}^{+1.50}$       &  $9.29_{-1.50}^{+0.54}$          & $-22.08_{-0.13}^{+0.26}$     & ...     & ...  & NIRSpec  \\
978062          &  $8.52_{-0.55}^{+0.44}$        &  $8.86_{-0.36}^{+0.18}$         &  $8.47_{-0.42}^{+0.36}$        &  $8.87_{-0.07}^{+0.06}$               &  $8.55_{-0.51}^{+0.89}$       &  $8.82_{-0.63}^{+0.66}$        & $-22.05_{-0.03}^{+0.03,}$\tnote{c}    & 762/Y1  & ...  & 7.675/R18 NIRSpec  \\
1055131         &  $8.69_{-0.66}^{+0.58}$        &  $8.36_{-0.35}^{+0.28}$         &  $8.86_{-0.16}^{+0.14}$        &  $8.40_{-0.29}^{+0.14}$                       &  $8.72_{-0.90}^{+0.95}$       &  $8.67_{-0.50}^{+0.44}$          & $-21.74_{-0.09}^{+0.10}$     & 839     & ...  & ...  \\
1103149         &  $8.50_{-0.85}^{+0.48}$        &  $7.78_{-0.16}^{+0.32}$         &  $8.72_{-0.19}^{+0.17}$        &  $7.67_{-0.07}^{+0.08}$               &  $8.65_{-0.88}^{+0.82}$       &  $8.25_{-0.38}^{+1.00}$        & $-21.31_{-0.10}^{+0.13}$      & 879     & ...  & 7.369/R19  \\
1151531         &  $8.25_{-0.53}^{+0.46}$        &  $8.32_{-0.37}^{+0.26}$         &  $7.79_{-0.16}^{+0.63}$        &  $8.57_{-0.09}^{+0.07}$               &  $8.49_{-0.55}^{+0.73}$       &  $8.65_{-0.61}^{+0.27}$        & $-22.02_{-0.05}^{+0.05,}$\tnote{c}    & 914/Y2  & ...  & 7.677/R36  \\
1212944         &  $8.08_{-0.88}^{+0.72}$        &  $8.60_{-0.40}^{+0.27}$                 &  $8.12_{-0.43}^{+0.43}$        &  $8.76_{-0.17}^{+0.11}$                 &  $8.21_{-4.70}^{+1.00}$       &  $8.89_{-0.43}^{+0.81}$          & $-21.10_{-0.25}^{+0.28}$     & 953/Y15 & ...  & ...  \\
1274544         &  $7.95_{-0.58}^{+0.75}$        &  $7.36_{-0.12}^{+0.12}$         &  $8.00_{-0.44}^{+0.58}$        &  $7.34_{-0.07}^{+0.06}$               &  $7.99_{-6.10}^{+1.10}$       &  $7.67_{-0.15}^{+0.10}$        & $-21.94_{-0.18}^{+0.20}$      & ...     & D    & ...  \\
1352064         &  $8.09_{-4.20}^{+1.00}$        &  $8.88_{-0.41}^{+0.20}$                 &  $7.65_{-0.42}^{+0.98}$        &  $7.51_{-0.06}^{+0.15}$                 &  $6.78_{-5.30}^{+2.40}$       &  $9.49_{-1.50}^{+0.35}$          & $-21.86_{-0.19}^{+0.38}$     & ...     & 2    & ...  \\
1409328         &  $9.40_{-0.92}^{+0.84}$        &  $8.57_{-1.30}^{+0.57}$                 &  $9.52_{-0.18}^{+0.23}$        &  $8.93_{-1.60}^{+0.11}$                         &  $7.52_{-3.70}^{+2.30}$       &  $7.77_{-0.31}^{+1.80}$          & $-21.43_{-0.55}^{+0.41}$     & ...     & ...  & ...  \\

\hline
\multicolumn{9}{l}{Blended galaxies (gold sample)}\\
\hline

428351          &  $1.41_{-0.24}^{+0.38}$        &  $7.59_{-0.11}^{+0.14}$         &  $1.52_{-0.09}^{+0.12}$   &  $7.55_{-0.05}^{+0.05}$           &  $1.15_{-0.65}^{+0.82}$         &  $7.81_{-0.11}^{+0.74}$        & $-21.79_{-0.13}^{+0.14}$         & 301/Y4 & DB  & 7.090/R27\\
442053          &  $2.47_{-0.81}^{+1.80}$        &  $8.97_{-6.70}^{+0.50}$                 &  $2.50_{-0.59}^{+0.59}$        &  $9.17_{-0.10}^{+0.16}$                 &  $2.22_{-0.92}^{+1.10}$       &  $8.38_{-6.70}^{+1.20}$          & $-22.13_{-0.47}^{+6.39}$     & ... & DB     & ... \\
859061          &  $2.01_{-0.30}^{+1.00}$        &  $7.88_{-0.17}^{+0.28}$         &  $3.38_{-0.21}^{+0.08}$   &  $7.65_{-0.08}^{+0.08}$           &  $2.03_{-0.41}^{+1.40}$         &  $8.12_{-0.27}^{+1.30}$        & $-20.93_{-0.17}^{+0.28}$         & Y11 & B      & ...\\
1313521         &  $2.53_{-0.60}^{+6.60}$        &  $8.07_{-0.32}^{+0.42}$         &  $2.05_{-0.26}^{+0.30}$        &  $1.57_{-0.03}^{+0.03}$               &  $2.27_{-0.88}^{+2.80}$       &  $7.98_{-6.40}^{+0.32}$        & $-21.47_{-0.16}^{+0.21}$      & ... & 2B     & ...\\

\hline
\multicolumn{9}{l}{Potential low-redshift galaxies (based on \eazy{} photometric redshift, gold sample)}\\
\hline

1209618         &  $9.48_{-0.68}^{+0.72}$        &  $8.38_{-0.37}^{+0.34}$         &  $0.07_{-0.03}^{+0.02}$ &  $8.37_{-0.27}^{+0.28}$             &  $8.16_{-2.70}^{+1.90}$         &  $8.39_{-0.34}^{+0.61}$        & $-21.47_{-0.24}^{+0.18}$         & Y8  & ... & ...\\
1371152         &  $8.49_{-1.50}^{+1.10}$        &  $9.33_{-0.20}^{+0.19}$                 &  $8.72_{-0.27}^{+0.30}$        &  $0.39_{-0.01}^{+0.01}$                 &  $7.75_{-5.30}^{+1.70}$       &  $9.72_{-0.25}^{+0.21}$          & $-21.89_{-0.14}^{+0.16}$     & Y12 & ... & ...\\

\hline
\multicolumn{9}{l}Potential low-redshift galaxies (based on \lephare{} photometric redshift)\\
\hline

720309          &  $9.71_{-0.40}^{+0.52}$        &  $10.10_{-0.16}^{+0.26}$         &  $9.60_{-0.27}^{+0.32}$        &  $11.20_{-0.40}^{+0.23}$                 &  $10.00_{-2.10}^{+0.51}$   &  $10.40_{-0.27}^{+0.33}$       & $-22.42_{-0.11}^{+0.12}$      & ...     & ...  & ...  \\
720309\tnote{b}         &  ...   &  $2.89_{-0.43}^{+6.52}$      &  ...   &  ...             &  ...       &  $9.46_{-1.6}^{+0.75}$         & ...  & ...     & ...  & ... \\

1356755 & $9.11_{-6.73}^{+0.41}$ & $2.55_{-0.18}^{+0.19}$ & $9.62_{-0.41 }^{+0.35}$ & $9.46_{-0.05}^{+0.06}$ & $8.89 _{-6.63}^{+0.97}$ & $2.51 _{-0.1}^{+0.25}$ & ... & ... & ... & NIRSpec \\

\hline
\multicolumn{9}{l}{Potential $z<6$ dusty galaxies}\\
\hline

241443          &  $7.09_{-5.00}^{+0.82}$        &  $7.70_{-0.21}^{+0.63}$                 &  $2.09_{-0.37}^{+5.10}$        &  $11.90_{-4.30}^{+0.06}$                         &  $2.23_{-1.10}^{+5.50}$       &  $8.05_{-0.25}^{+0.85}$          & $-21.86_{-0.19}^{+8.77}$     & ... & 1  & ...\\
984164          &  $8.15_{-3.60}^{+0.94}$        &  $8.47_{-0.44}^{+0.38}$         &  $7.84_{-5.60}^{+1.00}$        &  $8.37_{-0.34}^{+0.25}$               &  $4.51_{-3.40}^{+4.40}$       &  $8.37_{-0.33}^{+0.51}$        & $-21.54_{-0.19}^{+0.19}$      & ... & 2  & ... \\
1412106         &  $7.88_{-5.70}^{+1.10}$        &  $8.10_{-0.28}^{+0.37}$                 &  $8.34_{-0.39}^{+0.36}$        &  $7.95_{-0.13}^{+0.25}$                 &  $4.37_{-3.70}^{+4.60}$       &  $8.47_{-0.46}^{+0.65}$          & $-21.81_{-0.11}^{+0.11}$     & ... & 2  & ... \\           

\hline
\multicolumn{9}{l}{Potential stars}\\
\hline
485056          &  $7.81_{-4.10}^{+0.88}$        &  $7.45_{-0.11}^{+0.11}$         &  $8.17_{-0.43}^{+0.40}$        &  $7.34_{-0.08}^{+0.07}$               &  $7.46_{-5.90}^{+1.60}$       &  $7.65_{-0.18}^{+1.10}$        & $-21.03_{-0.15}^{+0.14}$      & 356 & ...  & ...\\
545752          &  $7.79_{-0.31}^{+0.94}$        &  $7.59_{-0.05}^{+0.05}$         &  $7.51_{-0.10}^{+0.09}$  &  $7.54_{-0.03}^{+0.03}$            &  $8.41_{-0.86}^{+0.93}$         &  $7.74_{-0.04}^{+0.04}$  & $-22.11_{-0.06}^{+0.07}$         & ... & 1    & ...\\
1346929         &  $7.83_{-0.35}^{+0.59}$        &  $8.22_{-0.41}^{+0.90}$                 &  $7.67_{-0.11}^{+0.14}$        &  $9.10_{-0.03}^{+0.05}$                 &  $1.23_{-0.74}^{+7.90}$       &  $9.34_{-1.30}^{+0.12}$          & $-21.37_{-0.13}^{+0.14}$     & 1032 & ... & ...\\

\hline
\multicolumn{9}{l}{Potential local noise in HSC DR3 images}\\
\hline
441697          &  $9.10_{-0.67}^{+0.35}$        &  $9.51_{-0.16}^{+0.11}$         &  $9.22_{-0.21}^{+0.18}$        &  $9.50_{-0.09}^{+0.06}$               &  $9.11_{-7.60}^{+0.71}$       &  $9.94_{-0.22}^{+0.09}$   & $-22.47_{-0.09}^{+0.10}$         & ...     & 1    & NIRSpec  \\
1297232         &  $7.72_{-1.10}^{+0.77}$        &  $7.60_{-0.07}^{+0.06}$         &  $7.57_{-0.55}^{+0.74}$        &  $7.62_{-0.04}^{+0.04}$               &  $7.79_{-6.50}^{+0.91}$       &  $7.80_{-6.20}^{+0.06}$        & $-21.91_{-0.07}^{+0.08}$      & ... & 2  & ...\\

\hline
\end{tabular}
\begin{tablenotes}
\item[a] The flags are the following. 1: outside the HSC ultra-deep region, 2: outside the IRAC SPLASH region, B: blended, D: inside the UltraVISTA deep stripes.
\item[b] \lephare{} \photoz{} computed using the UltraVISTA photometry uncontaminated by the cross-talk (see Sect.~\ref{sec:ID720309}).
\item[c] Absolute magnitudes computed using the ALMA \specz{}. We report the error associated with the $H$-band apparent magnitude.
\end{tablenotes}
\label{tab:candidates_photoz}
\end{threeparttable}
\end{table*}

\subsection{Robustness of the selected candidates} 
\label{sec:robustness}

We produce several tests to establish the robustness of our candidates. Based on this step, we define a gold sample of 22 sources satisfying all our criteria as $z>7.5$ galaxies (see Table~\ref{tab:candidates_photoz}). But we cannot exclude a possible contaminant for the remaining sources, based on these additional criteria.\\

First, we compare several \photoz{} estimates to assess the sensitivity of our selection to the template-fitting approach. Table~\ref{tab:candidates_photoz} presents the six photometric redshift estimates for each candidate, computed using the three different template-fitting procedures (\lephare{}, \eazy{} and \lephare{} BC03) applied to both the \classic{} and \farmer{} catalogues, as described in Sect.~\ref{sec:photoz_selection} and \ref{sec:photoz_confirmation}. All template-fitting codes consolidate our selection for the first 16 candidates listed in Table~\ref{tab:candidates_photoz}. They represent the core of our gold sample. We find some discrepancies for the remaining candidates, separated by categories in Table~\ref{tab:candidates_photoz}.

{\it Blended galaxies:} for four of these candidates, the three \photoz{} estimates with the \classic{} catalogue are at $z<3$. These sources are flagged as blended in the images (flag B in Table~\ref{tab:candidates_photoz}), as a bright foreground source is clearly identified and contaminates the $2\arcsec$ aperture. Based on the \classic{} aperture photometry, all four of these candidates show a $3\sigma$ detection in at least one HSC optical band. The flux captured in the fixed aperture present a spatial offset. The \classic{} redshift probability distributions peak at $z\sim2$ for all candidates, although one of them presents a secondary $z>7$ solution. In contrast, the majority of the \zPDF{} weights are located at $z>7$ with \farmer{} catalogue. It is precisely in these cases that \farmer{} photometry is expected to be more reliable than the \classic{} aperture photometry, since these candidates are well deblended thanks to the profile-fitting photometry. Therefore, we consider these estimates with \farmer{} as robust and we include these candidates in our gold sample. We note that ID859061 falls in the CANDELS region and \citet[][]{stefanon_brightest_2019} confirmed the lack of emission in the visible from HST data (see their Figure~5).

{\it Low-redshift galaxies based on \eazy{} \photoz{}:} for two sources, we find one solution at $z<3$ using \eazy{}. Since \eazy{} produces a high-redshift solution in either the \classic{} or \farmer{} catalogue, we decide to consider these candidates as robust and keep them in our gold sample.

{\it Low-redshift galaxies based on \lephare{} \photoz{}:} \eazy{} points towards a robust $z>9$ solution with a single peak for ID1356755. However, all the solutions obtained with \lephare{} present a significant peak at $z<3$, including more than $30\%$ of the \zPDF{}. Hence, we do not include this candidate in our gold sample. The configuration \farmer{}/\lephare{} is our reference to derive the UVLF, but corresponds to a \photoz{} of $2.55_{-0.18}^{+0.19}$. Therefore, we do not indicate the corresponding absolute magnitude in Table~\ref{tab:candidates_photoz} and this source is not used to derive the UVLF. For the source ID720309, all \photoz{} indicate a $z>9$ solution with the original COSMOS2020 photometry. However, our photometry could be contaminated by a cross-talk artefact (see Sect.~\ref{sec:ID720309}). We recompute the \photoz{} with \lephare{} using the new stack not contaminated by the cross-talk. The redshifts which minimise the $\chi^2$ are $9.46_{-6.59}^{+0.05}$ and $9.47_{-0.18}^{+1.53}$ for the COSMOS and BC03 templates, respectively. We obtain $2.89_{-0.43}^{+6.52}$ and $9.46_{-1.6}^{+0.75}$ using the median of the \zPDF{}. This galaxy remains a high-redshift candidate in our sample, but with a significant peak in the \zPDF{} at $z\sim 2.8$. 

{\it Dusty star-forming galaxies at intermediate redshift:} we checked that none of the selected candidates has $H-[3.6]>2$ which would indicate a low-redshift galaxy with strong Balmer-break or high dust obscuration \citep{oesch_probing_2013}. None of the candidates has coincident \textit{Spitzer}/MIPS $24$\,$\mu$m emission nor at longer wavelengths in the COSMOS Super-deblended catalogue \citep{jin_super-deblended_2018}, and are thus unlikely to be contaminated by low-redshift dusty galaxies. We also checked that none of these sources were detected in SCUBA2 \citep{Simpson19} at less than $6\arcsec$. Three of the candidates present a lower redshift $z<6$ solution when fitted using the \lephare{} BC03 configuration. In this case, we allow for a possible attenuation reaching $A_V\sim 8$, which could reveal a possible solution associated with a dusty star-forming galaxies. These three candidates have also secondary peaks in their \zPDF{} using other template-fitting configurations. They are located outside the ultra-deep HSC region, and outside the SPLASH coverage. Their mid-infrared photometry relies on S-COSMOS, which is about $1$\,mag shallower than in the centre of the field. The use of shallower data may explain the lower constraint on the \zPDF{} resulting in multiple peaks. In addition, the width of the \zPDF{} is systematically larger when using the \classic{} catalogue compared to \farmer{}, because of larger flux errors for the same objects (as explained in \citealt{2021arXiv211013923W}). While our fits favour a high-redshift solution for these sources, we flag them as possible intermediate redshift contaminants and we do not include them in our gold sample.\\

Secondly, we also assess the robustness of the star-galaxy classification. We already rejected all candidates with a better $\chi^2$ using the brown dwarf templates than with the galaxy templates. However, such a sharp cut does not quantify the risk of degeneracy in the classification. Therefore, we generate 500~realisations of the candidates, and for each realisation we add noise to the observed flux corresponding to the photometric uncertainty in each band\footnote{We note that even if the source is undetected in a given image, a flux is always provided with an associated uncertainty at the position indicated in the combined \texttt{CHI\_MEAN} image.}. We compute $\Delta \chi^2$ for each realisation. The analysis of these 500~realisations allows us to quantify the robustness of the star-galaxy classification for a given source. For three sources ID485056, ID545752, ID1346929, we find that 31\%, 28\%, 23\% (27\%, 33\%, 43\%) of the $\Delta \chi^2$ distribution falls at $\Delta \chi^2<0$ with \farmer{} (\classic{}). Therefore, these sources present a significant probability to be brown dwarf contaminants, and we do not include them in our gold sample.\\

Thirdly, we inspect the HSC-SSP DR3 images in $g$, $r$, $i$ and $z$ for the 32 candidates. The increase in depth for DR3 compared DR2 is minor in the COSMOS field \citep[][]{Aihara21}. DR3 correspond to the final images taken on COSMOS and the sky subtraction has been improved in the SSP pipeline. In the $z$-band image at the position of each of the 32 candidates, we measure the photometric flux in a $2$\arcsec{} diameter aperture and the corresponding sky flux in a $2-4$\arcsec{} diameter annulus using the python package \texttt{photutils}. To minimise the impact of objects in the sky annulus, we compute a sigma-clipped mean value. The noise per pixel is computed by aggressively detecting and removing all objects in the postage-stamp around the object and computing the standard deviation of the remaining pixels. We note that the precise signal-to-noise (hereafter S/N) values reported here are dependent on the exact details of object thresholding and background computation. However, we have verified that the conclusions presented below are largely robust to the exact parameter choice. In addition to the S/N values presented below, we examine both the pixels present in the measurement aperture and in the sky annulus around the object. Six candidates have $\text{S/N}>1$. One of them is already identified as potential $z<6$ dusty galaxy (ID241443). Two galaxies (ID1297232 and ID441697) have a $z$-band S/N of 1.4 and 2.7, respectively, and are listed separately in Table~\ref{tab:candidates_photoz}. The three remaining $\text{S/N}>1$ candidates are not considered since the bright pixels are shifted by $1$\arcsec{} with the expected source position, or the signal comes from only one pixel.\\

To summarise, the first 16 candidates satisfy all the criteria to be selected as $z>7.5$ galaxies. Six additional candidates present discrepancies between the different \photoz{} procedures, explained by blending in the \classic{} catalogue and one discrepant \photoz{} in the \eazy{} run. We still consider them as robust and include them in our gold sample. For the last 10 candidates in Table~\ref{tab:candidates_photoz}, there is a significant probability that these sources are either intermediate redshift galaxies or brown dwarf contaminants. 

\subsection{Comparison with $z>7.5$ candidates from the literature}
\label{sec:candidate_ancillary}

We find six~candidates with a match in the A$^3$COSMOS ALMA catalogue\footnote{https://sites.google.com/view/a3cosmos/data} \citep{Liu19_A3COSMOS} from the Reionisation Era Bright Emission Line Survey \citep[REBELS,][]{Bouwens21_REBELS}. Five of them present [CII]$_{ 158\mu m }$ line detection with a spectroscopic redshift from Schouws et al. (in prep.) provided in Table~\ref{tab:candidates_photoz}. One of them was independently confirmed using a detection of the Lyman-$\alpha$ emission line \citep[][]{Valentino22}. Only one ALMA source (ID564423/REBELS-24) does not present a [CII]$_{ 158\mu m }$ line detection, either because of too weak SFR, or observations which are still missing for this source \citep[][]{Bouwens21_REBELS}. These five \specz{} confirm the high-redshift nature $z>7$ of our candidates. We find a systematic over-estimate of the \photoz{}. For the \classic{} catalogue, these \specz{} are consistent with the \zPDF{} (with a PIT\footnote{Probability Integral Transform \citep[PIT, ][]{Dawid84} defined as ${\rm PIT} = \int^{z_{{\rm s}}}_0 {\mathcal P}(z) \, {\rm d} z.$} of 0.3, 0.05, 0.10, 0.13 for ID336101, ID978062, ID1103149, ID1151531, respectively). 
For \farmer{} catalogue, their PIT is below 0.02 and the uncertainties for these five sources are underestimated. While this comparison could suggest a systematic over-estimate of the \photoz{}, we note that these sources could be really specific, which could explain their detection of [CII]$_{ 158\mu m }$ with ALMA. Therefore, we do not extrapolate this trend to the full population at this stage. In Sect.~\ref{sec:UVLF}, we replace the \photoz{} of these five sources by their \specz{}.\\

We compare our candidates to those previously identified by \citetalias{stefanon_brightest_2019} and \citetalias{bowler_lack_2020} in the COSMOS field based on ground-based imaging. 
\citetalias{stefanon_brightest_2019} (see also \citealt{stefanon_hst_2017}) used the near-infrared broad and narrow bands from UltraVISTA DR3, all the available CFHT/MegaCam, Subaru/Suprime-Cam optical bands and \textit{Spitzer}/IRAC channels 1 to 4 in the mid-infrared. This study also benefited from HST/WFC3 coverage from the Drift And SHift mosaic (DASH; \citealt{momcheva_3d-hst_2016,mowla_cosmos-dash_2019}), with an improved spatial resolution. The HSC imaging was not available at the time of their initial sample selection but the authors used the HSC-SSP DR1 data to validate their candidate list, checking that there was no significant detection in the optical. \citetalias{stefanon_brightest_2019} identified $16$ galaxy candidates in the COSMOS field, including $10$ at $z\sim8$ and $6$ at $z\sim9$. 

\citetalias{bowler_lack_2020} included
the UltraVISTA DR4 data and the \textit{Spitzer}/IRAC $[3.6]$ and $[4.5]$ images from SPLASH, SEDS, and SMUVS in the infrared. Optical data consisted of the CFHT/MegaCam $u^*,g,r,i,z$ broad bands from CFHTLS, and the HSC-SSP DR1 $g,r,i,z,y$ broad bands. The Suprime-Cam $z'$ band, deeper than the HSC/$z$ band in DR1 release, was also used. The search was performed in the HSC ultra-deep area. \citetalias{bowler_lack_2020} recovered seven candidates from \citetalias{stefanon_brightest_2019} and selected nine new candidates for a total of $16$ LBG at $7.5<z<9.1$ in the COSMOS field, including $14$ at $z\sim8$ and $2$ at $z\sim9$. 

We select $15$ out of the $25$ high-redshift candidates from \citetalias{stefanon_brightest_2019} and \citetalias{bowler_lack_2020}. The majority of these candidates present single-peaked redshift probability distributions located at $z>7$, and are classified as galaxies in both COSMOS2020 catalogues. The identifiers from \citetalias{stefanon_brightest_2019} and \citetalias{bowler_lack_2020} are shown in Table~\ref{tab:candidates_photoz}. The $10$ rejected candidates from \citetalias{stefanon_brightest_2019} and \citetalias{bowler_lack_2020} are described in details in Appendix~\ref{sec:appendix_rejected_candidates}. Four of those candidates include strong low-redshift solutions, and six are not detected in the combined $izYJHK_s$ image. Since these galaxies are not expected to be visible in the $i$ and $z$ bands, the signal may be diluted in the combined $izYJHK_s$ detection image. To test the impact of such approach, we matched these six sources with the official release catalogues from UltraVISTA DR4 \citep[][]{mccracken_ultravista_2012} with a detection performed in each VIRCAM band. These sources do not have any counterpart in the $K_s$ catalogue. One source from \citetalias{stefanon_brightest_2019} is detected only in the $H$ selected catalogue (but not identified in COSMOS2020 or \citetalias{bowler_lack_2020} because of the proximity of a bright source $J=20.7$). 
Another source from \citetalias{stefanon_brightest_2019} is only detected in the $J$-band selected catalogue with a magnitude of $J=26.3\pm0.1$. At best, we could have retrieved one of these six candidates with a detection performed in each VIRCAM band, rather than our combined $izYJHK_s$ detection image.  \\

Finally, we checked the $z\geq7$ galaxy candidates identified in CANDELS \citep{bouwens_uv_2015}. None of these sources are in our sample considering our selection criteria at $z\geq7.5$. Although deeper, CANDELS covers a smaller area of $151.9$\,arcmin$^2$, which is less efficient when selecting the brightest sources at $z>7.5$. 

\subsection{Lensing magnification}
\label{sec:magnification_candidates}

Any high-redshift galaxy sample may be subject to gravitational lensing, and in particular lensing magnification, from massive, low-redshift galaxies \citep[e.g.][]{mason_correcting_2015,barone-nugent_impact_2015,roberts-borsani_z_2016}. While gravitational lensing preserves surface brightness, the apparent solid angle of the background source may increase, leading to an increased apparent flux. 

We investigate the possible impact of lensing magnification on our selected galaxy candidates. We search the full COSMOS2020 catalogue for massive low-redshift galaxies within a $20\arcsec{}$ radius for each candidate galaxy\footnote{We identify lenses leading to magnifications $\mu\geq1.1$ up to a $15\arcsec{}$ angular distance.}. Lens galaxies are modelled as singular isothermal spheres (SIS). 
Stellar velocity dispersions are estimated from photometry through the Faber-Jackson relation \citep[FJR;][]{faber_velocity_1976} of \citet{barone-nugent_impact_2015}, based on the rest-frame $B$-band absolute magnitude and calibrated using early-type galaxies with redshifts spanning $0<z<1.6$.
Velocity dispersion uncertainties are dominated by the intrinsic scatter in the FJR, estimated at $46$\,km\,s$^{-1}$. 
We checked that these velocity dispersion estimates are consistent with the FJR of \citet{bernardi_early-type_2003} using $i$-band absolute magnitudes. 
Photometric redshifts and absolute magnitudes in the rest-frame Suprime-Cam/$B$ are taken from \farmer{} catalogue. We restrict the lens selection to galaxies with a velocity dispersion of at least $\sigma_v=200$\,km\,s$^{-1}$, because the spectroscopic samples used to calibrate the FJR become incomplete at lower values \citep{barone-nugent_impact_2015}. In addition, we only include lenses with a magnification of $\mu\geq1.1$. 

We find that eight candidates out of $32$ are probably magnified with $1.1<\mu<1.2$, and $5$ galaxies with $\mu\geq1.2$. This includes five already identified candidates from \citetalias{stefanon_brightest_2019} and \citetalias{bowler_lack_2020}. However, we find no evidence of strongly lensed galaxies with multiple images. 
The most magnified candidate is ID441697, with a cumulative magnification of $\mu=2.36\pm0.80$ from five lenses within $14\arcsec{}$, representing a boost of $0.9$\,mag. The main contribution comes from a $z=0.50_{-0.01}^{+0.01}$ galaxy located within $4.6\arcsec$ and with a $\sigma_v=231$\,km\,s$^{-1}$ velocity dispersion, leading to a $\mu=1.37\pm0.20$ magnification. 
The second most magnified candidate is ID442053 because of two sources, one $z=1.50_{-0.04}^{+0.03}$ galaxy at $6.0\arcsec$ with $\sigma_v=218$\,km\,s$^{-1}$ and $\mu=1.14\pm0.07$, and one $z=1.74_{-0.04}^{+0.04}$ galaxy at $4.0\arcsec$ with $\sigma_v=207$\,km\,s$^{-1}$ and $\mu=1.17\pm0.09$. 
The candidate Y8 (ID1209618) has a $z=1.26_{-0.02}^{+0.02}$ galaxy at a $4.4\arcsec$ distance with a velocity dispersion of $\sigma_v=215$\,km\,s$^{-1}$, which gives a $\mu=1.21\pm0.11$ magnification. This candidate was already identified as magnified by \citetalias{stefanon_brightest_2019}, who found a $40$\,km\,s$^{-1}$ higher velocity dispersion and a $0.3\arcsec$ smaller angular separation\footnote{We find an astrometric shift of $0\farcs34$ in declination compared to \citetalias{stefanon_brightest_2019} for the candidate Y8 (ID1209618). This is more than the shift of about $0\farcs15$ induced by the change of astrometric calibration from UltraVISTA DR3 to DR4. We note that this candidate remains faint with $H>25.6$, which may have led to uncertain coordinates in \citetalias{stefanon_brightest_2019} who used DR3.}, resulting in a higher magnification $\mu=1.39$. 
Three other candidates (Y1, Y10, Y12) from \citetalias{stefanon_brightest_2019} are moderately magnified, with $1.1<\mu<1.2$.

Consequently, every object in our $z\geq7.5$ sample is potentially affected by lensing magnification. However, as we shall see in Sect.~\ref{sec:magnification}, the impact of this magnification remains limited on our luminosity function measurements. 

The velocity dispersion values quoted above are extracted from the FJR  and their uncertainties reach $46$\,km\,s$^{-1}$ due to the scatter of the relation \citep{barone-nugent_impact_2015}. Moreover, the velocity dispersion may be overestimated, because of the calibration based on early-type galaxies, leading to overestimated magnifications. Inversely, we note that the velocity dispersion lower limit of $\sigma_v=200$\,km\,s$^{-1}$ significantly restricts the number of lens galaxies, so that the reported total magnifications may be underestimated. 
This will need to be further investigated with more precise velocity dispersion estimates at fainter luminosities.

\section{The UV luminosity function}
\label{sec:UVLF}

In this section, we use our $z\geq7.5$ candidates to make a new measurement of the bright end of the UVLF. We select objects brighter than $H=25.6$, equivalent to the $5\sigma$ depth in the ultra-deep stripes. We use the $0.812$\,deg$^2$ of the HSC-masked UltraVISTA ultra-deep area, so we do not include the $4$ candidates selected in the deep area. This ensures a homogeneous selection function across the field. We take the physical parameters estimated with the \lephare{}/\farmer{} configuration. This is necessary because the blended candidates are not included in the \classic{} selection. 

We split our galaxy sample into three redshift bins, centred at $z=8,9,10$ with a $\Delta z=1$ width. Since the photometric redshift estimates mainly rely on infrared broad-band imaging, the redshift probability distributions are relatively broad in the interval $7<z<10$, and so are the photometric redshift uncertainties. As a result, the candidates with photometric redshifts at the limit between two adjacent bins may be scattered in one bin or the other. Therefore, we develop a Monte Carlo simulation to propagate the photometric uncertainties into the absolute magnitude estimates. We generate 500~realisations of the high-redshift candidates. For each realisation, we add noise to the observed flux, according to the photometric uncertainty measured in the considered band. We base our estimate of the UVLF on the 500 catalogues for which the \photoz{} and absolute magnitudes have been derived.

We adopt the \specz{} derived with REBELS, when possible. As a result, two sources among the five are included in our first redshift range, for which we recompute the FUV absolute magnitudes.

\subsection{Completeness correction}
\label{sec:completness}

To estimate the completeness of our sample of high-redshift galaxies, we simulate point-like sources with a Moffat profile \citep[][]{Moffat69} with parameter $\beta=3$ and a FWHM of $0.9\arcsec$. We generate these objects with a uniform distribution in magnitude over $22<H<27$ and in position over the ultra-deep stripes area and add them to the $J,H,K_s$ bands, avoiding placing objects on masked areas or where object are present in the segmentation map derived from the \texttt{CHI\_MEAN} image (see Sect.~\ref{photometry}). We then generate a new \texttt{CHI\_MEAN} detection image from these simulated $J,H,K_s$ images combined with the original $i,z,Y$ images. Object detection is repeated using the same parameters as the main catalogue and using this new \texttt{CHI\_MEAN} image. The resulting catalogue is cross-correlated with the input simulated catalogue to the find the fraction of recovered sources as a function of magnitude. 

We must assume colours for the simulated galaxies. We use \lephare{} and the BC03 templates to predict the expected colours. For $E(B-V)=0.1$, we find average colours of $J-H=0.05$, $0.49$, $4.94$ at $z=8$, $9$, $10$, respectively, and $K_s-H=0.00$, $-0.10$, $-0.09$ at $z=8$, $9$, $10$, respectively. The predicted $K_s-H$ colours do not vary by more than $0.25$\,mag at $z>7$, depending on the assumed attenuation. We simplify the simulation by assuming $J=H=K_s$ for galaxies at $z\sim 8$, and $H=K_s$ for galaxies at $z\sim 9$ and $10$, with no flux contribution in the other bands. 

We estimate the completeness as a function of the $H$-band magnitude, defined as the fraction of recovered sources in the \texttt{CHI\_MEAN} image. At $z\sim 8$, we find a drop in completeness from $84\%$ to $60\%$ between $H=25$ to $25.6$, the latter being our selection limit. At $z\sim 9$ and $10$, the completeness drops from $72\%$ to $37\%$ in the same magnitude range. 

We do not attempt to correct for contamination or incompleteness in our selection method, considering that possible biases in the redshift estimate (as shown by the comparison with ALMA) dominate the uncertainty budget and  cannot be captured by a simulation. Moreover, the use of the \zPDF{} to generate multiple realisations of the UVLF would make such correction difficult, since a single source is split in several redshift bins.

\subsection{Binned luminosity function}
\label{sec:binned_LF}

Absolute UV magnitudes are computed with \lephare{} in the GALEX far-UV filter\footnote{With a
central wavelength of $1526$\,\AA{} and a full-width at half maximum of $224$\,\AA{}.}, as follows:
\begin{equation}
M_\text{UV} =\; m_{f} - DM(z) - \text{KC}(z,\text{SED}) 
,\end{equation}
with $m_f$ being the observed magnitude in the filter~$f$, $DM$ the distance modulus. KC is the sum of the k-correction $k_{f}$ and a rest-frame colour term, all derived from the best-fit SED. The filter $f$ is chosen among $J, H, K_s, [3.6]$ to minimise the SED dependency of the KC term \citep{ilbert05}.
For each source, we list in Table~\ref{tab:candidates_photoz} the median of the $M_\text{UV}$ distribution, as well as the associated 0.16 and 0.84 quantiles. 

The binned luminosity function is calculated using the $V_\text{max}$ estimator \citep{schmidt_space_1968}. This estimator is non-parametric, although the number of bins and the bin widths are set, and implicitly assumes a uniform spatial distribution of galaxies. The number density in a given magnitude bin depends on the maximum volume $V_\text{max}$ in which each galaxy could have been selected. This volume, for a given galaxy $i$, is computed as:
\begin{equation}
V_{\text{max},i} = \int_\Omega\int_{z_{\text{min},i}}^{z_{\text{max},i}} \dfrac{dV}{d\Omega dz} d\Omega dz,
\end{equation}
where $z_{\text{min},i}$ and $z_{\text{max},i}$ are the lower and upper redshift limits in which a galaxy $i$ can be included in the sample, and $dV$ is the differential co-moving volume. The co-moving volume is defined as the shell between the limits of the redshift bins. The maximum redshift $z_{\text{max},i}$ is the redshift at which a galaxy with the same intrinsic properties would not be selected in our sample, with a flux limit set to $H=25.6$. 
Thus, the luminosity function $\phi$ can be expressed as:
\begin{equation}
\phi(M)\Delta M = \frac{1}{N_{\rm real}}\sum_{i=1}^{N(M)} \dfrac{ w_i}{V_{\text{max},i}},
\end{equation}
where $N(M)$ is the number of galaxies in the magnitude bin centred at $M$ and of width $\Delta M$, considering all the $N_{\rm real}=500$ realisations of the catalogues.  $w_i$ is the inverse of the completeness estimated at the  $H$-band magnitude of the galaxy $i$. The associated Poisson uncertainties $\sigma_\phi$ are computed as \citet{marshall_evolution_1985}:
\begin{equation}
\sigma_\phi(M)\Delta M = \sqrt{\frac{1}{N_{\rm real}} \sum_{i=1}^{N(M)} \dfrac{ w_i}{V_{\text{max},i}^2}}
.\end{equation}
The total uncertainties are computed as the quadratic sum of the Poisson errors and cosmic variance errors, estimated following \citet{trenti_cosmic_2008}. 
%
Since the galaxy samples become incomplete at faint magnitudes, the LFs are computed brightward of $M_\text{UV}=-21.5$.

\begin{figure}
  \centering
  \includegraphics[width=\hsize]{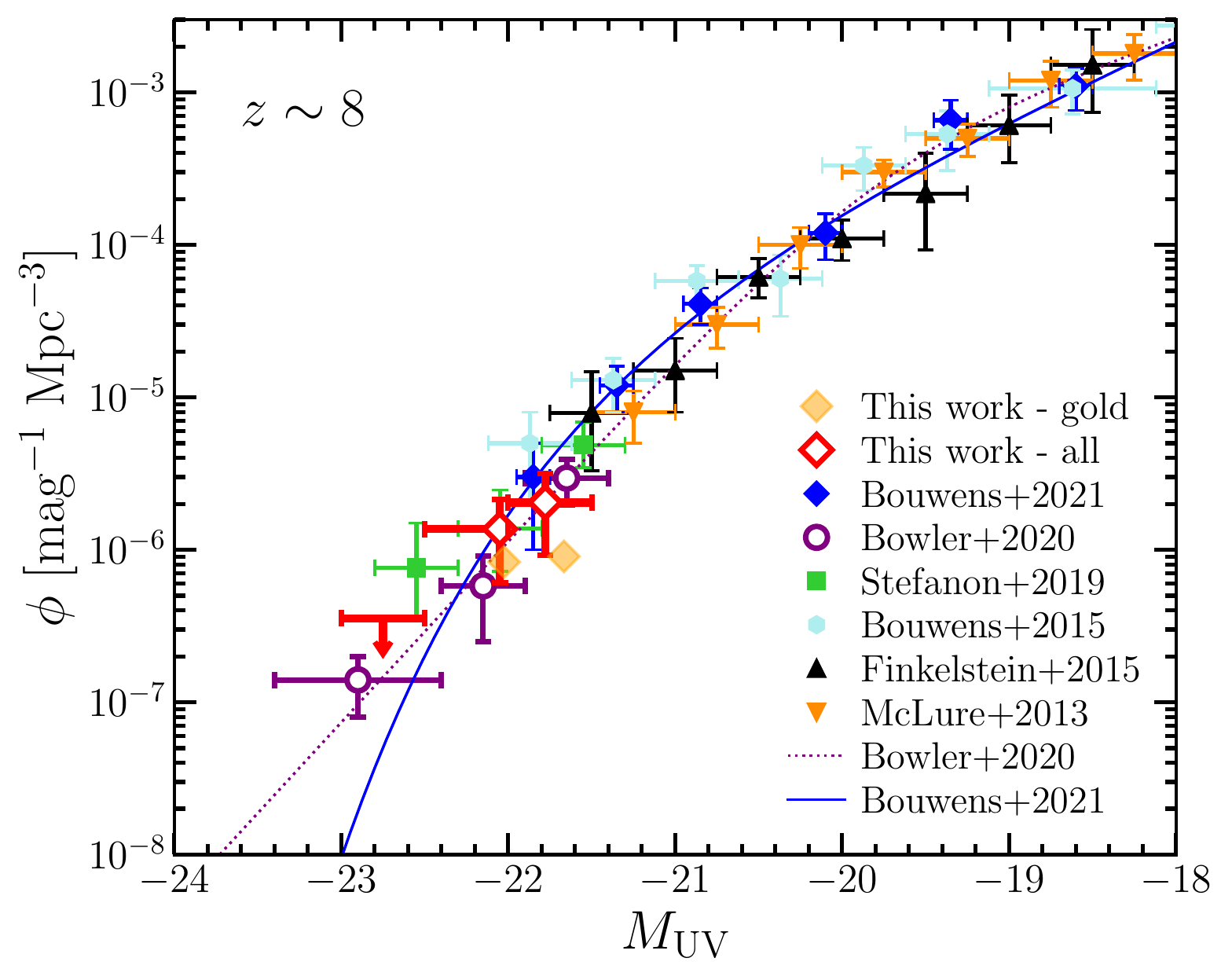}
  \caption{Galaxy UV luminosity functions at $z=8$. The red symbols show number densities from the galaxy sample presented in this work (uncorrected for incompleteness). The upper limits are at $1\sigma$. The orange diamonds correspond to the measurement only based on the gold sample. The data points from  \citetalias{bowler_lack_2020,stefanon_brightest_2019}; \citet{bouwens_2021,bouwens_uv_2015,finkelstein_evolution_2015,mclure_new_2013} are represented. The best-fitting double power-law function from \citetalias{bowler_lack_2020} and the Schechter function from \citet{bouwens_2021} are displayed. }
  \label{fig:LF_z8}
\end{figure}

\begin{figure}
  \centering
  \includegraphics[width=\hsize]{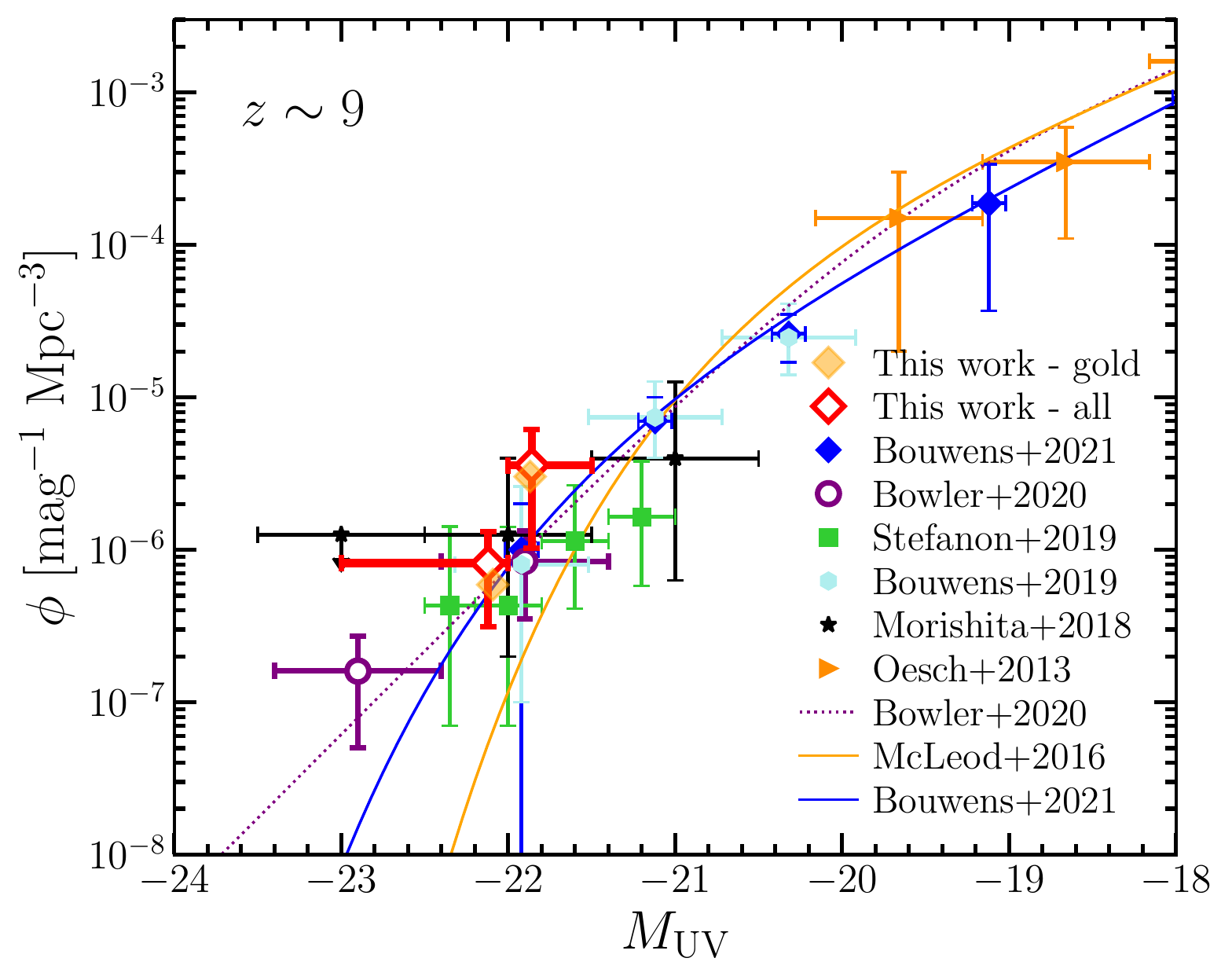}
  \caption{Galaxy UV luminosity functions at $z=9$ as in Fig.~\ref{fig:LF_z8}. Data points from \citetalias{bowler_lack_2020,stefanon_brightest_2019}; \citet{bouwens_2021,bouwens_newly_2019,morishita_bright-end_2018,oesch_probing_2013} are shown. The best-fitting double power-law function from \citetalias{bowler_lack_2020} and the Schechter function from \citet{mcleod_z_2016} are displayed. }
  \label{fig:LF_z9}
\end{figure}

\begin{figure}
  \centering
  \includegraphics[width=\hsize]{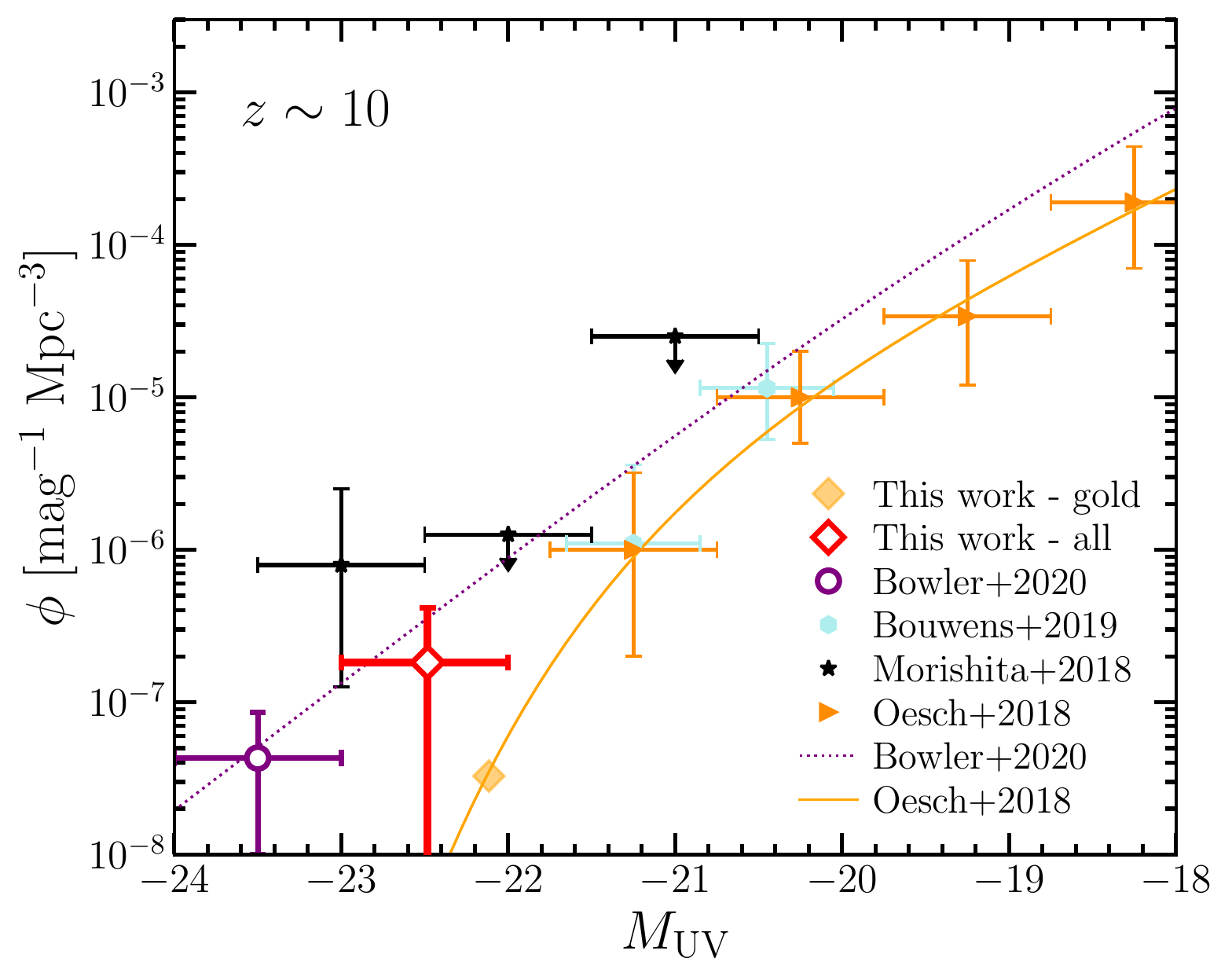}
  \caption{Galaxy UV luminosity functions at $z=10$, as in Fig.~\ref{fig:LF_z8}. Data points from \citetalias{bowler_lack_2020}; \citet{bouwens_newly_2019,morishita_bright-end_2018,oesch_dearth_2018} are shown. The best-fitting double power-law function from \citetalias{bowler_lack_2020} and the Schechter function from \citet{oesch_dearth_2018} are displayed. }
  \label{fig:LF_z10}
\end{figure}

The galaxy UV luminosity functions at $z=8,9,10$ estimated from the selected galaxy candidates are represented in Fig.~\ref{fig:LF_z8}, \ref{fig:LF_z9}, and \ref{fig:LF_z10}, and tabulated in Table~\ref{tab:LF_data}. In magnitude bins with no galaxies, we put an upper limit with the number density computed from one galaxy. 
Cosmic variance represents about $14$\,\% of the Poisson uncertainties at $M_\text{UV}<-23$, $20$\,\% at $M_\text{UV}=-22.75$, and $30$\,\% at $M_\text{UV}=-22.25$. 

\begin{figure}
  \centering
  \includegraphics[width=\hsize]{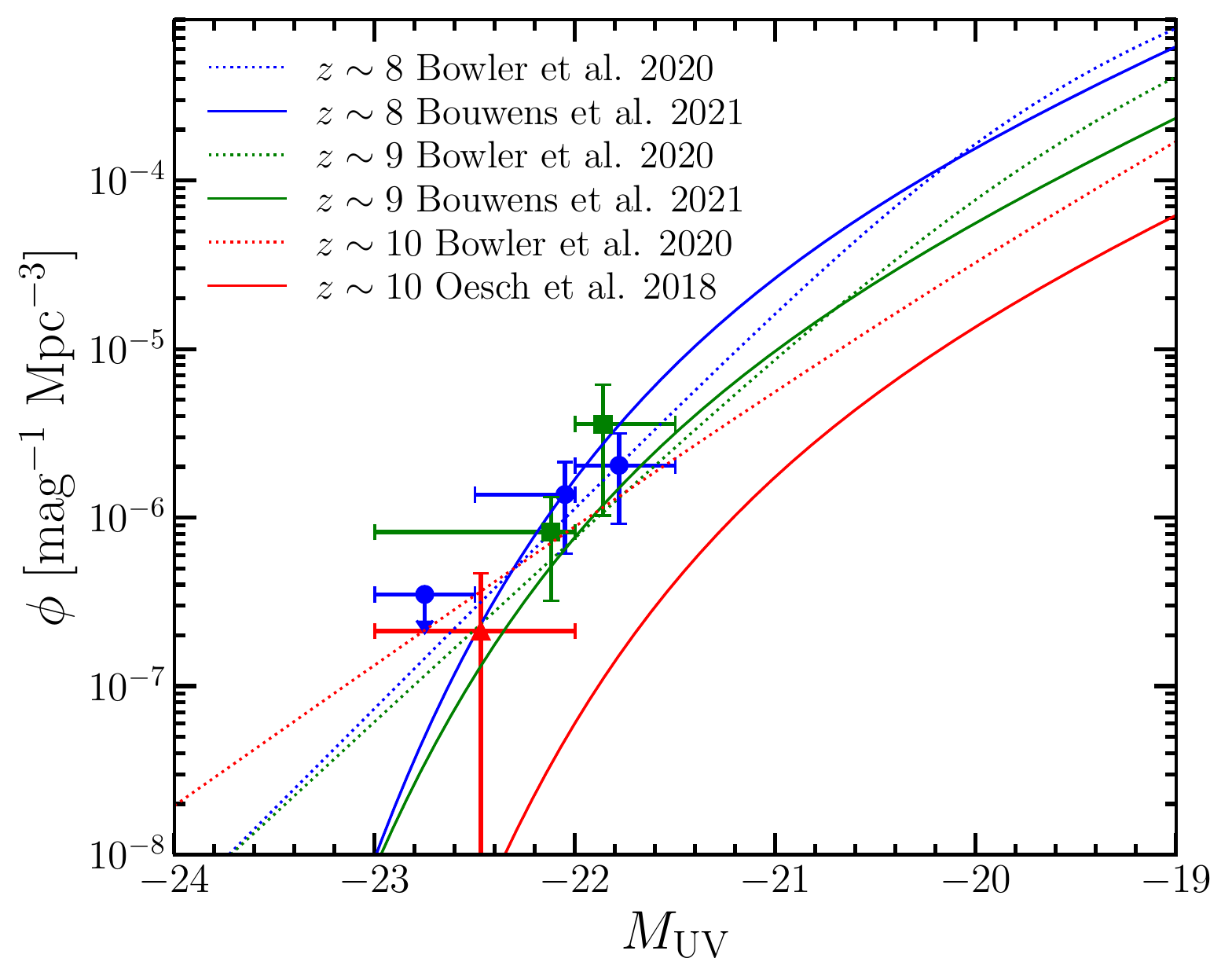}
  \caption{Galaxy UV luminosity functions from $z=8$ to $z=10$. Our results are shown with blue, green, and red squares for $z=8$, $9$, $10$, respectively. The dotted and solid lines are results from the literature using the same colours, for double power-law functions and Schechter functions, respectively.}
  \label{fig:LF_allz}
\end{figure}

We show in Fig.~\ref{fig:LF_allz} the three UVLFs. Given the magnitude limit at $H<25.6$, we cover the brightest end of the UVLF at $-23<M_\text{UV}<-21.5$. There is no clear evolution of the number densities from $z=8$ to $z=9$, with an equivalent number of candidates at $-22.5<M_\text{UV}<-21.5$ in both redshift bins. At $z=10$, the density in the single absolute magnitude bin is lower than at $z=9$, but the median $M_\text{UV}$ is shifted to brighter magnitudes and could follow a simple extrapolation of the LF at lower redshift. However, the constraint on the UVLF at $z\sim 10$ remains weak.

We provide a constraint on the brightest part of the UVLF and over 1.5\,mag range, thus the shape of the UVLF cannot be constrained from our data alone. Hence, we do not attempt a fit with a power law or a Schechter function. Nevertheless, we discuss the consistency with the various fitting forms when comparing with the literature in Sect.~\ref{LFlitt}.

\begin{table}
\centering
\caption{Galaxy luminosity functions derived from all the $z\geq7.5$ candidates. The columns indicate redshifts, central absolute magnitudes, magnitude bin widths, median absolute magnitude in the bin, number of galaxies averaged over the 500 realisations (see Sect.~\ref{sec:binned_LF}) and co-moving number densities.}
\footnotesize
\renewcommand{\arraystretch}{1.1}
\setlength{\tabcolsep}{5pt} 
\begin{threeparttable}
\begin{tabular}{cccccc}
\hline\hline
$z$ & $M_\text{UV}$ & $\Delta M_\text{UV}$  & $M_\text{median}$ & $N$ & $\phi$ \\
 & [mag] & [mag] & [mag] & & [$10^{-6}$\,mag$^{-1}$Mpc$^{-3}$] \\
\hline
8 & $-22.75$ & $0.5$ &          & $0.0$ & $<0.35$ \\
  & $-22.25$ & $0.5$ & $-22.05$ & $3.312$ & $1.37\pm0.76$ \\
  & $-21.75$ & $0.5$ & $-21.78$ & $4.02$ & $2.04\pm1.12$ \\

\hline
9 & $-22.50$ & $1.0$ & $-22.12$ & $2.714$ & $0.82\pm0.50$ \\
  & $-21.75$ & $0.5$ & $-21.86$ & $3.32$ & $3.58\pm2.55$ \\

\hline
10  & $-22.50$ & $1.0$ & $-22.47$ & $0.702$ & $0.21\pm0.26$ \\

\hline
\end{tabular}
\end{threeparttable}
\label{tab:LF_data}
\end{table}

\subsection{Comparison with UVLFs from the literature}
\label{LFlitt}

The calculated UVLFs are in good agreement with results from the literature in particular at $M_\text{UV}<-22$. This suggests that our sample is complete for the brighter magnitude bins. 
The most straightforward comparison is with \citetalias{bowler_lack_2020} who used the same near-infrared images as this work (see Sect.~\ref{sec:candidate_ancillary}). In the COSMOS field, we have $12$ galaxy candidates in common. However, the authors analysed about $6$\,deg$^2$ of imaging data from both the COSMOS and the XMM-LSS fields, sampling larger co-moving volumes than in this work. 

All candidates from \citetalias{bowler_lack_2020} at $z<8.5$ and $M_\text{UV}<-22.5$ are outside the COSMOS field, and this is the reason for their better constraint on the bright end of the UVLF. Nonetheless, our number densities at $M_\text{UV}<-22.5$ are in excellent agreement with \citetalias{bowler_lack_2020} and  \citetalias{stefanon_brightest_2019}. 

At redshift $z\sim10$, the number densities computed from the three candidates selected in this work are in agreement with the double power-law evolution from \citetalias{bowler_lack_2020}. 
The candidate XMM3-3085 from \citetalias{bowler_lack_2020}, identified in the XMM-LSS field with a photometric redshift of $z_\text{phot}=10.8\pm1.0$, is extremely bright with an absolute magnitude of $M_\text{UV}=-23.7$ and $H=23.9$. It is the brightest $z>7$ galaxy candidate ever found in the literature, although spectroscopic confirmation is still required. 
The $z\sim10$ candidate 2140+0241–303 from \citet{morishita_bright-end_2018} has an HST/F160W flux of $24.4$\,mag and an absolute magnitude of $M_\text{UV}=-22.6$. The authors used the Brightest of Reionising Galaxies (BoRG[z9]) survey, including HST optical and near-infrared imaging in five broad bands over $370$\,arcmin$^2$, in addition to IRAC/$[3.6]$ imaging. The resulting number density from that paper is an order of magnitude higher than our results at $M_\text{UV}<-22.5$. 

We compare our results with \citet{bouwens_2021}, who used the most comprehensive compilation of HST data taken on deep fields. Our results at $z\sim 8$ and $9$ are consistent within the uncertainties. The faintest of our points at $z\sim 8$ falls below their fit by a factor two in density. At $z\sim 10$, they considered the results from \citet[][]{oesch_dearth_2018} which indicate a sharp decline of the galaxy density at $M_\text{UV}<-22$. Our data probe brighter absolute magnitudes (our only point is brighter than $M_\text{UV}<-22$ while their brightest point is at $M_\text{UV}=-21.25$). However, the lower density of our fit is unconstrained and consistent with zero. Therefore, we cannot draw clear conclusions regarding differences with \citet[][]{oesch_dearth_2018}.

\section{Discussion}
\label{sec:discussion}

\subsection{Sources of contamination}
\label{contamination}

The most obvious explanation for the high density of bright sources at $z\geq7.5$ is the contamination by low-redshift galaxies or brown dwarfs wrongly classified as $z>9$ candidates. Even if we combine several photometric redshift codes, and two different methods to extract the photometry, contamination of our sample remains a possibility. In Sect.~\ref{sec:robustness}, we found 10 candidates with a significant probability of being either dusty star-forming galaxies at intermediate redshift or brown dwarf contaminants. Seven over ten of these candidates fall on the westernmost ultra-deep stripe with a lower HSC and IRAC coverage (see Fig.~\ref{fig:cand_coords}). The larger surface density of sources in this stripe compared to the others points out towards a significant population of contaminants among these ten galaxies.

In Fig.~\ref{fig:LF_z8},  Fig.~\ref{fig:LF_z9} and Fig.~\ref{fig:LF_z10}, we present the UVLF after having removed these sources (gold squares). We find that the measurement at $7.5<z<8.5$ is severely affected by this removal, as expected since 7~sources have a \photoz{} estimated in this redshift range. Therefore, the density could be affected by a factor~2 if all these sources are in fact contaminants. 

A systematic overestimate of the \photoz{} could also impact the UVLF. The 5 sources with a \specz{} have a systematically higher \photoz{} (see Sect.~\ref{sec:robustness}). If this trend affects the whole sample, galaxies could move in a lower redshift bin. Given the shape of the redshift distribution, it would lead to an overestimation of the UVLF density. A better modelling of the galaxy properties (e.g. dust attenuation law, emission lines) could alleviate these biases. We are particularly sensitive to the modelling because the major feature used to derive the \photoz{} at this epoch, namely the continuum break around 1216\,\AA{}, cannot be accurately located. Indeed, this break is redshifted within the gap between the ground-based $Y$ and $J$-bands for galaxies between $z\sim7.6$ and $z\sim8.7$. Spectroscopic confirmation of the candidate redshifts is the solution to alleviate these uncertainties \citep{2008MNRAS.386.2323M}, as well as future space mission without such gap between near-infrared filters. 

In addition, the uncertainties in the absolute UV magnitudes may affect the bright end through the Eddington bias \citep{eddington_formula_1913}. Because of the steep slope of the luminosity function, there are statistically more faint galaxies scattered in the brighter bins than the reverse, resulting in a flattened slope. The Eddington bias is also stronger for steep luminosity functions. In the selected galaxy sample, photometric redshift uncertainties can be relatively large, leading to large uncertainties once propagated to absolute magnitudes. This bias may be limited by using large magnitude bins. 

The presence of Active Galaxy Nuclei (AGN) in the selected galaxy sample may also affect the estimated galaxy UVLF at the bright end. High-redshift AGN have Lyman alpha break features similar to star-forming galaxies without an AGN component. At intermediate redshifts $4<z<6$, the contribution from faint AGN dominates the number densities at $M_\text{UV}<-23$, whereas it becomes negligible at $M_\text{UV}>-22$ \citep{ono_great_2018,Harikane22_goldrush}. 
At $z>6$, the number density of faint AGN is still uncertain. The evolution of the quasar spatial density is often parametrised as $\rho(z)\propto10^{k(z-6)}$, with $k\simeq-0.47$ from $z=3.5$ to $z=5$ \citep{fan_high-redshift_2001} and $k\simeq-0.72$ from $z=5$ to $z=6$ \citep{jiang_final_2016}. Recently, \citet{wang_exploring_2019} measured a consistent value $k\simeq-0.78$ from $z=6$ to $z=6.7$. With this accelerated redshift evolution, high-redshift AGN are sufficiently rare such that they have a negligible impact on the galaxy number density at $M_\text{UV}=-23$ in a survey of this size. The faint-end slope of the quasar UV luminosity function is nonetheless poorly constrained at high redshift \citep{matsuoka_discovery_2019}.

\subsection{Impact of magnification}
\label{sec:magnification}

As a consequence of gravitational lensing at very high redshifts, and in particular for steep luminosity functions \citep{mason_correcting_2015}, the bright end of the luminosity function is expected to be artificially higher. 
As discussed in Sect.~\ref{sec:magnification_candidates}, we find no evidence of strongly magnified galaxies in the selected sample; although, candidates are still affected by multiple lenses, leading to cumulative magnifications up to $\mu=2.4$. We estimate the UVLF after having corrected the UV absolute magnitudes from the magnification (not attempting to correct the density). The binned UVLF at $z=8$ and $z=9$ remain unchanged. Thus, magnification bias does not explain the lack of evolution of the UVLF bright end at the probed magnitudes. In the $z=10$ galaxy sample, the candidate ID441697 with a cumulative magnification of $\mu=2.36$ becomes fainter than $M_\text{UV}=-22$ after removing the effect of lensing. Therefore, it would even lower down the density found at $z\sim 10$.

\subsection{Shape of the UVLF}

It is well established that the UVLF at $z<6$ is well-fit with a Schechter function \citep[e.g.][]{Moutard20, bouwens_2021}. At the same time, the very different shape of the bright and massive ends of the dark matter (DM) halo mass function from cosmological models and the observed galaxy luminosity function is currently explained with ``quenching'', or any process which can halt star formation in these haloes and galaxies. This cessation of the star-forming activity can be due to numerous processes like AGN feedback \citep[e.g.][]{croton06,hopkins06} or halo quenching due to shock-heated gas in halos more massive than $10^{12}\Msol{}$ \citep[e.g.][]{somerville08,gabor15}. \citet[][]{peng10} showed that if the quenching rate is proportional to the SFR, it would naturally lead to a galaxy luminosity function with a Schechter shape.

While the Schechter form is well established at $z<6$, whether this holds at higher redshifts is unclear. \citet[][]{bowler_galaxy_2015} find a high density of sources at $M_\text{UV}<-22$ and conclude that the UVLF at $z>8$ is better fit with a power law than a Schechter function, without displaying an exponential cutoff at the bright end. We confirm the density measured by \citetalias{bowler_lack_2020} and our points are superposed on their extrapolation of the UVLF with a power law. Our measurements are inconsistent with the Schechter function obtained by \citet[][]{mcleod_z_2016}, and the density of galaxies we measure is ten times higher than expected by their extrapolation to bright magnitudes (or brighter by 0.4 mag). The brightest galaxies observed by \citet[][]{mcleod_z_2016} is at $M_\text{UV}=-20.7$, not allowing them to constrain the UVLF in the bright regime probed by our data. 

The latest estimates by \citet[][]{bouwens_2021} compile all HST measurements and find that the UVLF are consistent with a Schechter function at $z\sim 8-9$. The new Schechter parameters obtained by \citet[][]{bouwens_2021} shift the exponential cut-off at brighter magnitude than previous publications \citep[e.g.][]{mcleod_new_2015}. This is more consistent with the density of galaxies we find at $M_\text{UV}<-22$. At $z\sim 8$, our faintest point falls below their fit; however, such an effect could be explained by some residual incompleteness that is not corrected in our measurement. We find that our points are still consistent with their fit at $z\sim 9$. So, we cannot reject the Schechter fit by \citet[][]{bouwens_2021}.

\subsection{Lack of evolution at the bright end}
\label{sec:evolution}

The UVLF evolves rapidly at magnitudes fainter than $M_\text{UV}> -20.5$. At a given density of $\phi\sim10^{-4}\,\rm{mag}^{-1}\rm{Mpc}^{-3}$, we expect a brightening of about $0.7\,$mag between $z\sim 9$ and $z\sim 8$
following the latest compilation from \citet[][]{bouwens_2021}. Between $z\sim9$ and $z\sim10$, this trend is even more extreme with a brightening of $0.9$\,mag. This evolution is interpreted as the galaxy population following the growth of the dark matter halos with a constant star-formation efficiency \citep{oesch_dearth_2018}. 

In contrast with the faint end, the density of galaxies at $M_\text{UV}\sim -22$ is consistent with no evolution from $z\sim8$ to $z\sim9$ as shown in Fig.~\ref{fig:LF_allz}. This is consistent with the findings of \citetalias{bowler_lack_2020}. 

One interpretation is that quenching efficiency increases between $z=9$ and $z=8$ at high mass and as a consequence we observe the building of the exponential cut-off of the UVLF. \citet[][]{peng10} introduced a quenching rate proportional to the SFR to preserve a constant characteristic stellar mass around $10^{10.6}\,\Msol$. If we extrapolate the relation between $M_\text{UV}$ and stellar mass found by \citet[][]{stefanon21}, we do not reach this regime at $z=8$. Also, halo quenching is efficient above $10^{12}\,\Msol{}$ \citep[][]{cattaneo06}. According to \citet[][]{stefanon21}, galaxies in our sample are hosted by lower mass halos. So, mass quenching is not expected to be efficient yet. However, the physical conditions at $z>8$ may be different from those considered in lower redshift studies and the quenching mechanisms may be already effective for lower mass galaxies or DM halos.

Another possibility is that dust is generated in significant quantities between $z=9$ and $z=8$ \citep{finkelstein_evolution_2015}, sufficiently to decrease the light emerging from the brightest UV galaxies at $z=8$, leading to the wrong interpretation that the density of bright galaxies remains the same between $z=9$ and $z=8$. Indeed, if dust is increasing with decreasing $M_\text{UV}$, it would bend the bright end of the luminosity function, with a more pronounced dimming of the UV absolute magnitudes at bright magnitudes\footnote{We stress that dust was included in our modelling of the galaxy SED during the fitting process, but we did not correct the $M_\text{UV}$ absolute magnitudes for dust attenuation.}. The mean dust content is generally expected to decrease with increasing redshift in particular for bright galaxies (e.g. \citealt{bouwens_alma_2016}), because of the low metallicity of young stars in galaxies. Hence, the mean dust attenuation at $z>9$ is often assumed to be zero (e.g. \citealt{bouwens_uv_2015}), so that the bright end of the UVLF is expected to reflect the recent star formation in the galaxy population.

\section{Summary and conclusions}
\label{sec:conclu}

This paper presents one of the first science results from the COSMOS2020 catalogues: a search for candidate $z\geq7.5$ galaxies at the epoch of the reionisation of the Universe. The deep optical, near-infrared, and mid-infrared imaging over the 1.4~square-degree field of COSMOS enables the detection of rare and bright galaxies, a complementary population to the much fainter galaxies found in deep pencil beam surveys such as CANDELS or HUDF. 

COSMOS2020 uses the latest and deepest optical, mid-infrared, and near-infrared imaging in COSMOS. It includes the most recent UltraVISTA DR4 data together with new optical data from the HSC-SSP PDR2 release. We also used new images comprising all of the \textit{Spitzer}/IRAC $[3.6]$, $[4.5]$ bands' data ever taken on COSMOS from the Cosmic Dawn Survey. From these data, we extracted two photometric catalogues using independent approaches: the \classic{} catalogue measures colours using circular aperture photometry, whilst \farmer{} catalogue colours are measured based on a fit of surface brightness models. Sources were detected in a combined $izYJHK$ image.

The galaxy selection was primarily based on photometric redshifts estimated from SED fitting with \lephare{} and their associated \zPDF{}. We selected sources with a robust solution at $z\geq7.5$ with \lephare{} using either the \classic{} or \farmer{} catalogues. 
The final sample consists of $32$ candidates at $z\geq7.5$, including $17$ unpublished candidates. To assess the robustness of the COSMOS2020 \photoz{} estimates, we computed new sets of photometric redshifts with three different template-fitting procedures using \lephare{} and \eazy{}, with both the \classic{} and \farmer{} catalogues. We isolated a gold sample of $22$ out of $32$ candidates, which is more robust. Among them, four blended candidates were identified thanks to the profile-fitting photometry from \farmer{}, where the contamination of light from nearby sources can be identified. These candidates have low \photoz{} solutions of $z<3$ with the \classic{} catalogue because of the optical flux of the nearby sources. This illustrates the effectiveness of profile-fitting techniques in deblending confused sources typically found in the search of distant galaxies. The final sample of high-redshift galaxies is therefore more complete.

From this unique list of $z\geq7.5$ star-forming galaxies, we made a new determination of the bright end of the UV luminosity functions in three redshift bins, centred at $z=8,9,10$. There is no clear evolution of the number densities from $z=8$ to $z=9$ between $-23<M_\text{UV}<-21.5$, which is in excellent agreement with \citetalias{bowler_lack_2020}. 
One interpretation is that quenching efficiency increases between $z=9$ and $z=8$. Another possibility is that dust is generated in a significant quantity, which is sufficient enough to decrease the light emerging from the brightest UV galaxies at $z=8$. 
Another explanation for the high density of bright sources at $z\geq7.5$ is the contamination by low-redshift galaxies or brown dwarfs, or a systematic overestimation of photometric redshifts. Spectroscopic confirmation is thus essential. Follow-up observations with JWST-NIRSpec \citep{2021jwst.prop.2659W} are planned for five of our brightest candidates (see Table~\ref{tab:candidates_photoz}). We have also begun a systematic follow-up of these candidates with the WERLS Key Strategic Mission Support Program on Keck (Casey et al., in prep.).

This work has demonstrated the great potential of COSMOS2020 for finding rare, luminous objects in the distant Universe and the benefit from using multiple photometric extraction techniques and photometric redshift codes to assess the robustness of the results. In the longer term, deep Euclid near-infrared observations of COSMOS and other Cosmic dawn survey fields, together with our approved $218+80$\,h, $0.6$\,deg$^2$ COSMOS-Webb proposal, will provide the ultimate survey of bright, luminous objects in the early Universe. 


\begin{acknowledgements}

We are grateful to the referee for a careful reading of the manuscript and useful suggestions.

This paper is dedicated to Olivier Le F\`evre who initiated this work and devoted so much energy and passion to building the COSMOS survey.

We warmly acknowledge the contributions of the entire COSMOS collaboration consisting of more than 100 scientists. The HST-COSMOS program was supported through NASA grant HST-GO-09822. More information on the COSMOS survey is available at \url{https://cosmos.astro.caltech.edu}.

This research is also partly supported by the Centre National d'Etudes Spatiales (CNES).
OI acknowledges the funding of the French Agence Nationale de la Recherche for the project ``SAGACE''. 
HJMcC acknowledges support from the PNCG. 
ID has received funding from the European Union’s Horizon 2020 research and innovation program under the Marie Sk\l{}odowska-Curie grant agreement No. 896225. This work used the CANDIDE computer system at the IAP supported by grants from the PNC, DIM-ACAV and the CNES and maintained by S. Rouberol. 
BMJ is supported in part by Independent Research Fund Denmark grant DFF - 7014-00017. The Cosmic Dawn Center (DAWN) is funded by the Danish National Research Foundation under grant No. 140. 
ST, GB and JW acknowledge support from the European Research Council (ERC) Consolidator Grant funding scheme (project ConTExt, grant No. 648179). 
CMC thanks the National Science Foundation for support through grants AST-1814034 and AST-2009577, and the Research Corporation for Science Advancement from a 2019 Cottrell Scholar Award sponsored by IF/THEN, an initiative of Lyda Hill Philanthropies.
CMC thanks the National Science Foundation for support through grants AST-1814034 and AST-2009577, and the Research Corporation for Science Advancement from a 2019 Cottrell Scholar Award sponsored by IF/THEN, an initiative of Lyda Hill Philanthropies.
The authors wish to recognise and acknowledge the very significant cultural role and reverence that the summit of Mauna Kea has always had within the indigenous Hawaiian community. We are most fortunate to have the opportunity to conduct observations from this mountain.
This work is based on data products from observations made with ESO Telescopes at the La Silla Paranal Observatory under ESO program ID 179.A-2005 and on data products produced by CALET and the Cambridge Astronomy Survey Unit on behalf of the UltraVISTA consortium. This work is based in part on observations made with the NASA/ESA \textit{Hubble} Space Telescope, obtained from the Data Archive at the Space Telescope Science Institute, which is operated by the Association of Universities for Research in Astronomy, Inc., under NASA contract NAS 5-26555.
\end{acknowledgements}

\bibliographystyle{aa}
\bibliography{biblio.bib}

\FloatBarrier

\begin{appendix}

\section{Artefacts}
\label{sec:appendix_cross-talks}

\begin{figure}
  \centering
  \includegraphics[width=0.9\hsize]{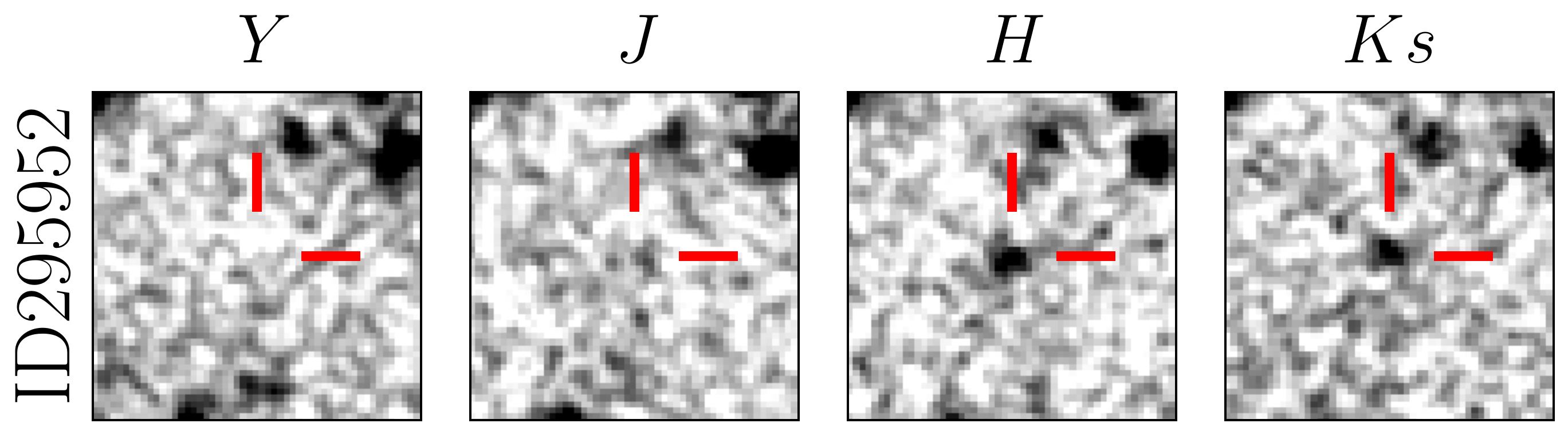}
  \includegraphics[width=0.9\hsize]{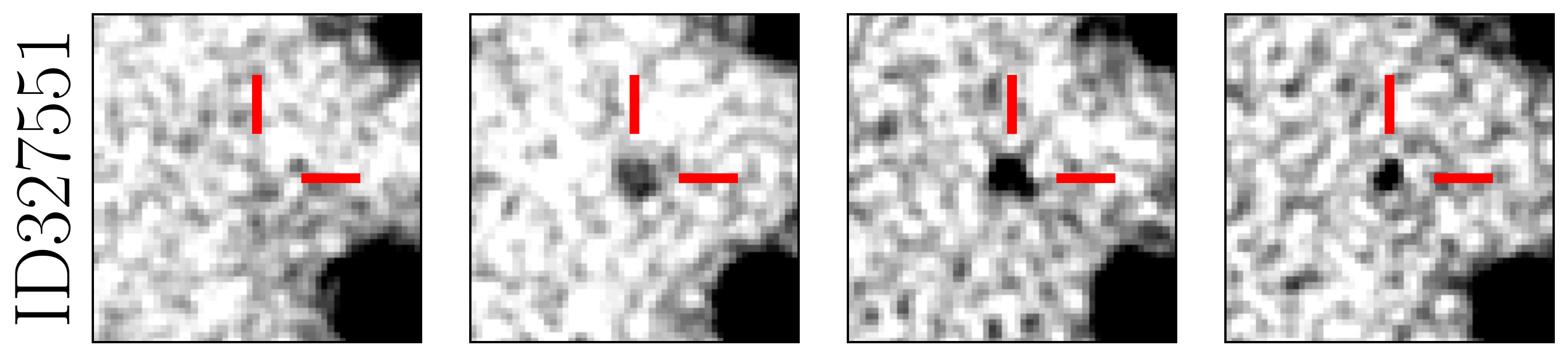}
  \includegraphics[width=0.9\hsize]{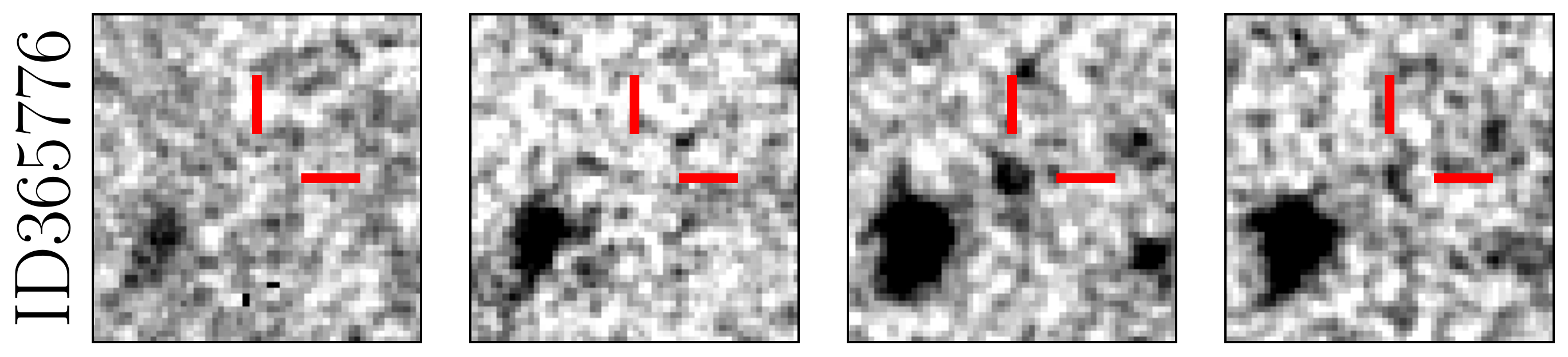}
  \includegraphics[width=0.9\hsize]{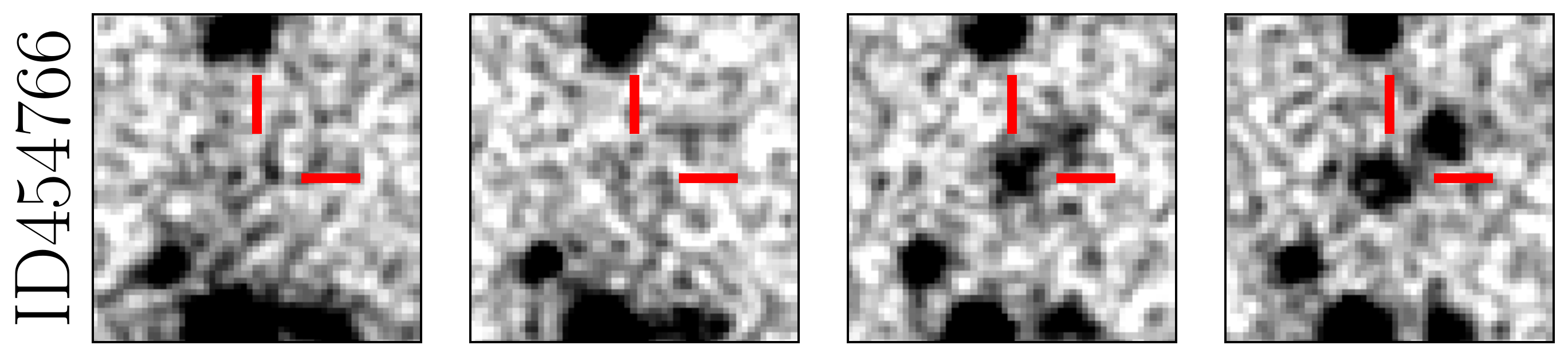}
  \caption{Same as Fig.~\ref{fig:snaps_unblended_1}, but for the cross-talk artefacts.}
  \label{fig:snaps_cross-talks}
\end{figure}

The cross-talk effect is an electronic artefact in which bright sources reappear at different locations in the detector focal plane \citep[e.g.][]{Bowler17}. Such artefacts are particularly problematic for high-redshift galaxy searches as they appear in the near-infrared channels only. We identified several high-redshift candidates which are most probably these inter-channel cross-talk artefacts in the UltraVISTA images. They are caused by bright sources located at the same RA with a Dec differing by $\pm k\,\times\,128\,\times\,0\farcs339$ with integers $k=1,...,15$ in the VIRCAM images. Here, $128$ is the number of detector pixels in each of the $16$ readout channels of the detector, and $0\farcs339$ is approximately the size of the original pixels. These sources are typically $11.9-13.8$\,mag brighter than the artefacts, and were identified comparing single pawprint stacked images with similar total exposure times, in the UltraVISTA bands. 
Moreover, we find no evidence of an optical counterpart with HSC for these sources, or in the mid-infrared with IRAC.
The coordinates of these candidates are reported in Table~\ref{tab:cross-talks}. Figure~\ref{fig:snaps_cross-talks} shows the UltraVISTA stamps at the corresponding coordinates. \\

\begin{table}
\small\centering
\renewcommand{\arraystretch}{1.2}
\setlength{\tabcolsep}{4pt} 
\begin{threeparttable}
\caption{Coordinates of the high-redshift galaxy candidates identified as cross-talk artefacts}
\begin{tabular}{ccc}
 \hline \hline
ID & R.A. & Dec. \\
\classic{} & [J2000] & [J2000] \\
 \hline
295952 & 10:01:56.01 & 01:42:08.37 \\
327551 & 09:57:48.08 & 01:44:01.39 \\
365776 & 10:02:16.98 & 01:46:16.88 \\
454766 & 10:00:57.43 & 01:51:27.89 \\
 \hline
\end{tabular}
\label{tab:cross-talks}
\end{threeparttable}
\end{table}

\section{Detailed description of each candidate}
\label{sec:appendix_seds_stamps}

\begin{figure}
  \centering
  \includegraphics[width=0.33\textwidth]{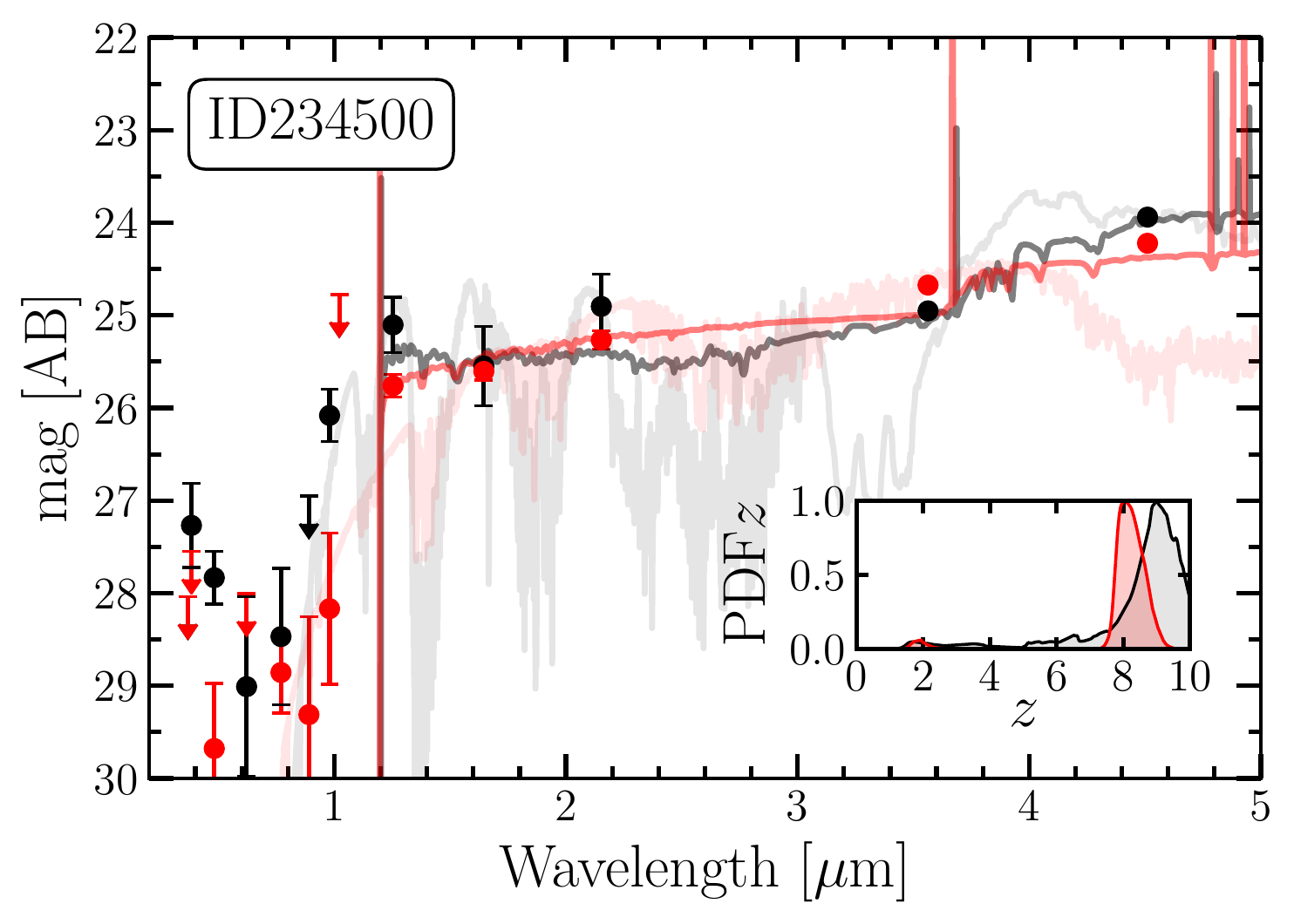}
  \includegraphics[width=0.33\textwidth]{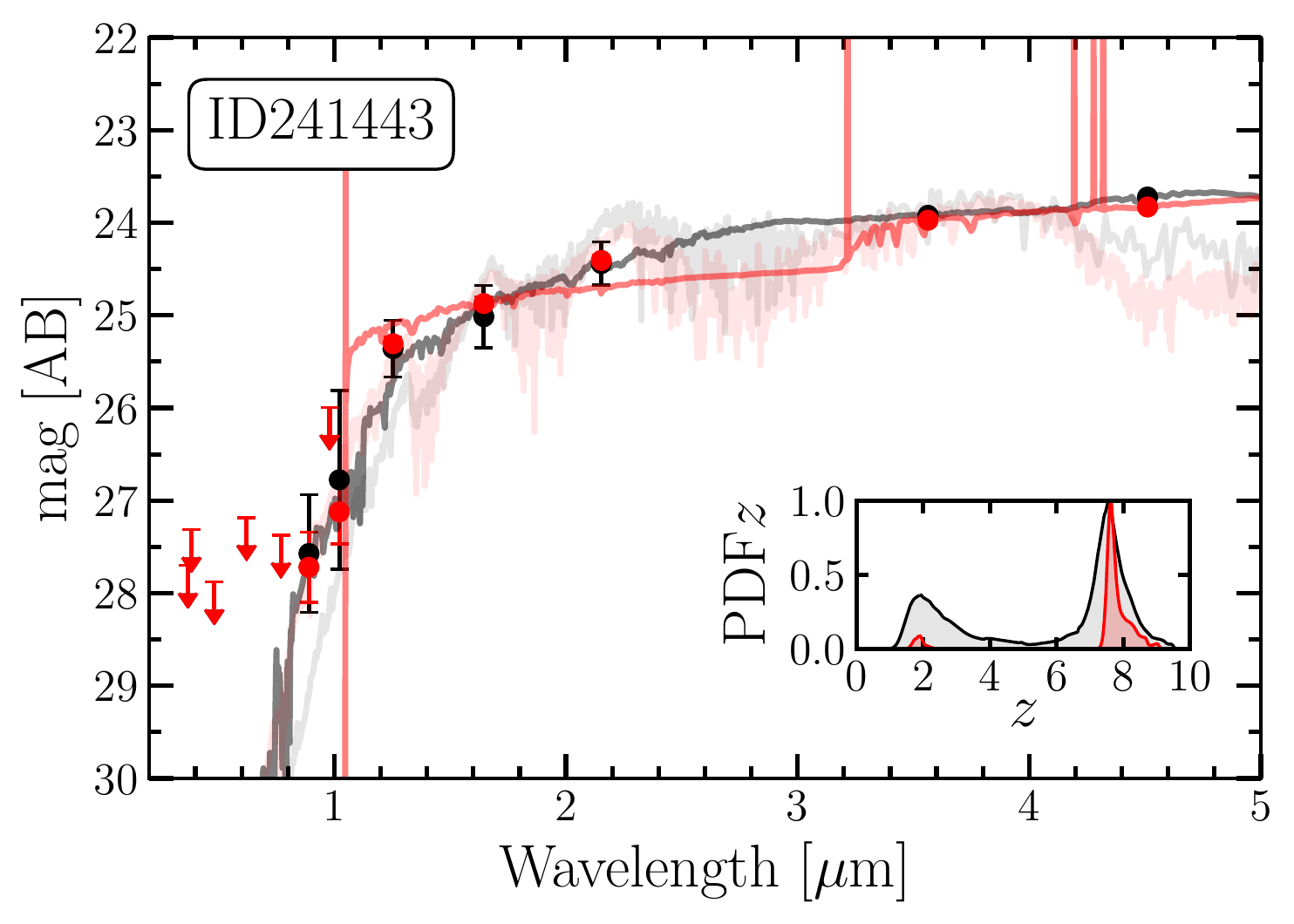}
  \includegraphics[width=0.33\textwidth]{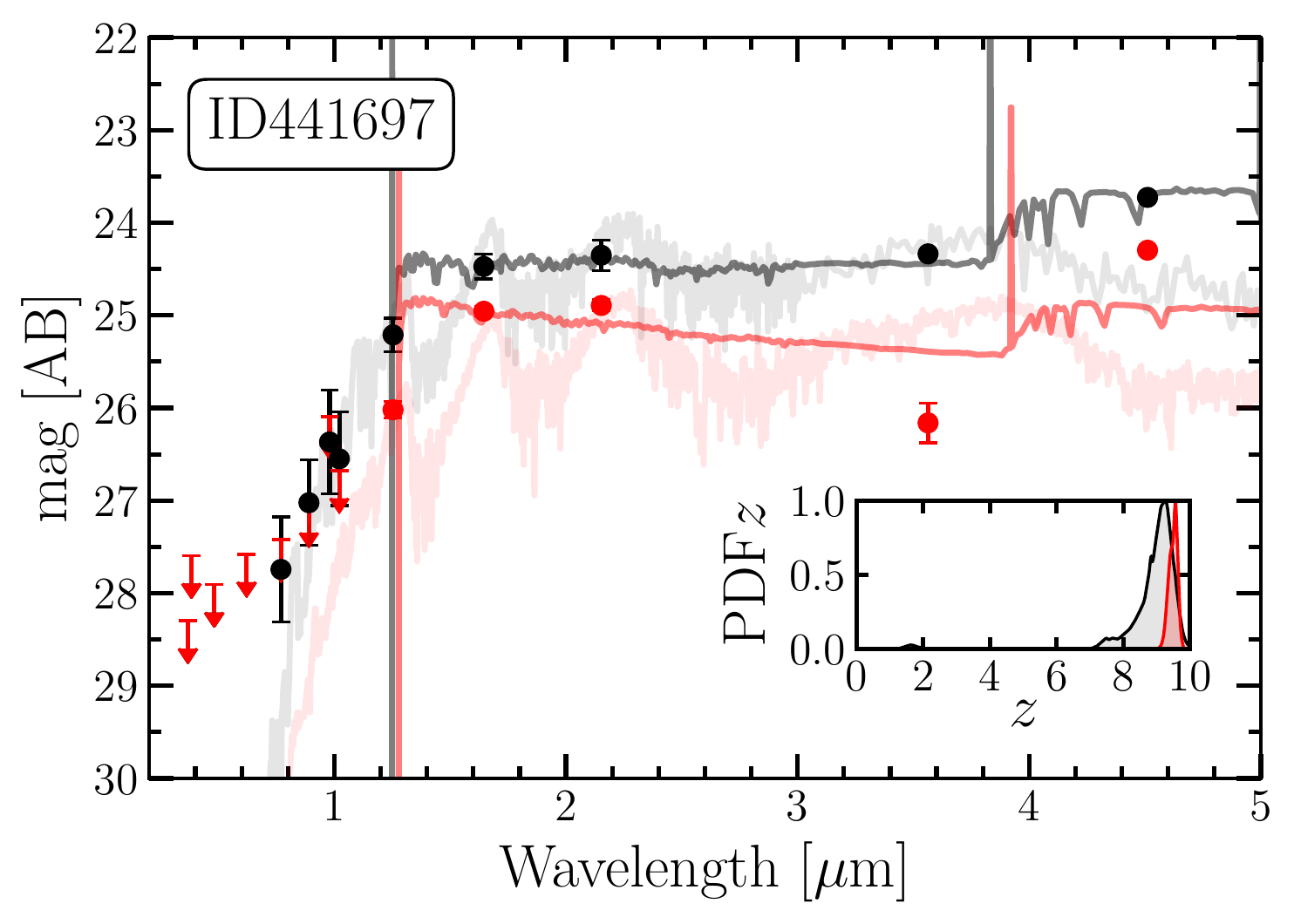}
  \includegraphics[width=0.33\textwidth]{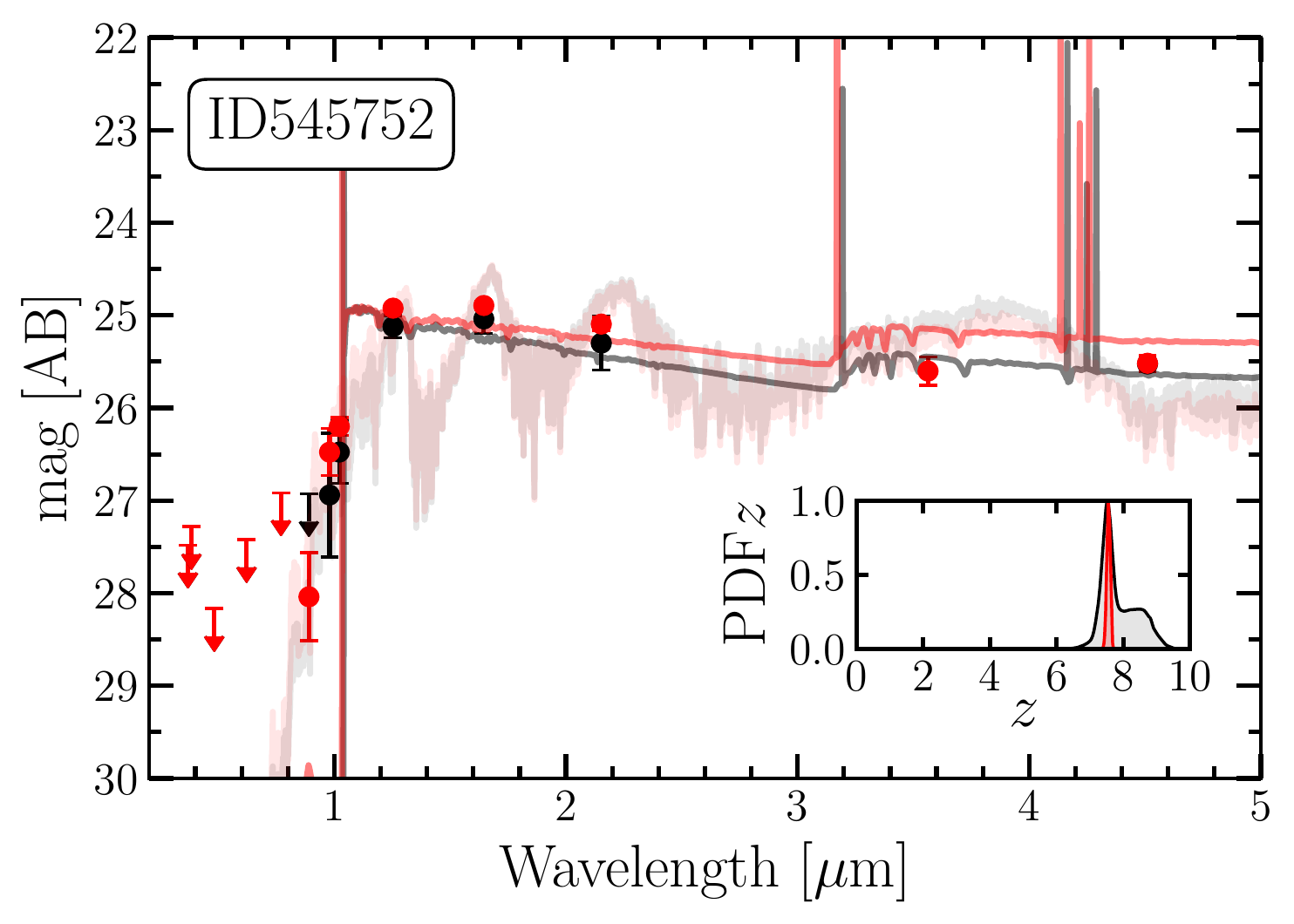}
  \includegraphics[width=0.33\textwidth]{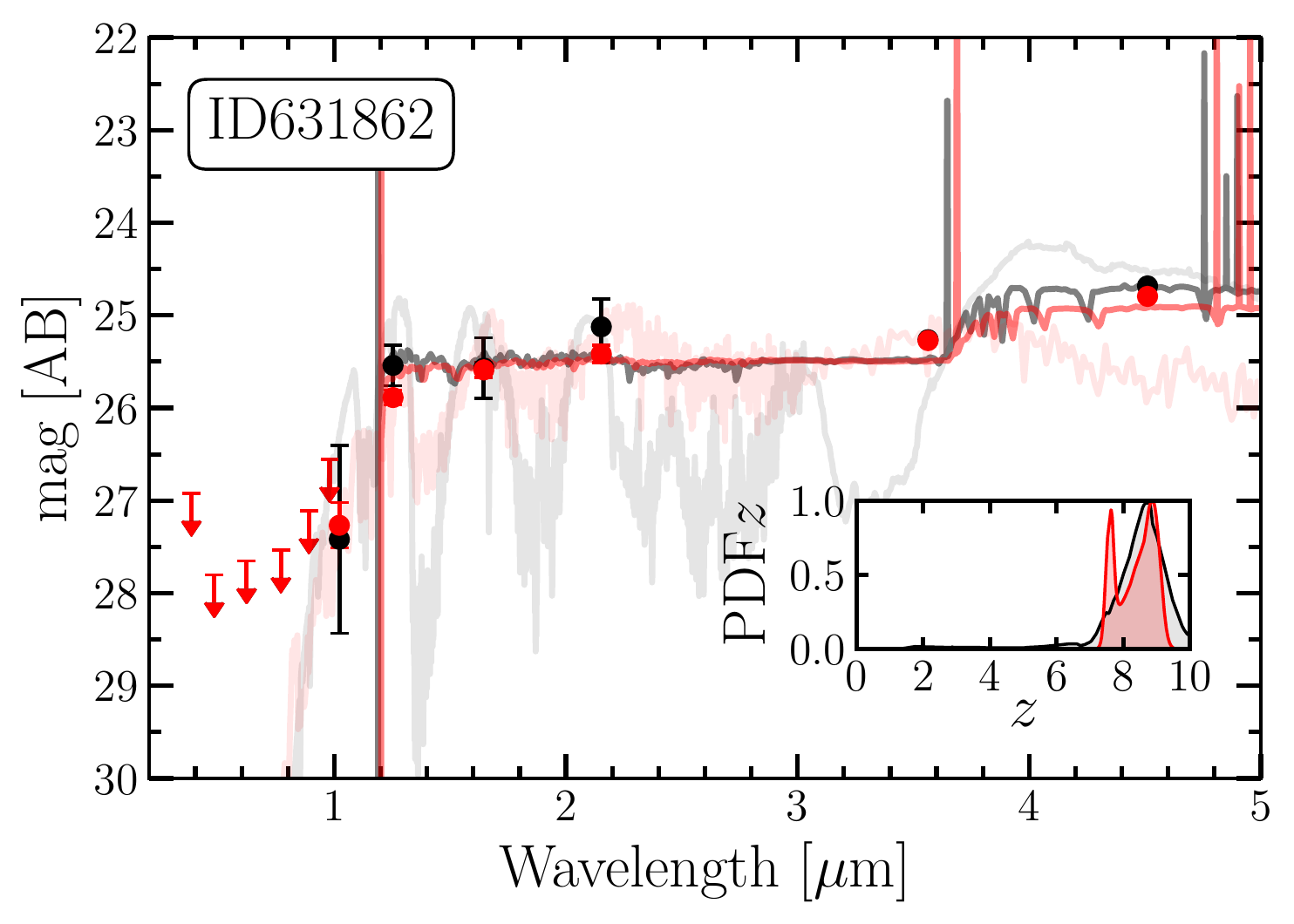}
  \includegraphics[width=0.33\textwidth]{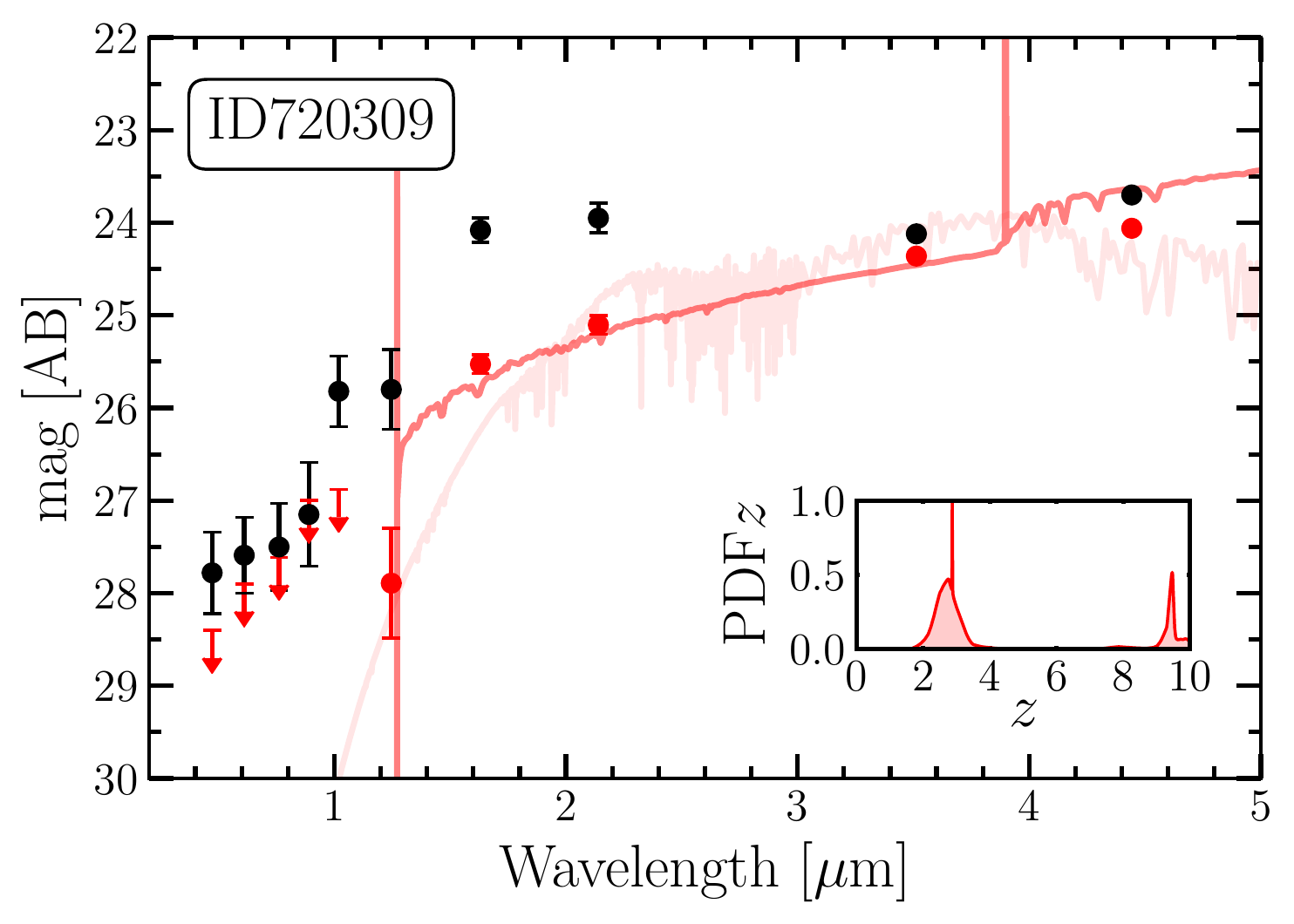}
  \includegraphics[width=0.33\textwidth]{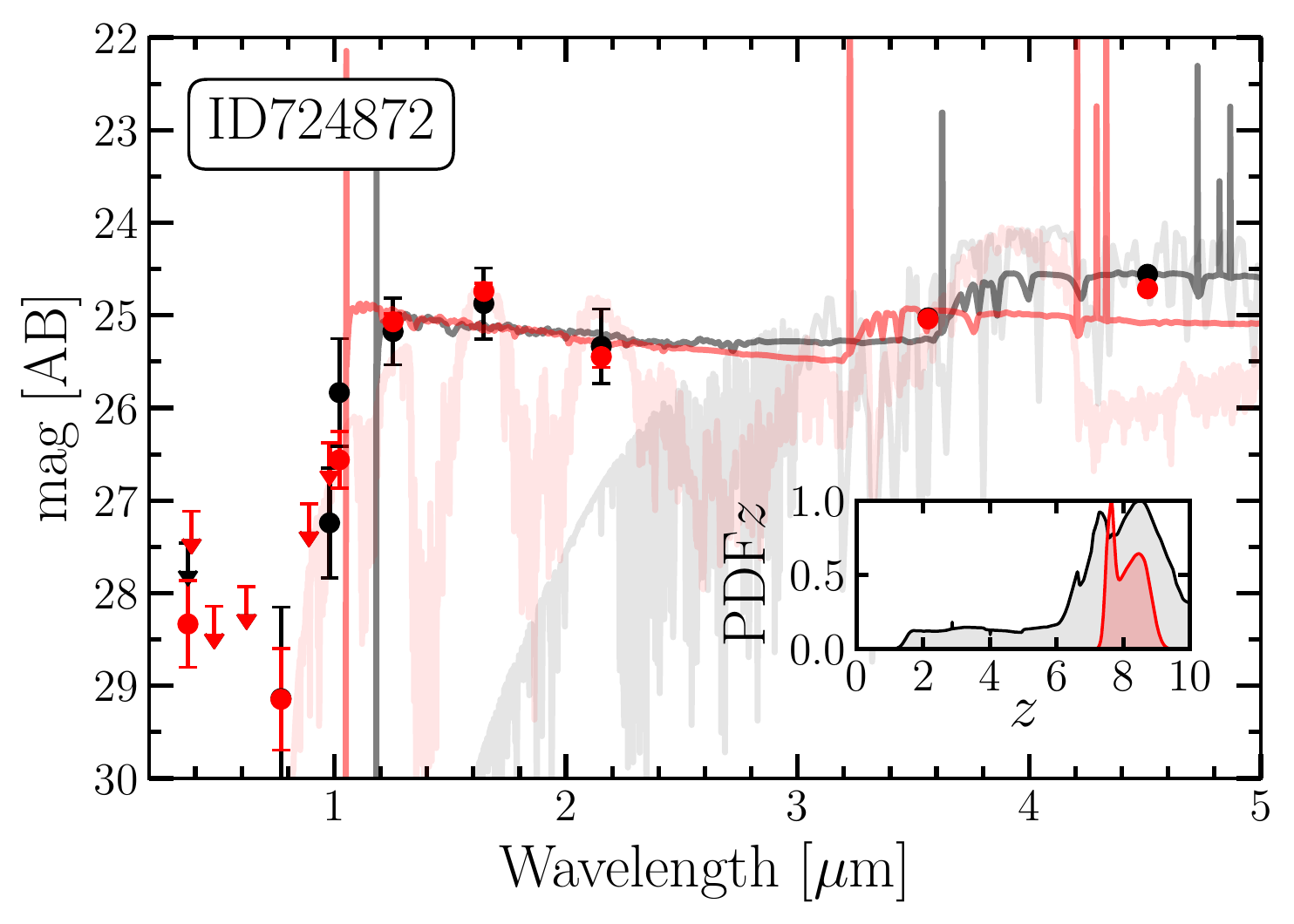}
  \includegraphics[width=0.33\textwidth]{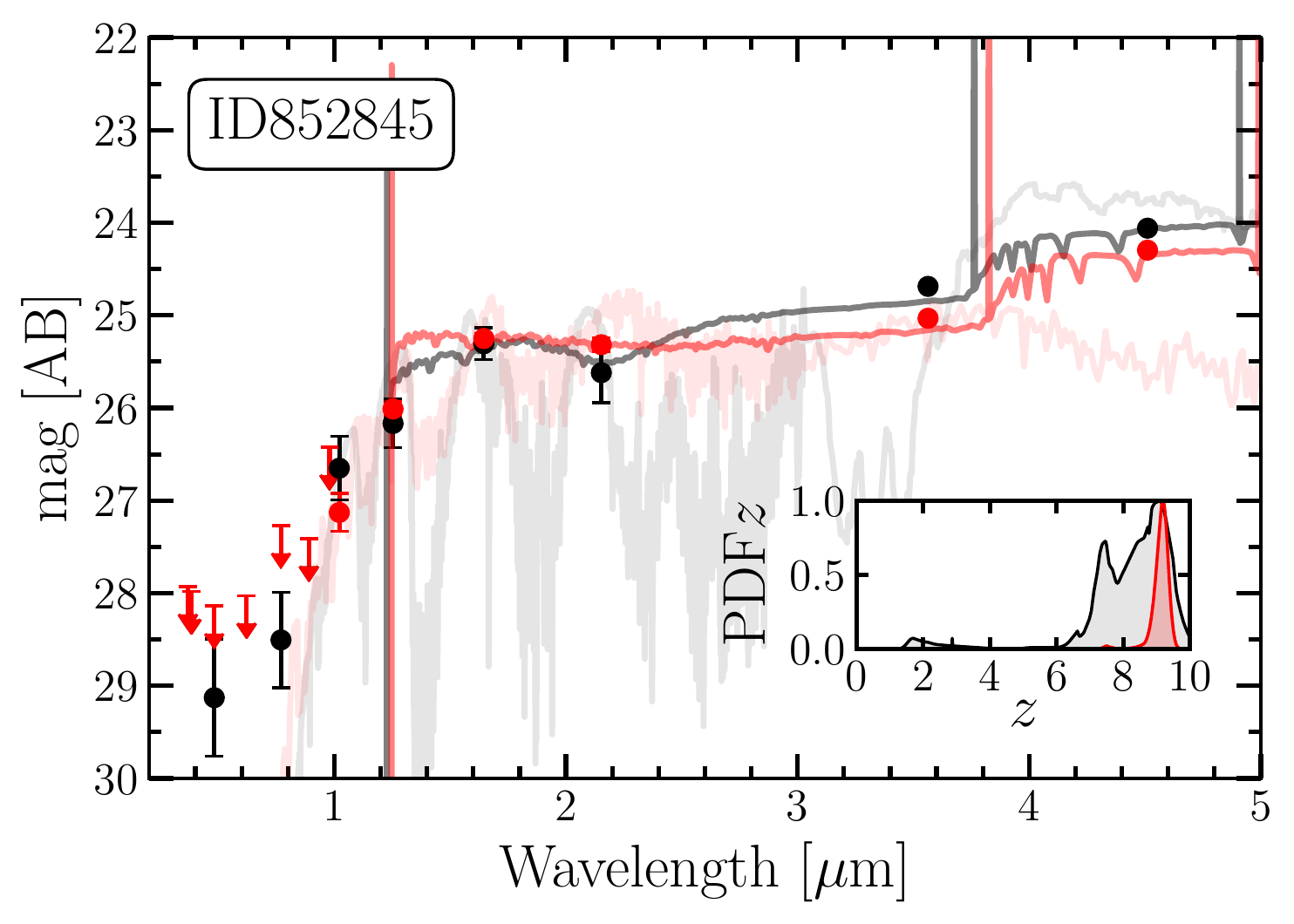}
  \includegraphics[width=0.33\textwidth]{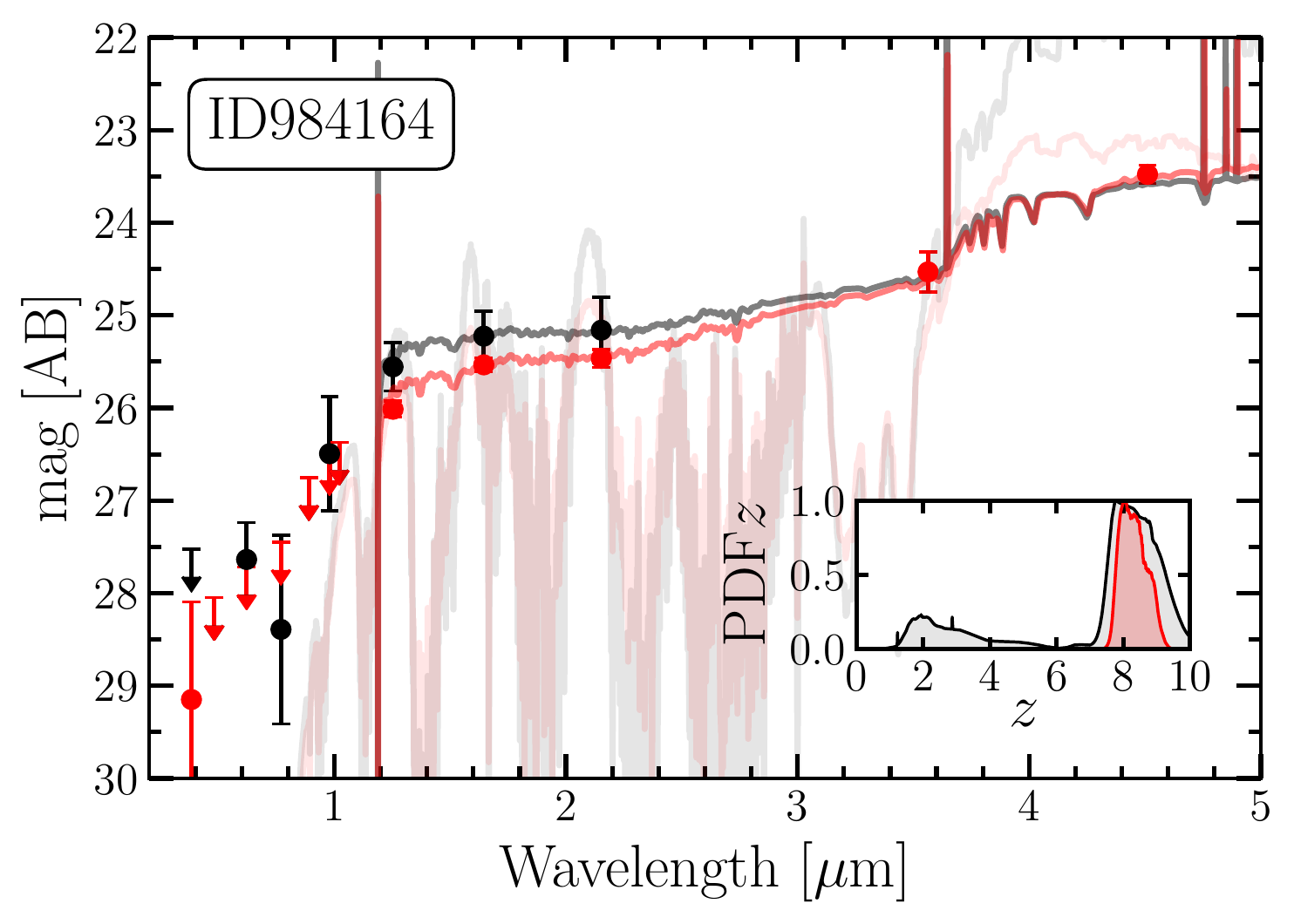}
  \includegraphics[width=0.33\textwidth]{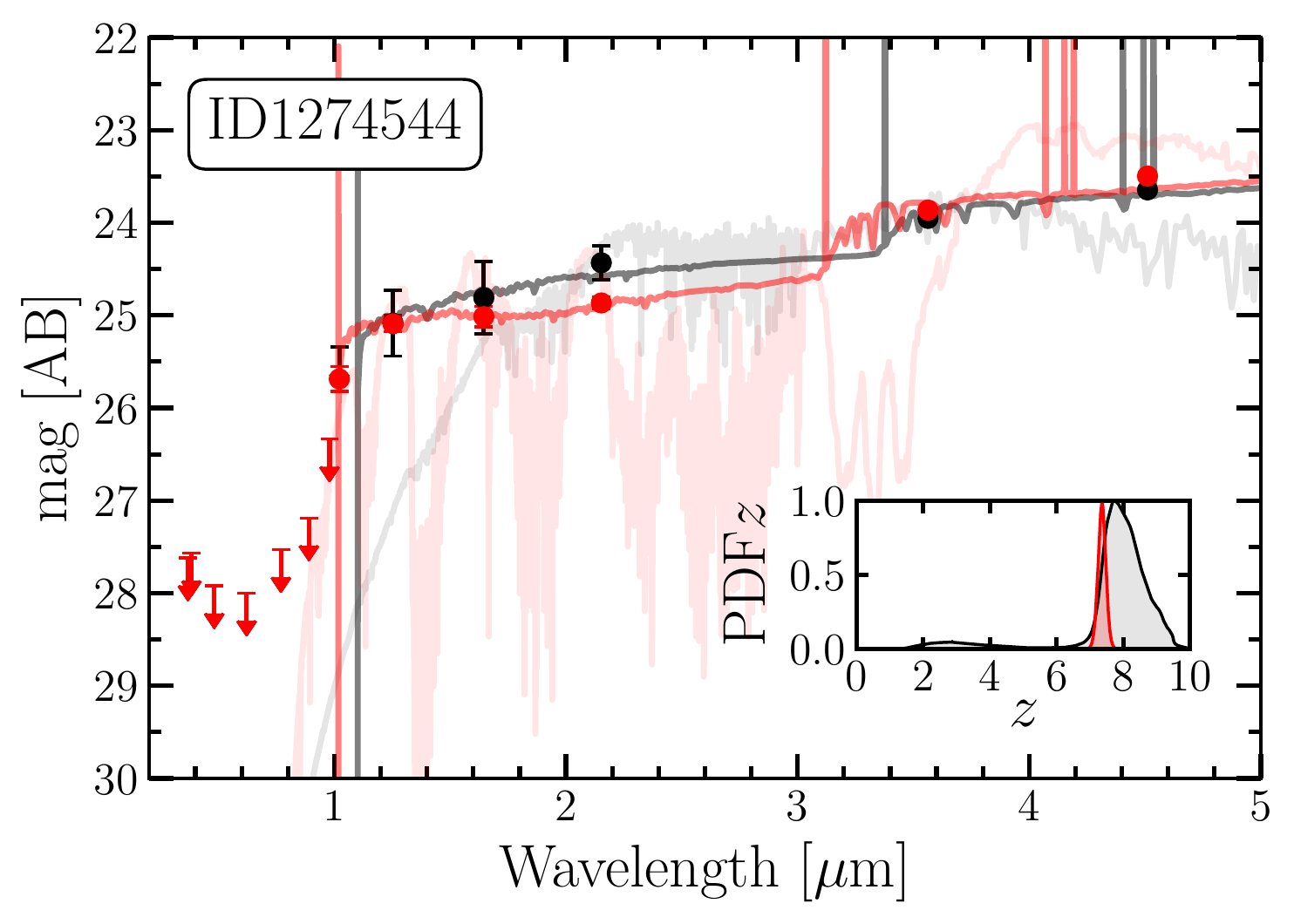}
  \includegraphics[width=0.33\textwidth]{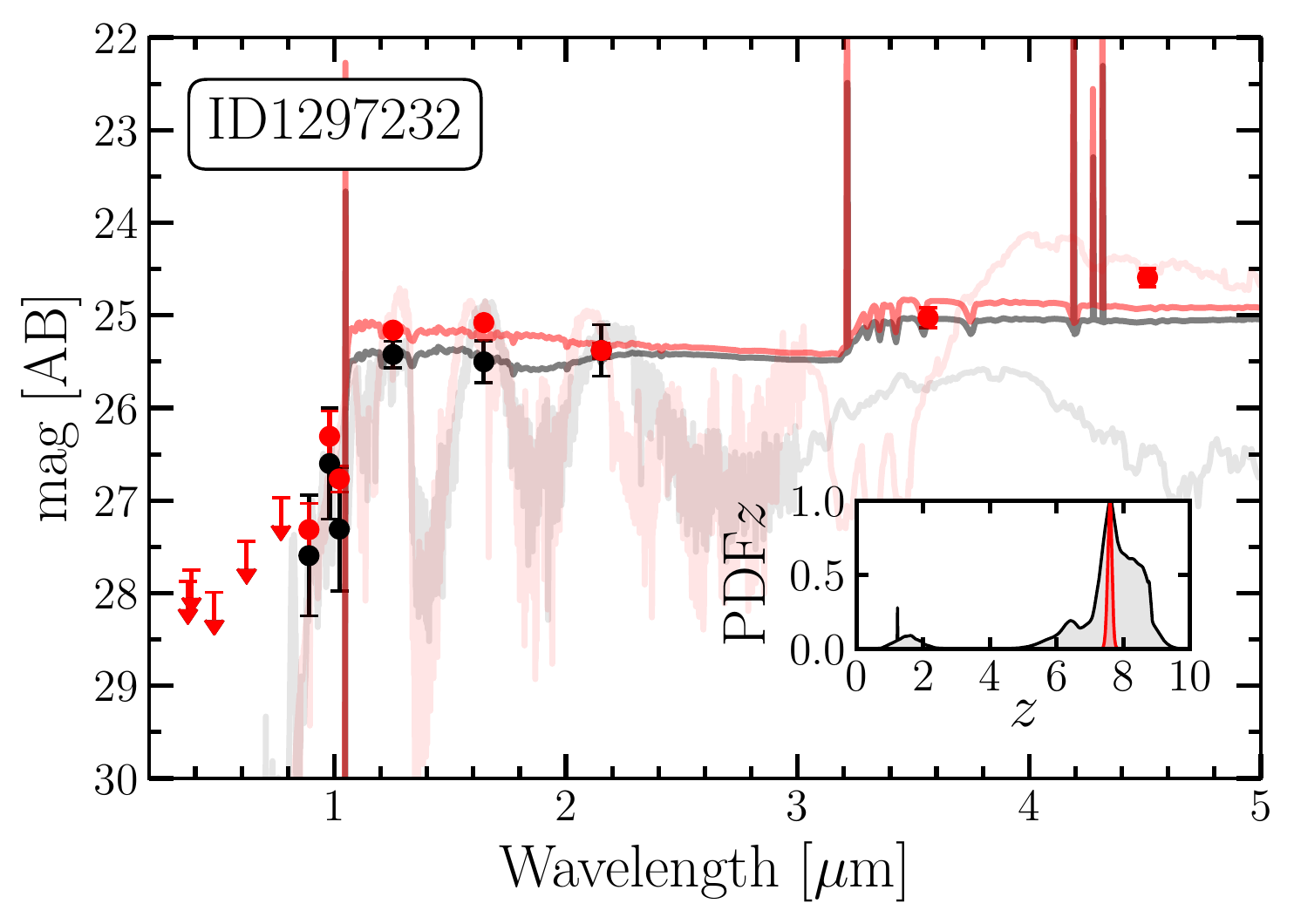}
  \includegraphics[width=0.33\textwidth]{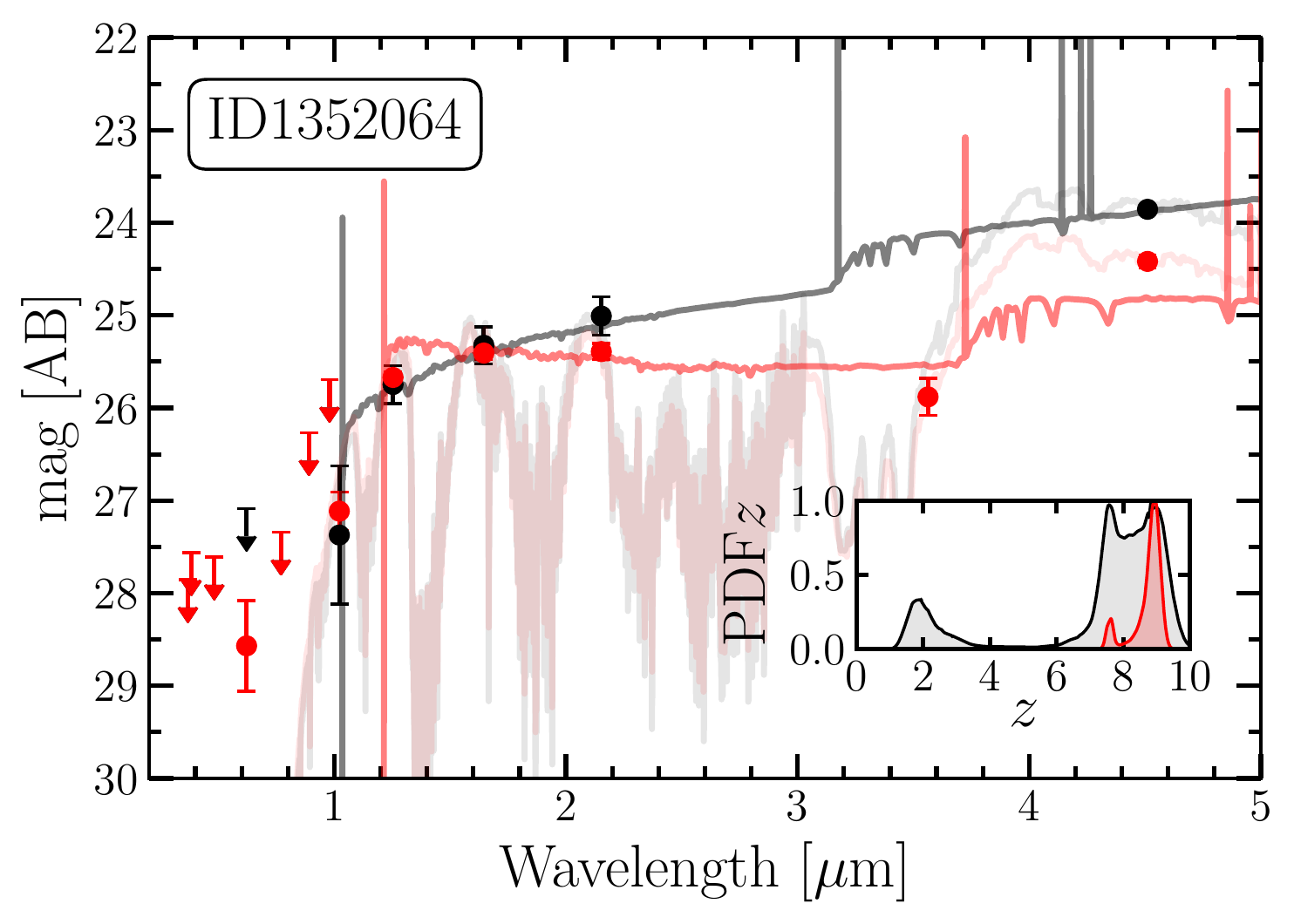}
  \includegraphics[width=0.33\textwidth]{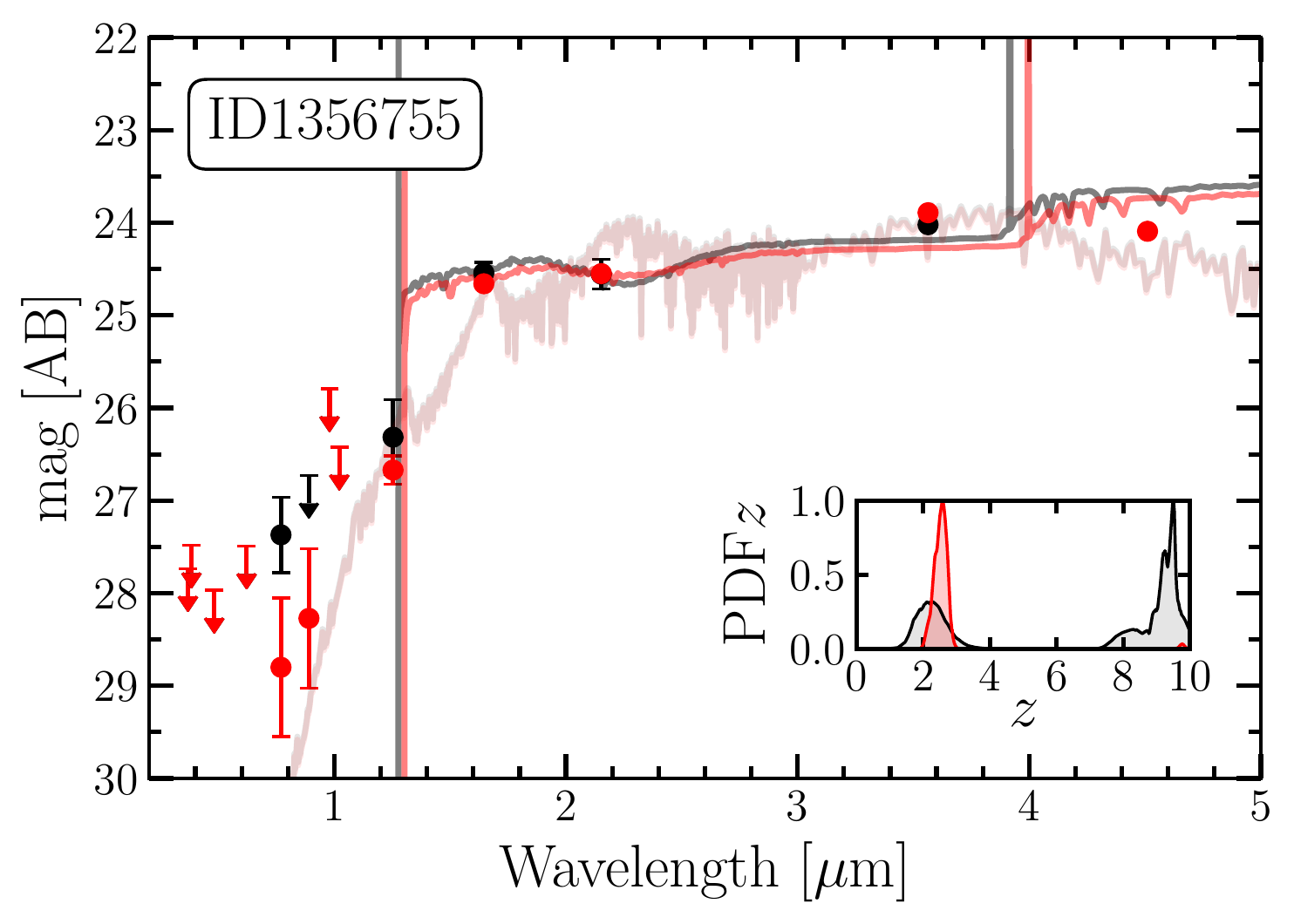}
  \includegraphics[width=0.33\textwidth]{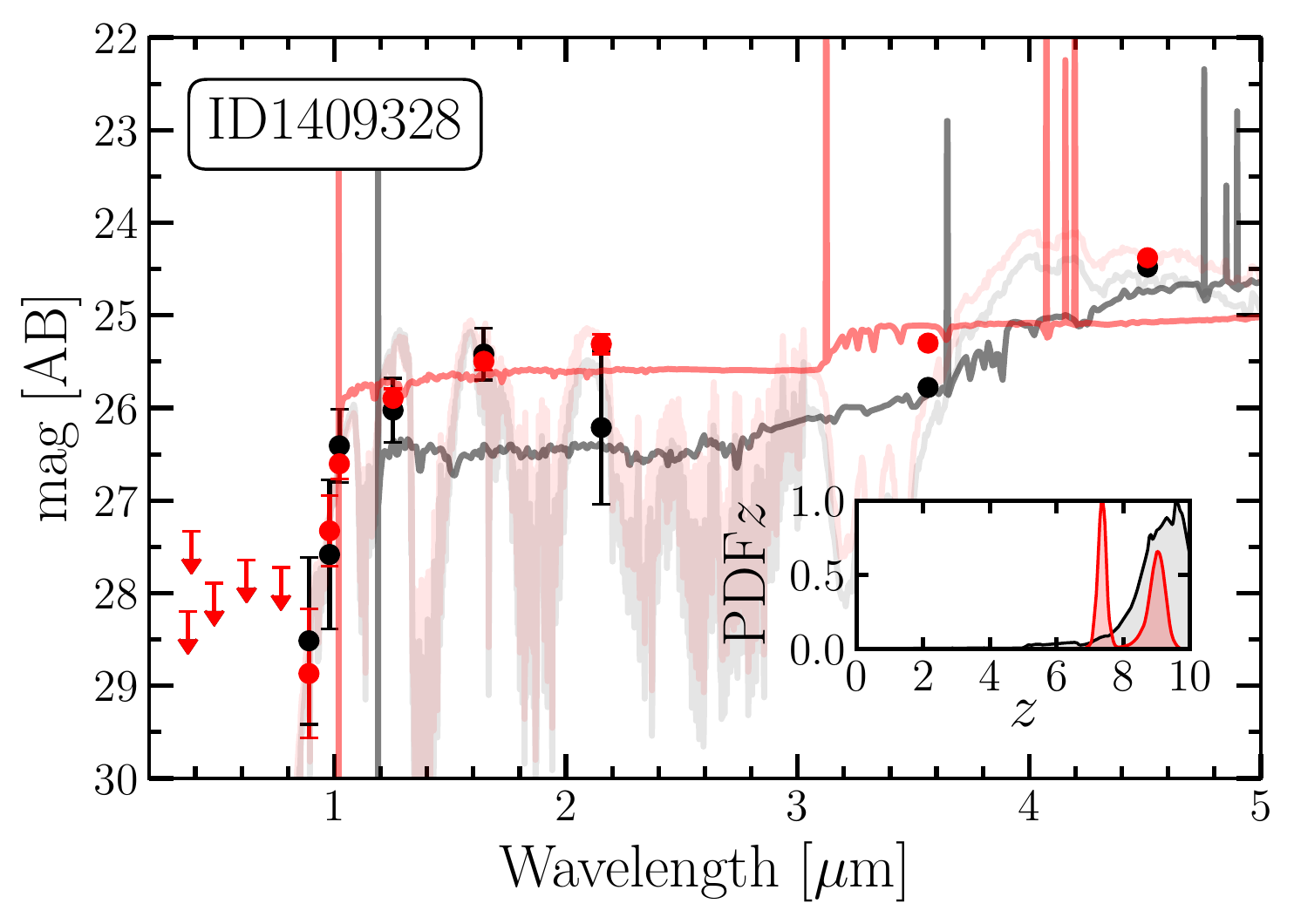}
  \includegraphics[width=0.33\textwidth]{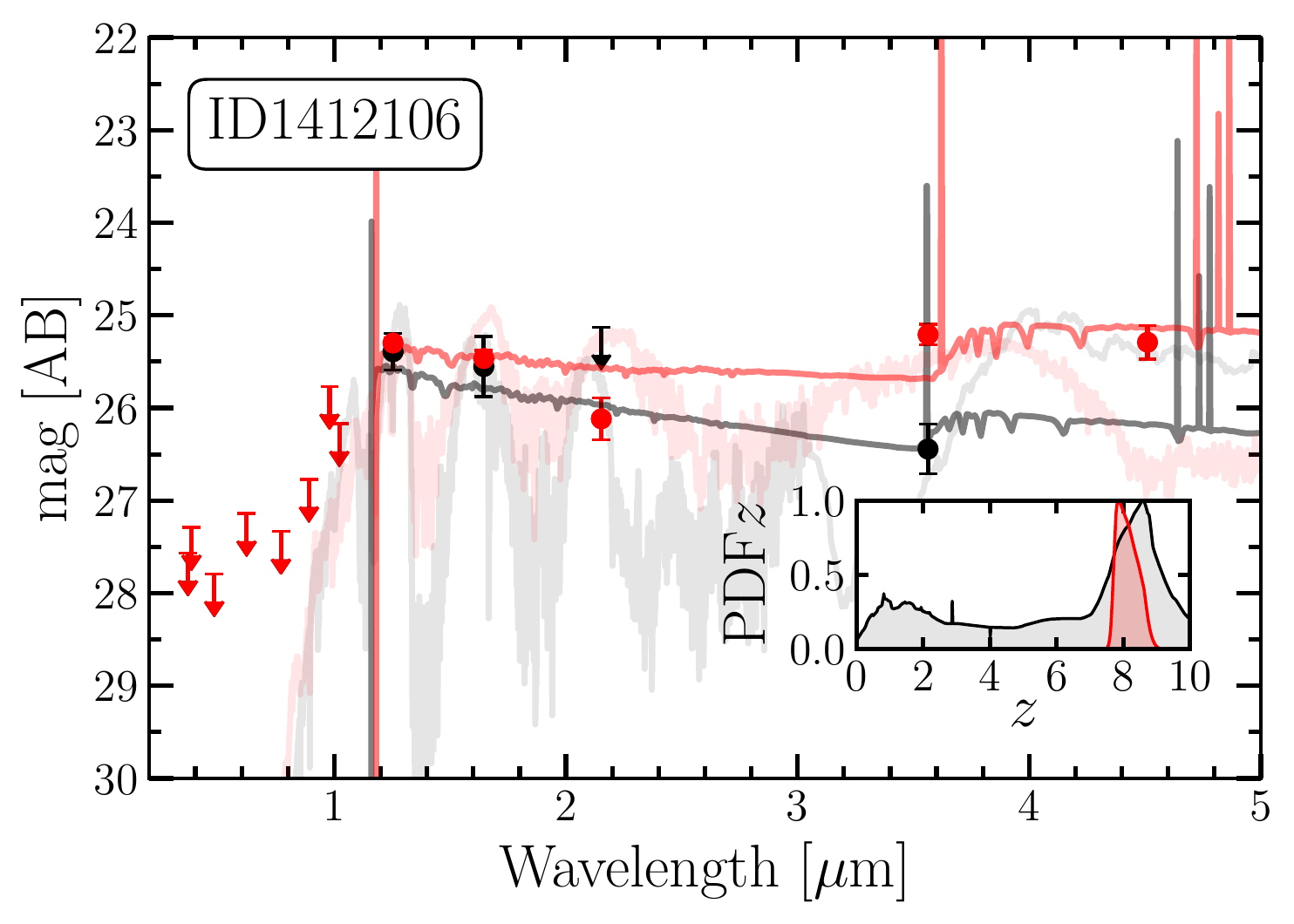}
  \caption{Observed photometry and best fitting templates for the new $z\geq7.5$ candidates in the COSMOS field. The markers and lines are based on the \classic{} (black) \farmer{} (red) photometry. The photometry is replaced by the $3\sigma$ upper limit for non-detection at $1\sigma$. The bright lines show the best-fitting galaxy templates. The faint lines show the best-fitting stellar templates. The insets give the redshift probability distributions. For ID720309, \farmer{} photometry is measured on the paw~2 stack, free from the potential cross-talk contamination. We show only the fit corresponding to this case (see Sect.~\ref{sec:ID720309}).
  }
  \label{fig:seds_unblended_1}
\end{figure}

From Fig.~\ref{fig:seds_unblended_1} to~\ref{fig:seds_ancillary}, we show the observed photometry of each candidate, together with the best-fitted galaxy and stellar templates, for both the \classic{} and \farmer{} catalogues. In the case where the flux is smaller than the flux uncertainty, for clarity the photometric measurement is replaced by a $3\sigma$ upper limit. 
The \classic{} photometry and posterior redshift distributions are also shown. Figures~\ref{fig:snaps_unblended_1}, \ref{fig:snaps_unblended_2}, \ref{fig:snaps_blended}, \ref{fig:snaps_ancillary_1} and \ref{fig:snaps_ancillary_2} display stamp images centred on the candidate coordinates. 

We note that there may be some offsets between the total apparent magnitudes between the \classic{} and \farmer{} photometric COSMOS2020 catalogues, even though the observed colours match each others. One of the main reasons for this is that \texttt{IRACLEAN} and \farmer{} provide total fluxes, whereas fixed aperture fluxes are used in the \classic{} catalogue. Hence, these aperture fluxes are rescaled using aperture-to-total corrections applied to all the aperture-extracted bands, and computed from the weighted mean difference between fixed aperture and pseudo-total fluxes (using Kron apertures) from \texttt{SExtractor}. While the corrections remain low (-0.29, -0.11, -0.01 mag for the 25\%, 50\%, 75\% percentiles), this procedure introduces some additional noise for faint sources. 

In IRAC, the flux is already total in both catalogues. Still, we find inconsistencies between the \texttt{IRACLEAN} and \farmer{} photometry (e.g. for ID720309 and ID1103149), partially explained by the two different approaches in the algorithms. \farmer{} assumes a parametric light-profile for a galaxy, convolved with the IRAC PSF before the fit. \texttt{IRACLEAN} repeatedly removes point-like source contributions from the residual map, until reaching a threshold with no pixel having a flux above a given signal-to-noise within the detection area, defined by the segmentation map. We identify that the presence of close-by objects induces large differences between the two sets of IRAC photometry. The different deblending procedures in \farmer{} and \classic{} introduce small differences in the segmentation map, resulting in this case in large variations of the IRAC flux. We are not able to conclude the superiority of one catalogue over the other, highlighting the importance of using several methods to assess the robustness of the results.

\subsection{New galaxy candidates}

We identify $17$ new candidates at $z\geq7.5$ from the selection in both COSMOS2020 catalogues. We describe them below.

We find two candidates with $z_\text{phot}>9.5$ according to \lephare{}/\farmer{} results. The candidate ID441697 is robustly detected in the $J$ band, its photometric redshift with \farmer{} is $z_\text{phot}=9.51_{-0.15}^{+0.12}$. In this case, the \classic{} photometric redshift $z_\text{phot}=9.09_{-0.68}^{+0.35}$ suggests a lower redshift, nonetheless all the \zPDF{} weight is at $z>8$ in both catalogues. The other candidate ID720309 is detected in $H$ and $K_s$ bands, and has $z_\text{phot}\sim 10.1$ for \farmer{} photometry and $z_\text{phot}\sim9.7$ in the \classic{} catalogue but see Sect.\ref{sec:ID720309}. 

Four sources (ID241443, ID984164, ID1412106) have a primary solution at $z<5$ when applying \eazy{} to the \classic{} catalogue or allowing for more attenuation in the fit (BC03 configuration). We note that these sources are located outside the ultra-deep HSC region which could explain a looser constraint on the \zPDF{} and multiple peaks in redshifts (in particular for the \classic{} catalogue). 

One new candidate, ID234500, is located in the deep stripe at the south-eastern edge of the field (at high RA). The south-eastern field edge has been covered by NIR data for UltraVISTA DR4 for the first time in COSMOS2020. This region had been previously masked because of the non-uniform quantum efficiency of the VISTA NIR detectors; this region has a higher noise, particularly in the $Y$-band. Moreover, new HSC DR2 data in this area allows us to select this candidate in a region which has already been studied by \citetalias{stefanon_brightest_2019} and \citetalias{bowler_lack_2020}. 

The observed photometry and images of best-fitted galaxy templates of two blended candidates ID442053 and ID1313521 are shown in Fig.~\ref{fig:seds_blended} and Fig.~\ref{fig:snaps_blended}, respectively.These candidates have magnitudes $25.0<H<25.5$. For these objects we find a \photoz{} solution at low redshift $z<3$ with the \classic{} catalogue. From aperture photometry alone, all of these candidates have a $3\sigma$ detection in at least one HSC optical band. The \classic{} redshift probability distributions peak at $z\sim2$ for all of them, although one shows a secondary $z>7$ solution. In contrast, the majority of the \zPDF{} weights are located at $z>7$ with \farmer{} catalogue. 

\onecolumn

\begin{figure}
  \centering
  \includegraphics[width=\textwidth]{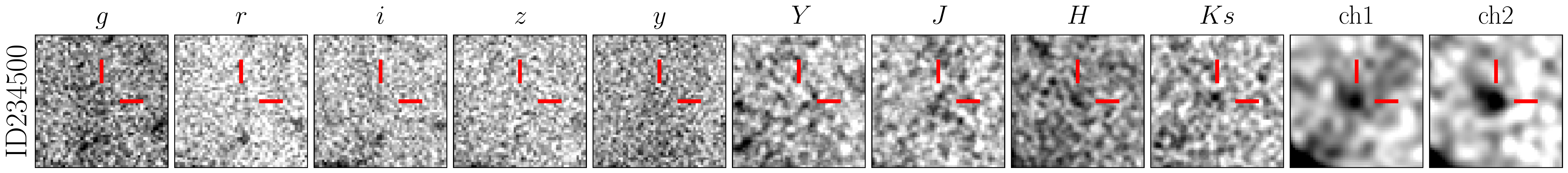}
  \includegraphics[width=\textwidth]{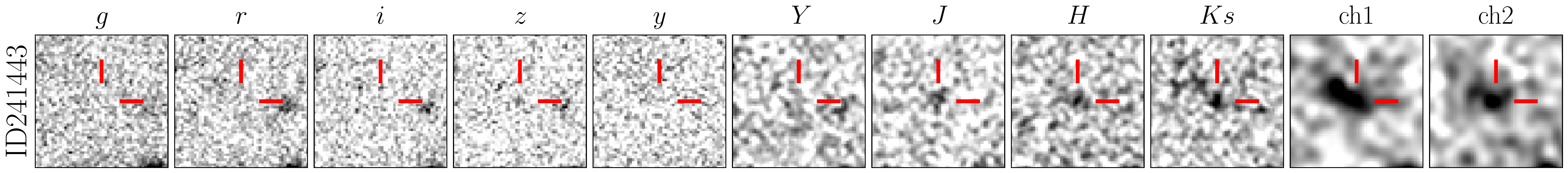}
  \includegraphics[width=\textwidth]{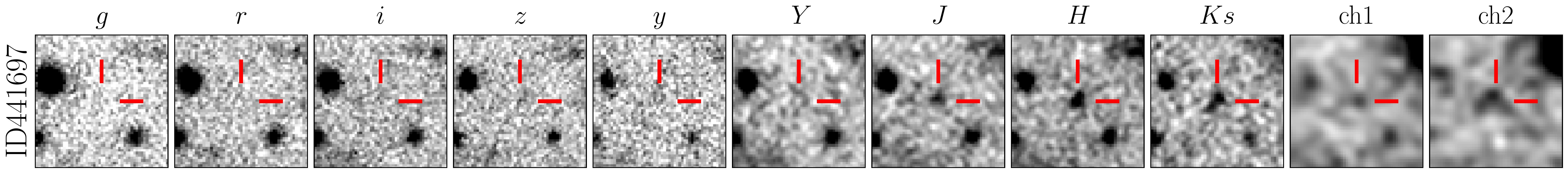}
  \includegraphics[width=\textwidth]{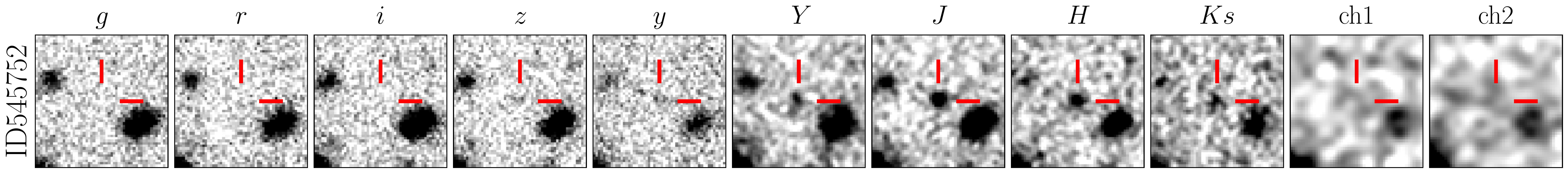}
  \includegraphics[width=\textwidth]{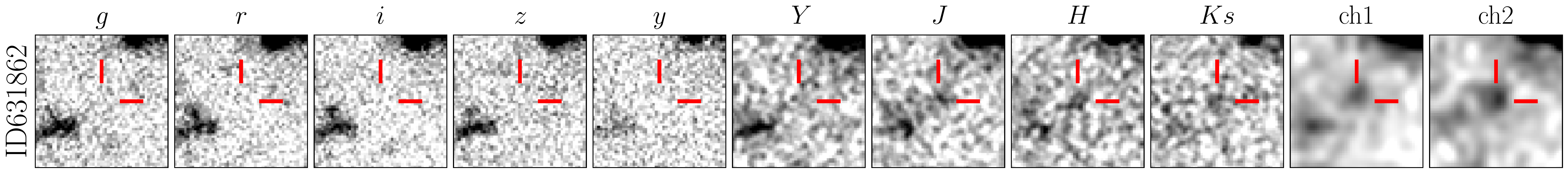}
  \includegraphics[width=\textwidth]{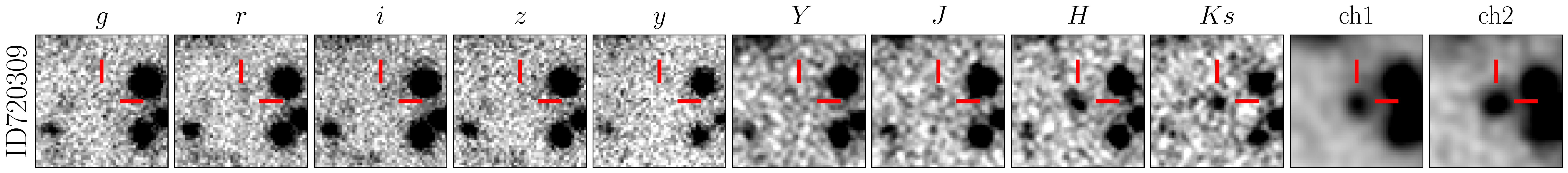}
  \includegraphics[width=\textwidth]{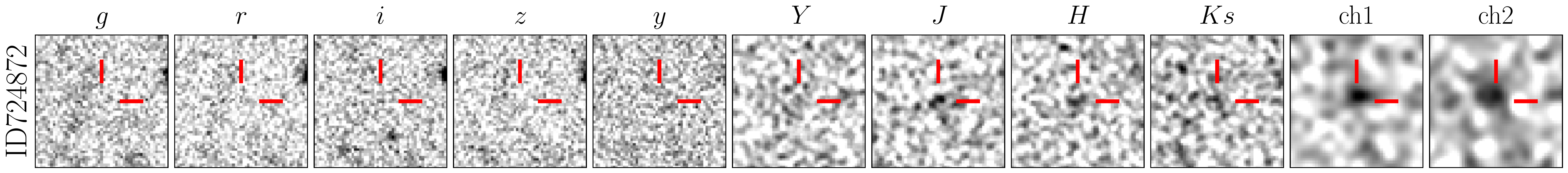}
  \includegraphics[width=\textwidth]{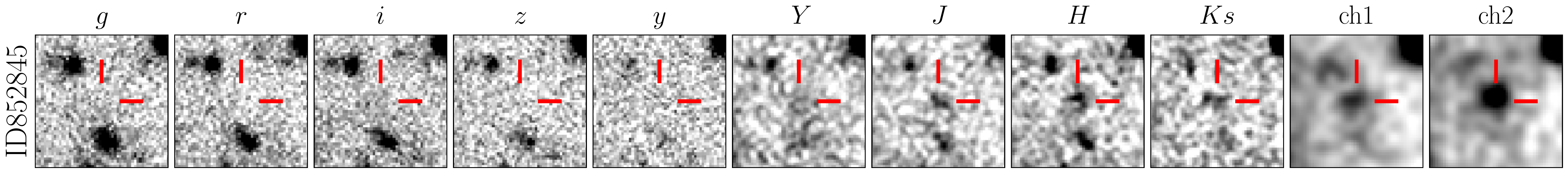}
  \includegraphics[width=\textwidth]{fig/cand_unblended/snap_ID852845.pdf}
  \includegraphics[width=\textwidth]{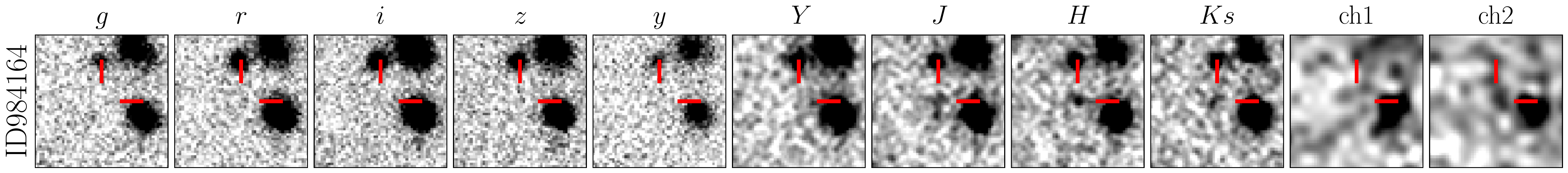}
  \caption{Stamp images of the $z\geq7.5$ new  candidates in the COSMOS field. Each candidate appears in one row of stamps. The stamps are $8\arcsec$ wide, with north to the top and east to the left. The stamps are saturated beyond the range $[-1,4]\sigma$, where $\sigma$ is the $3\sigma$ clipped standard deviation of the pixel values in the stamp. 
  }
  \label{fig:snaps_unblended_1}
\end{figure}

\begin{figure}
  \centering
  \includegraphics[width=\textwidth]{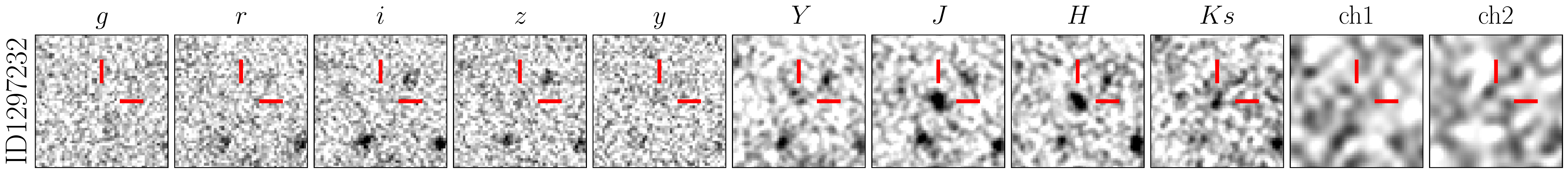}
  \includegraphics[width=\textwidth]{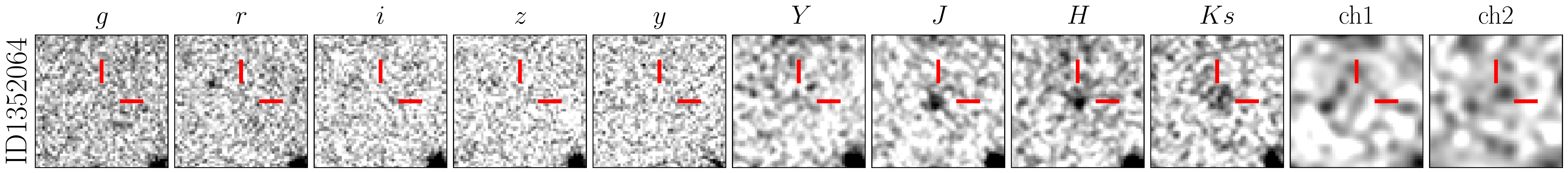}
  \includegraphics[width=\textwidth]{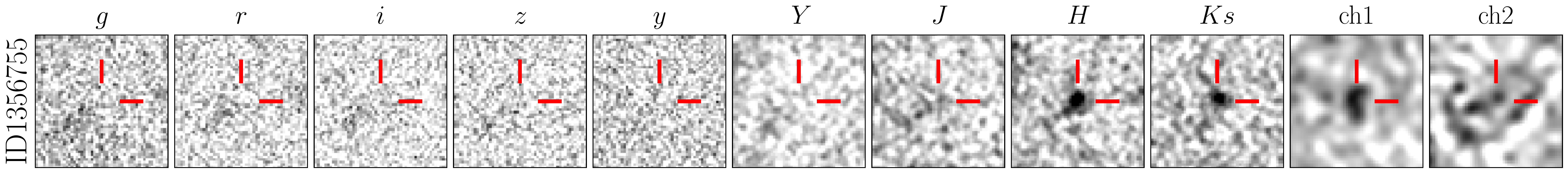}
  \includegraphics[width=\textwidth]{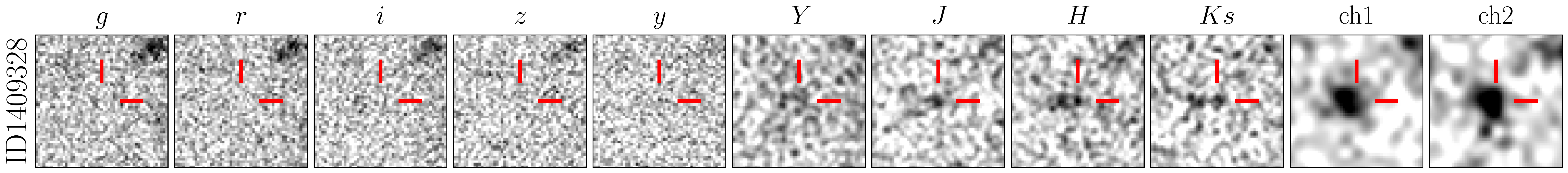}
  \includegraphics[width=\textwidth]{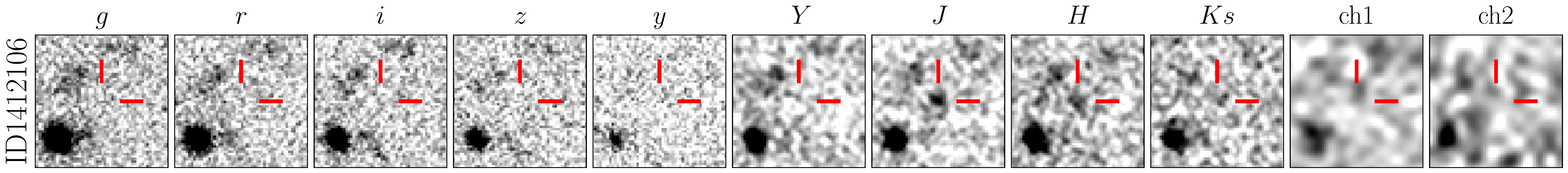}
  \caption{continued}
  \label{fig:snaps_unblended_2}
\end{figure}

\begin{figure}
  \centering
  \includegraphics[width=0.33\textwidth]{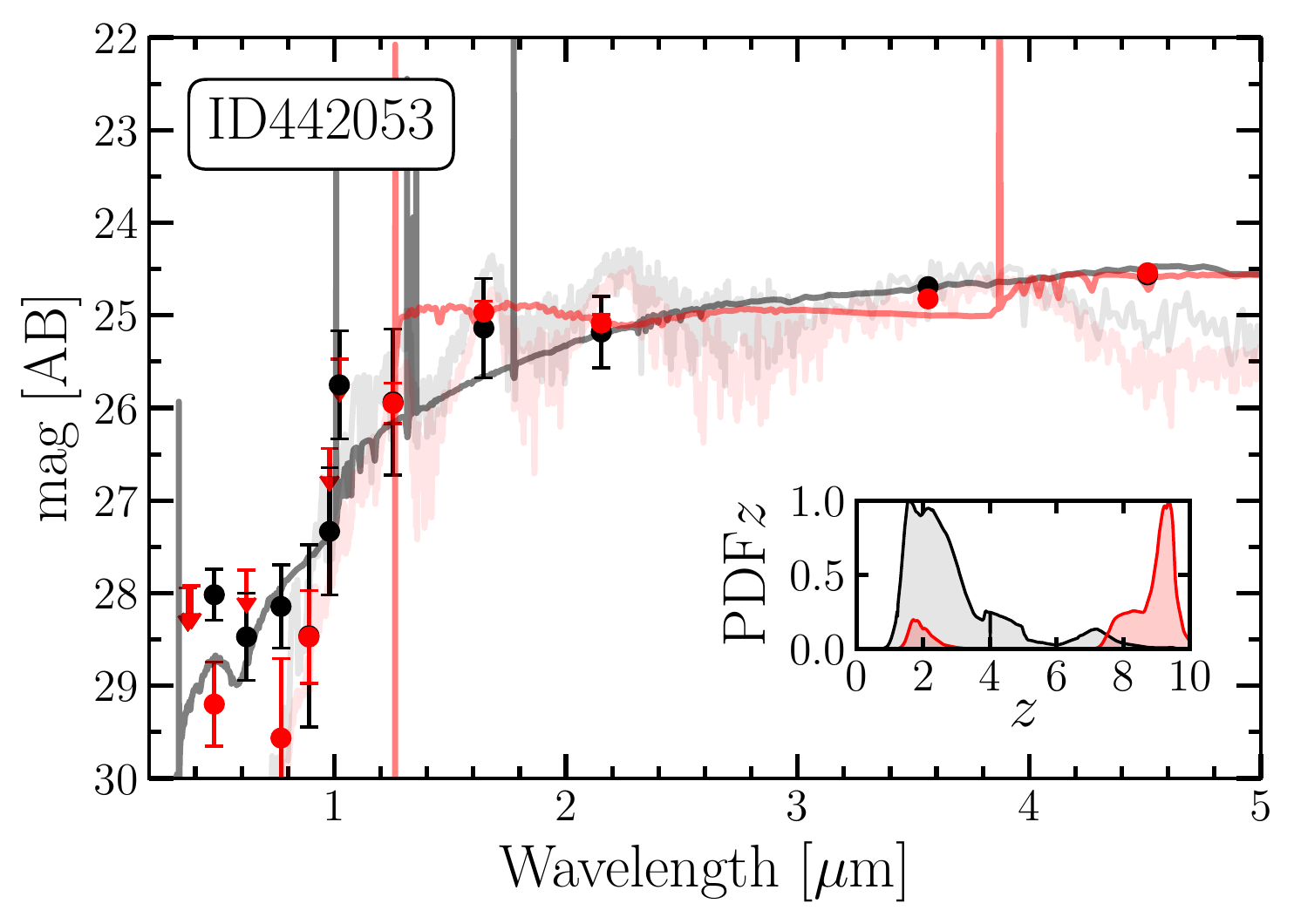}
  \includegraphics[width=0.33\textwidth]{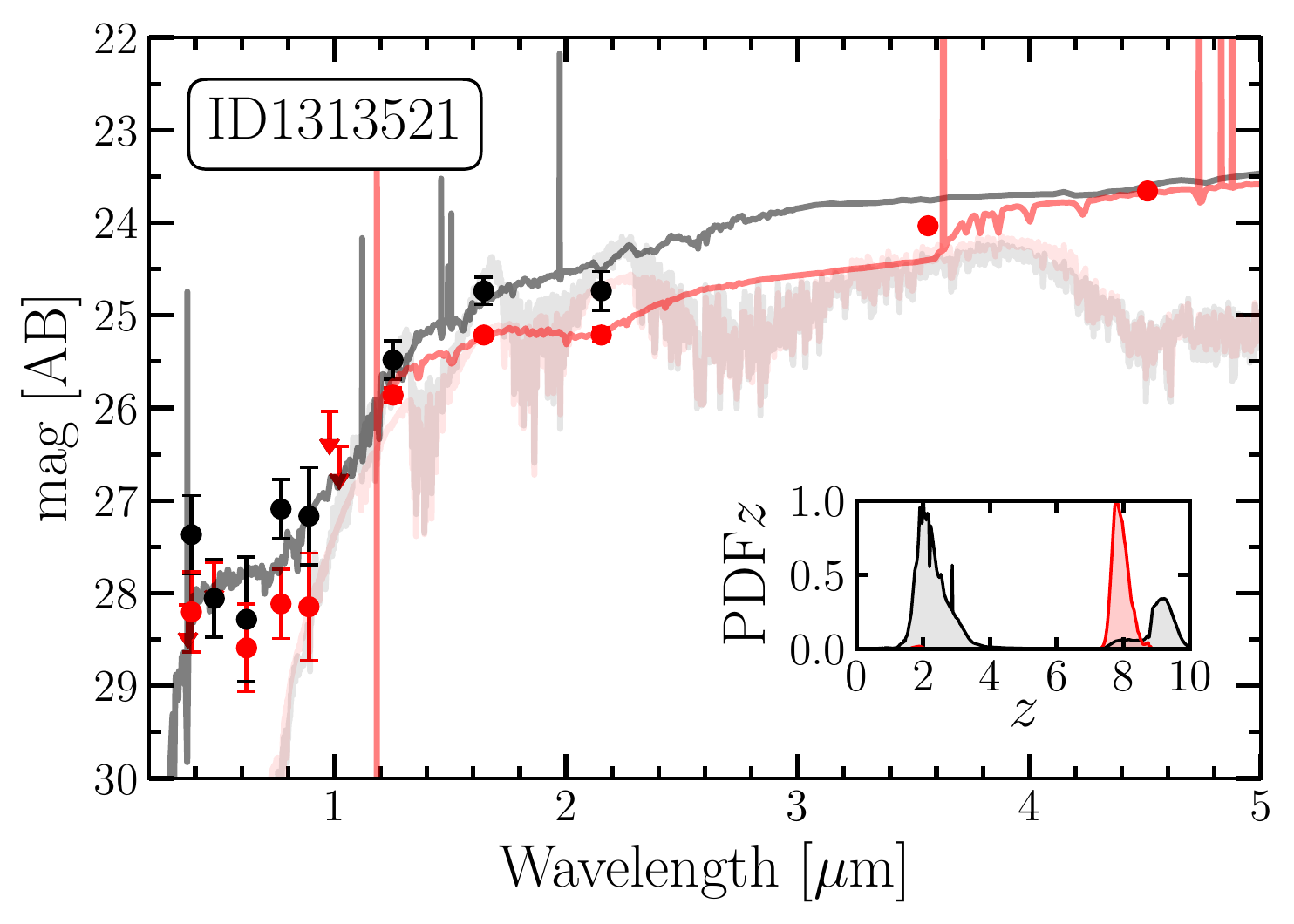}
  \caption{Same as Fig.~\ref{fig:seds_unblended_1}, but for the new blended $z\geq7.5$ galaxy candidates.}
  \label{fig:seds_blended}
\end{figure}

\begin{figure}
  \centering
  \includegraphics[width=\textwidth]{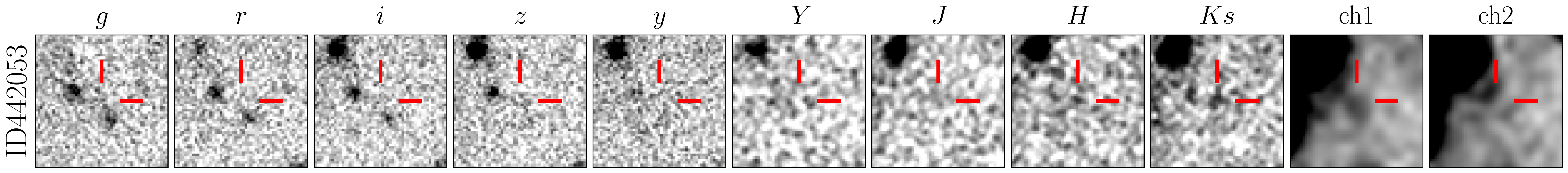}
  \includegraphics[width=\textwidth]{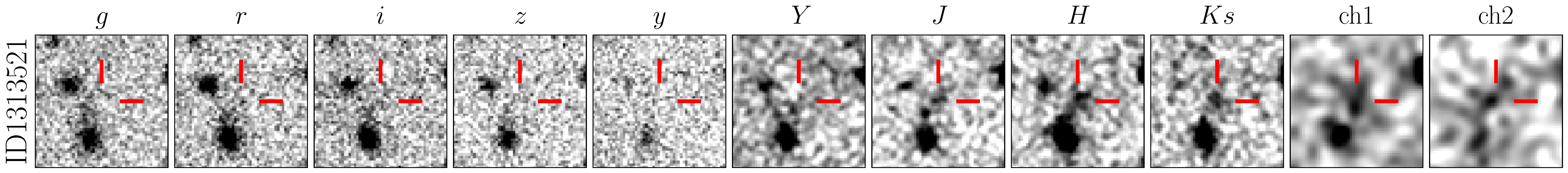}
  \caption{Same as Fig.~\ref{fig:snaps_unblended_1}, but for the new blended $z\geq7.5$ galaxy candidates.}
  \label{fig:snaps_blended}
\end{figure}
\twocolumn

\subsection{Previously published galaxy candidates}

\begin{figure}
  \centering
  \includegraphics[width=0.33\textwidth]{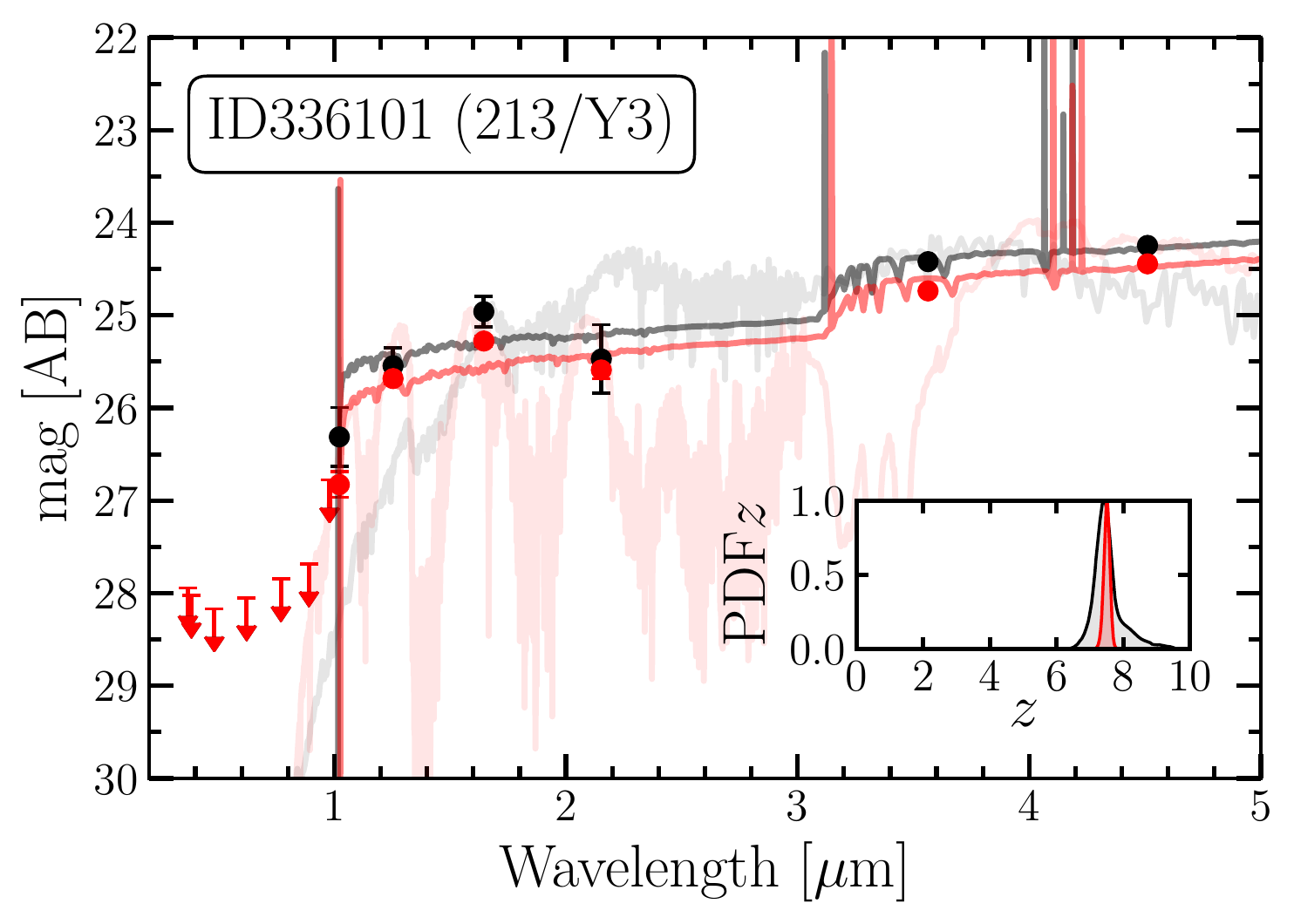}
  \includegraphics[width=0.33\textwidth]{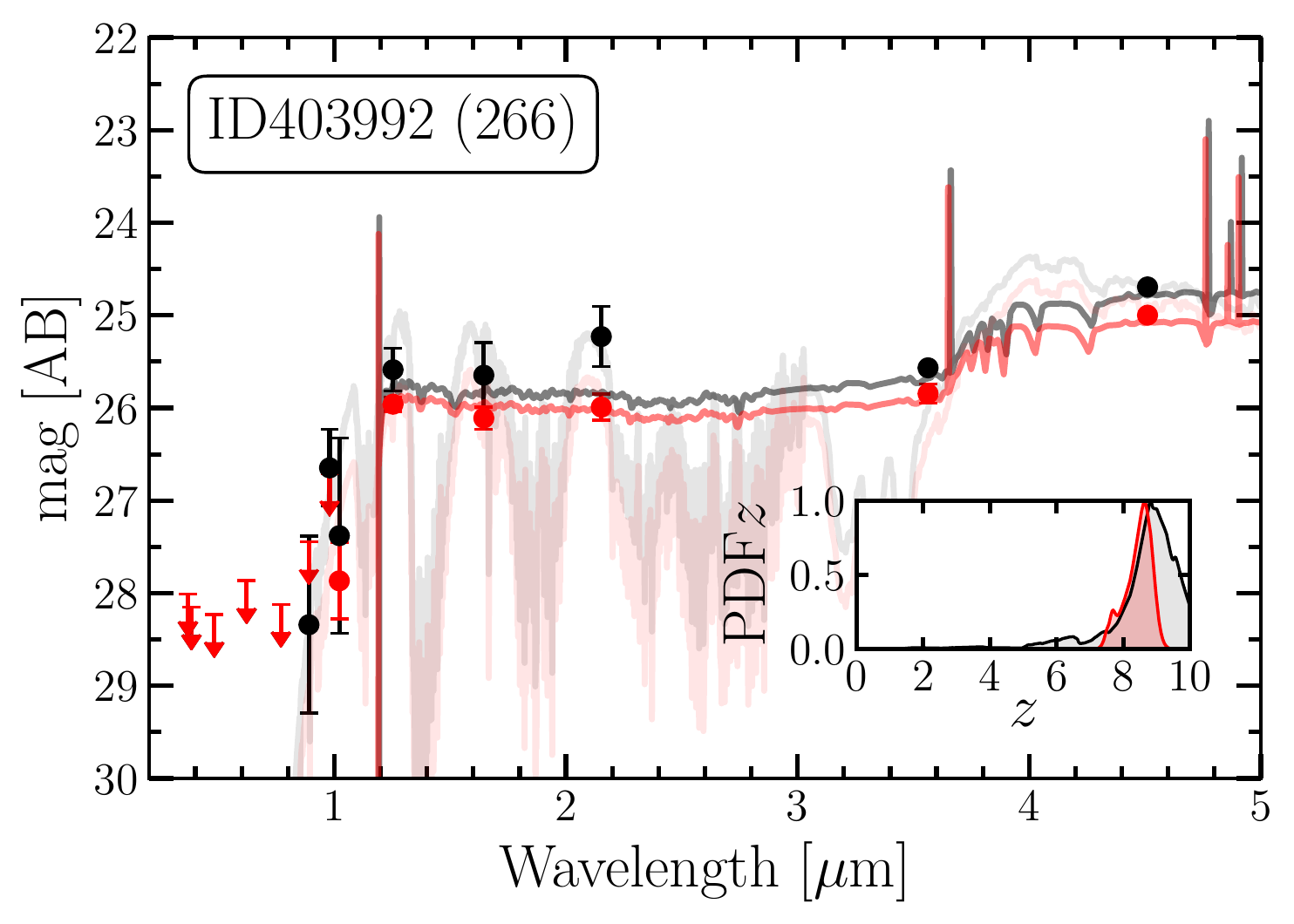}
  \includegraphics[width=0.33\textwidth]{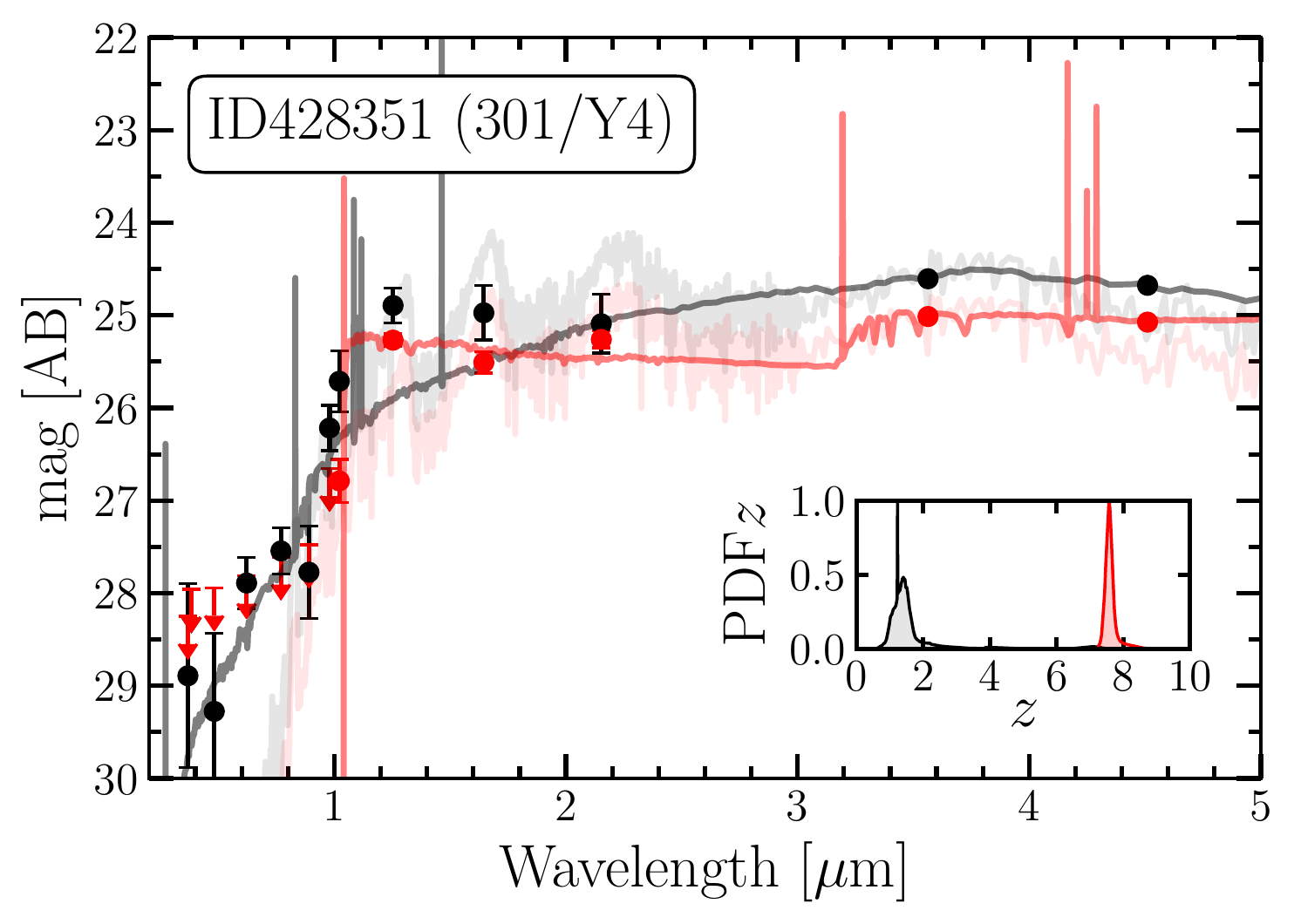}
  \includegraphics[width=0.33\textwidth]{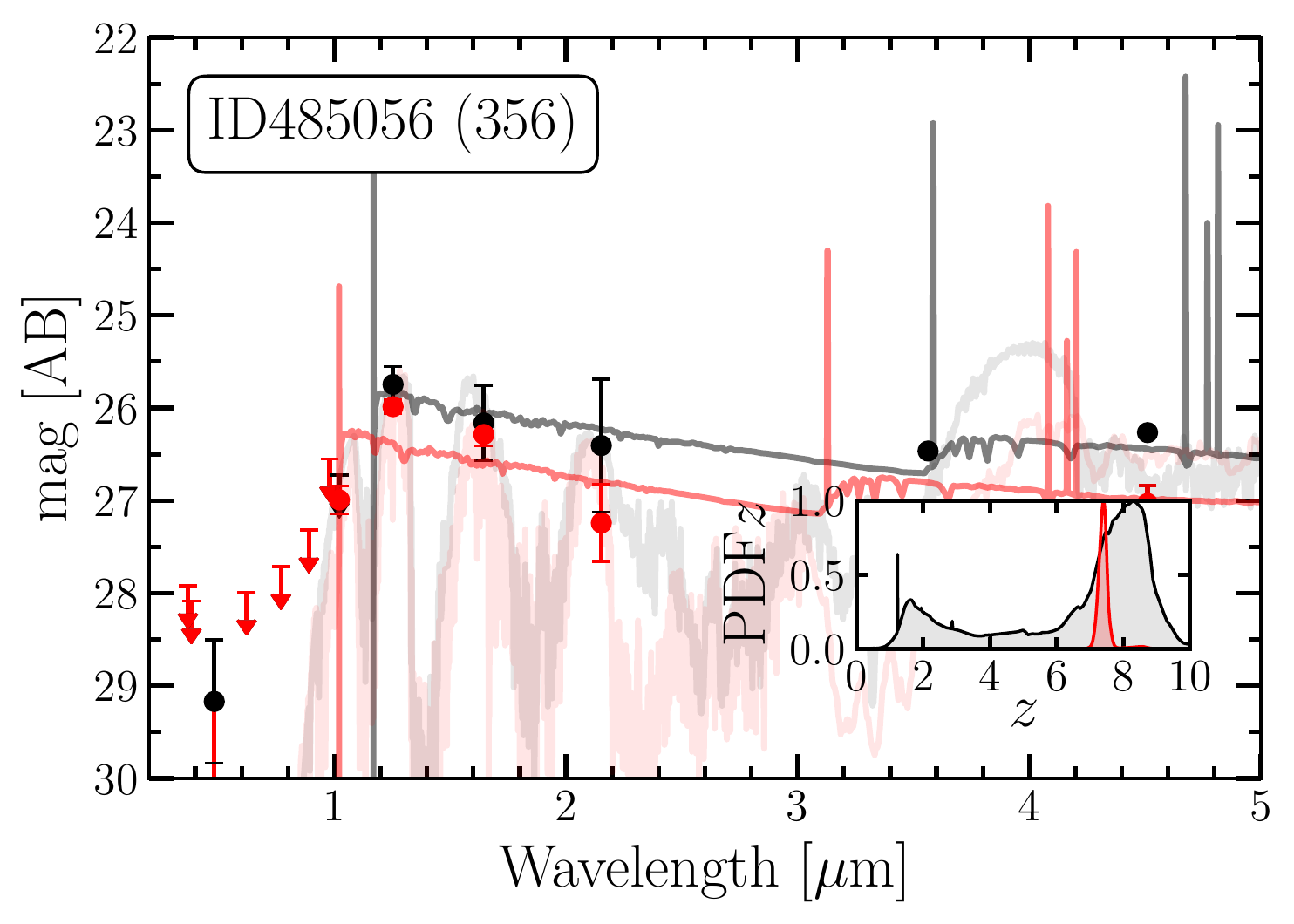}
  \includegraphics[width=0.33\textwidth]{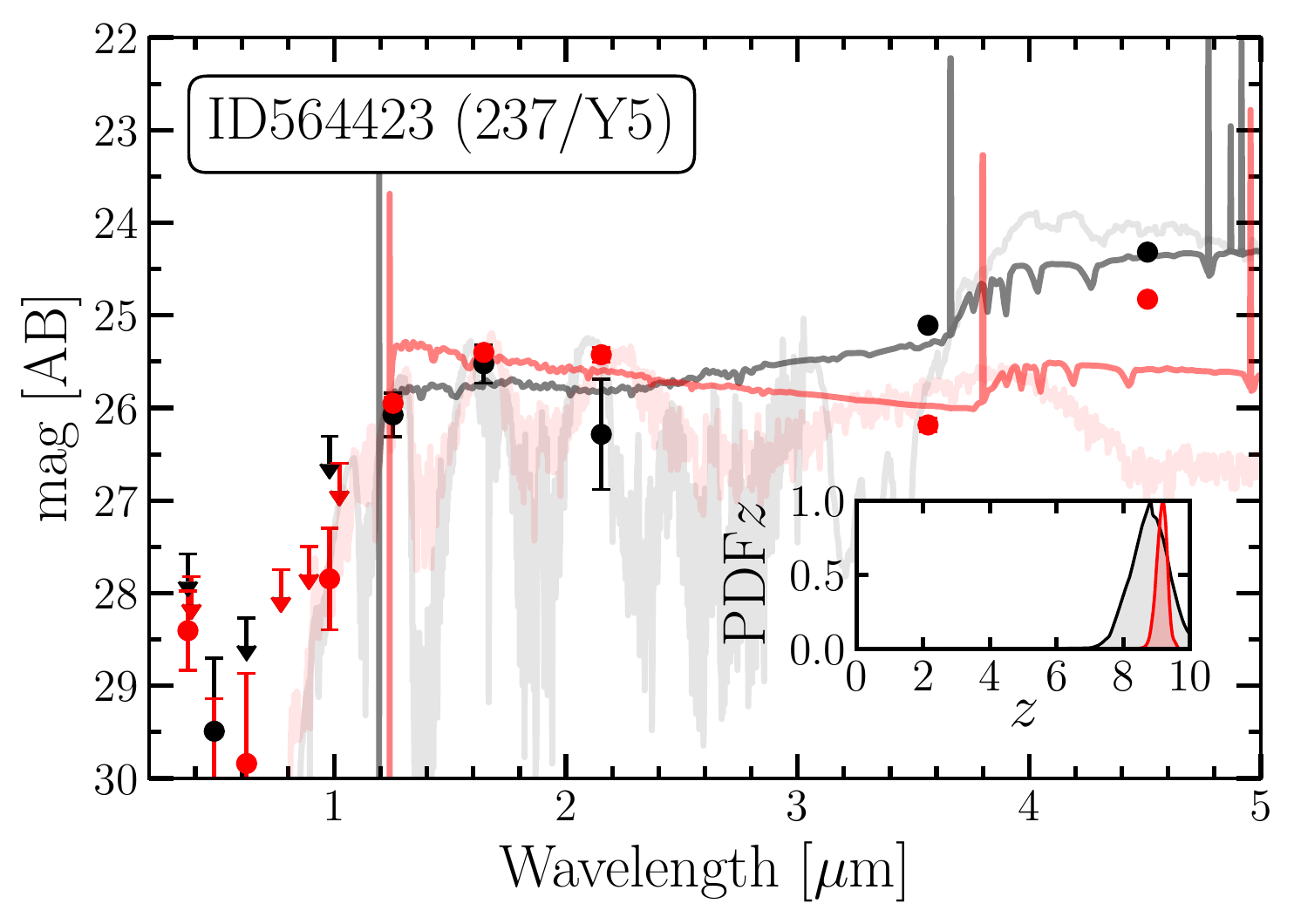}
  \includegraphics[width=0.33\textwidth]{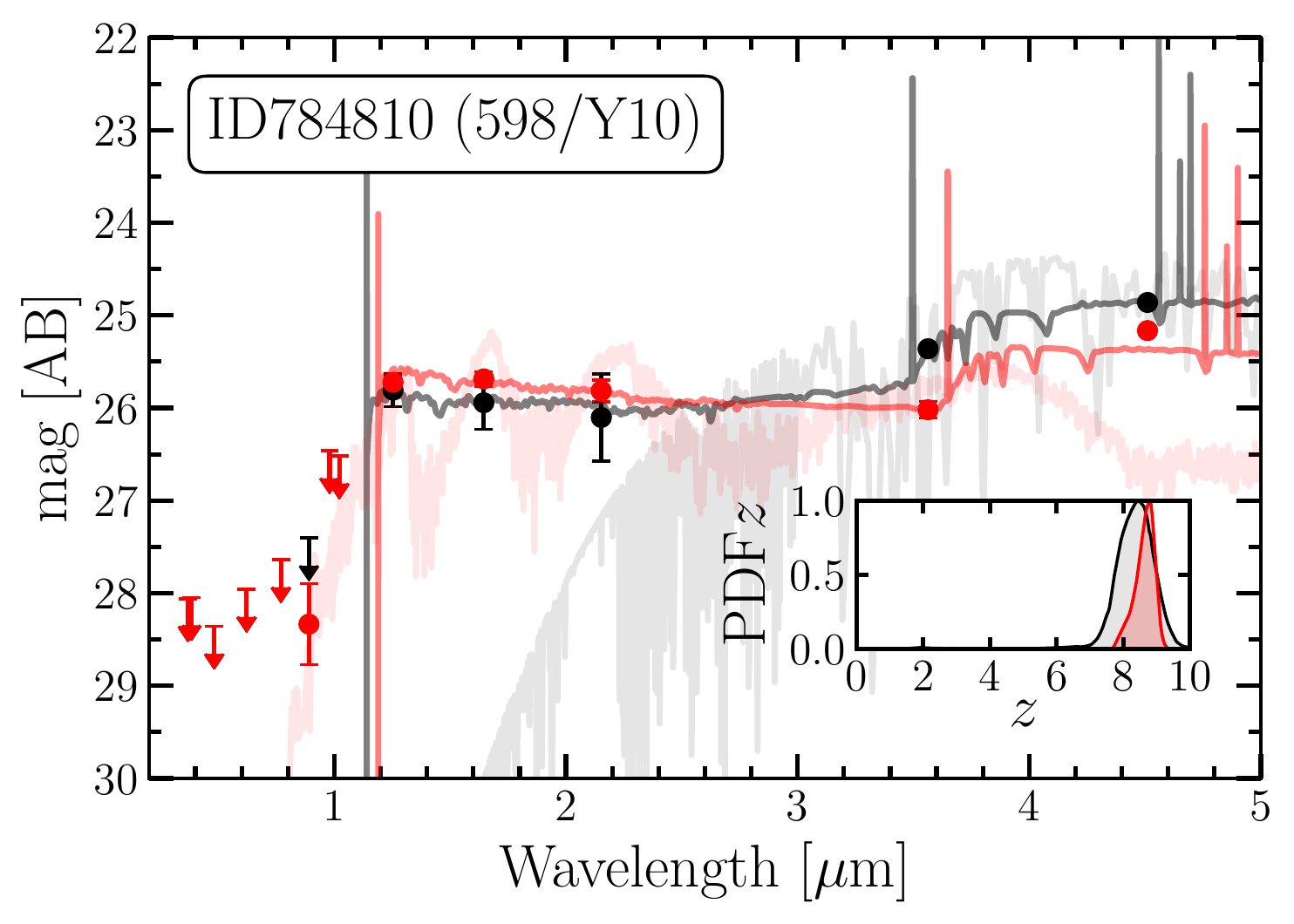}
  \includegraphics[width=0.33\textwidth]{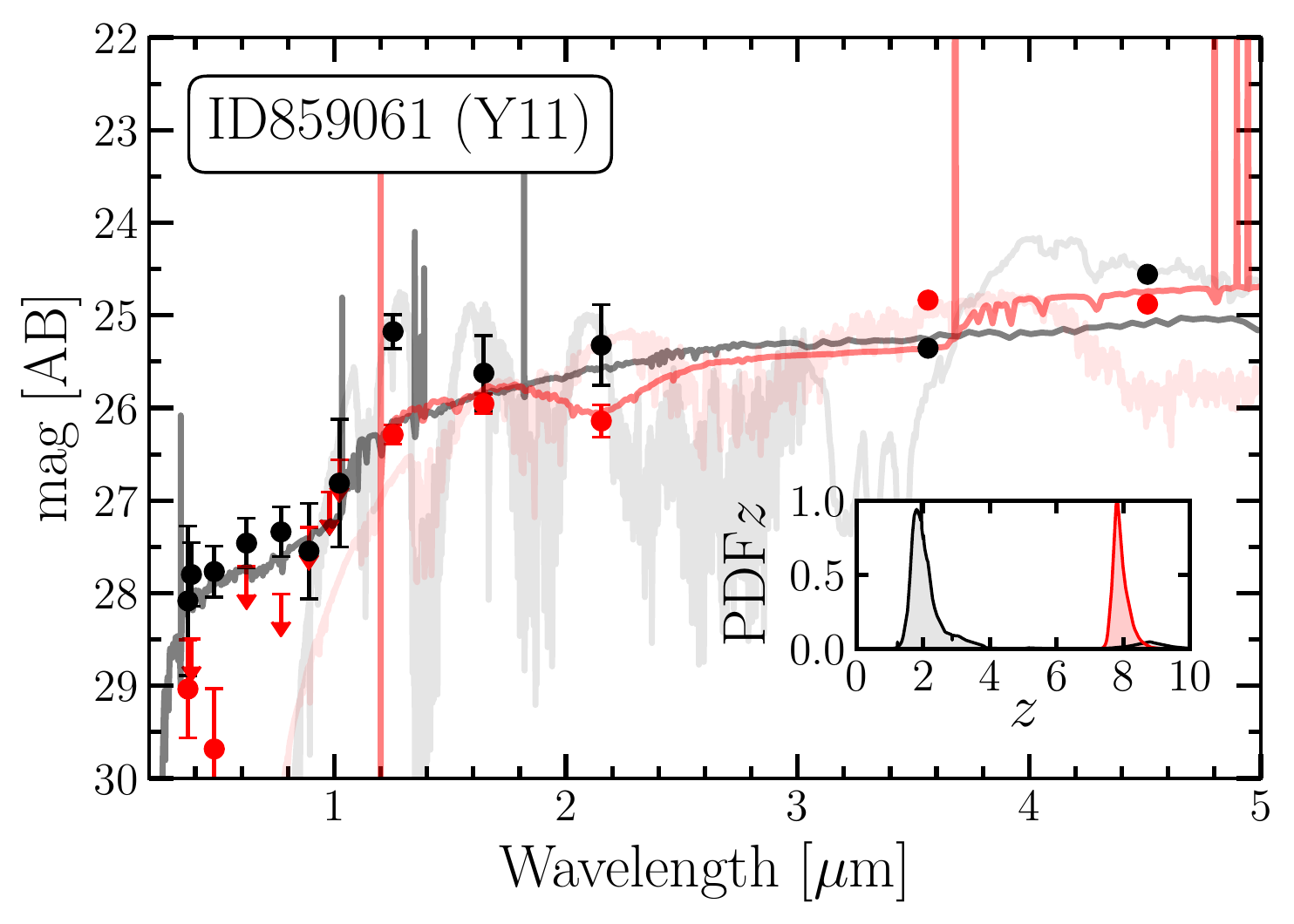}
  \includegraphics[width=0.33\textwidth]{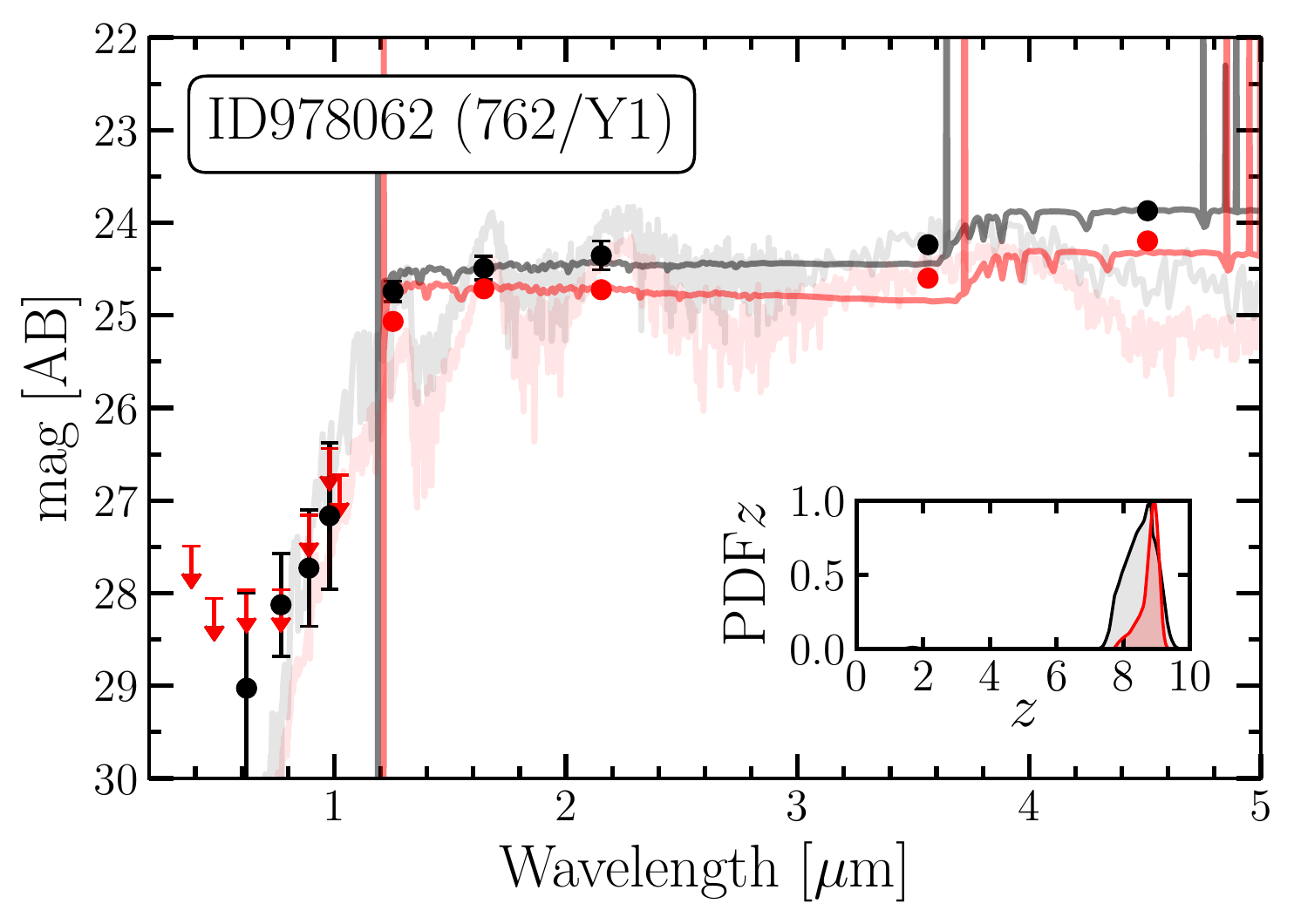}
  \includegraphics[width=0.33\textwidth]{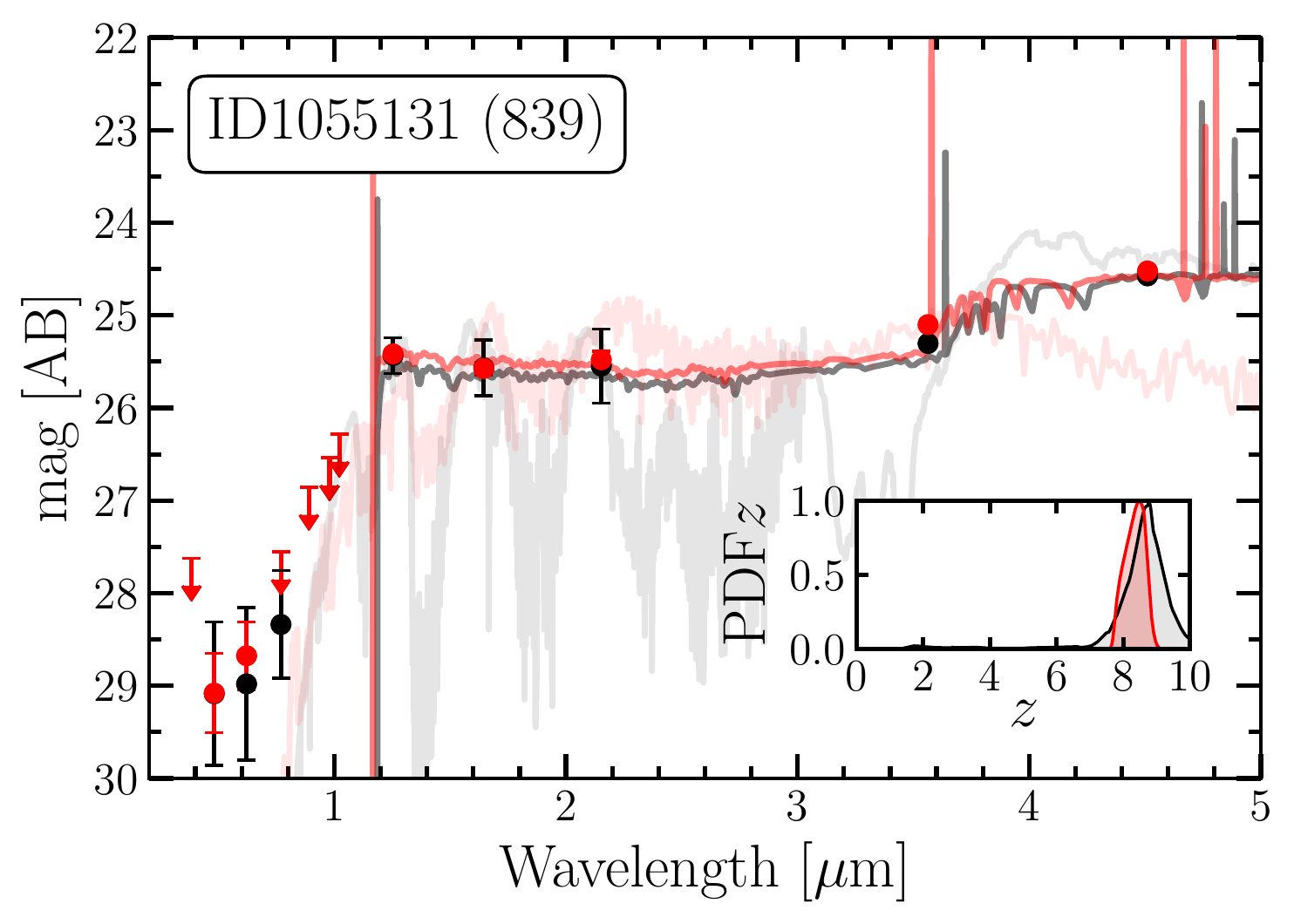}
  \includegraphics[width=0.33\textwidth]{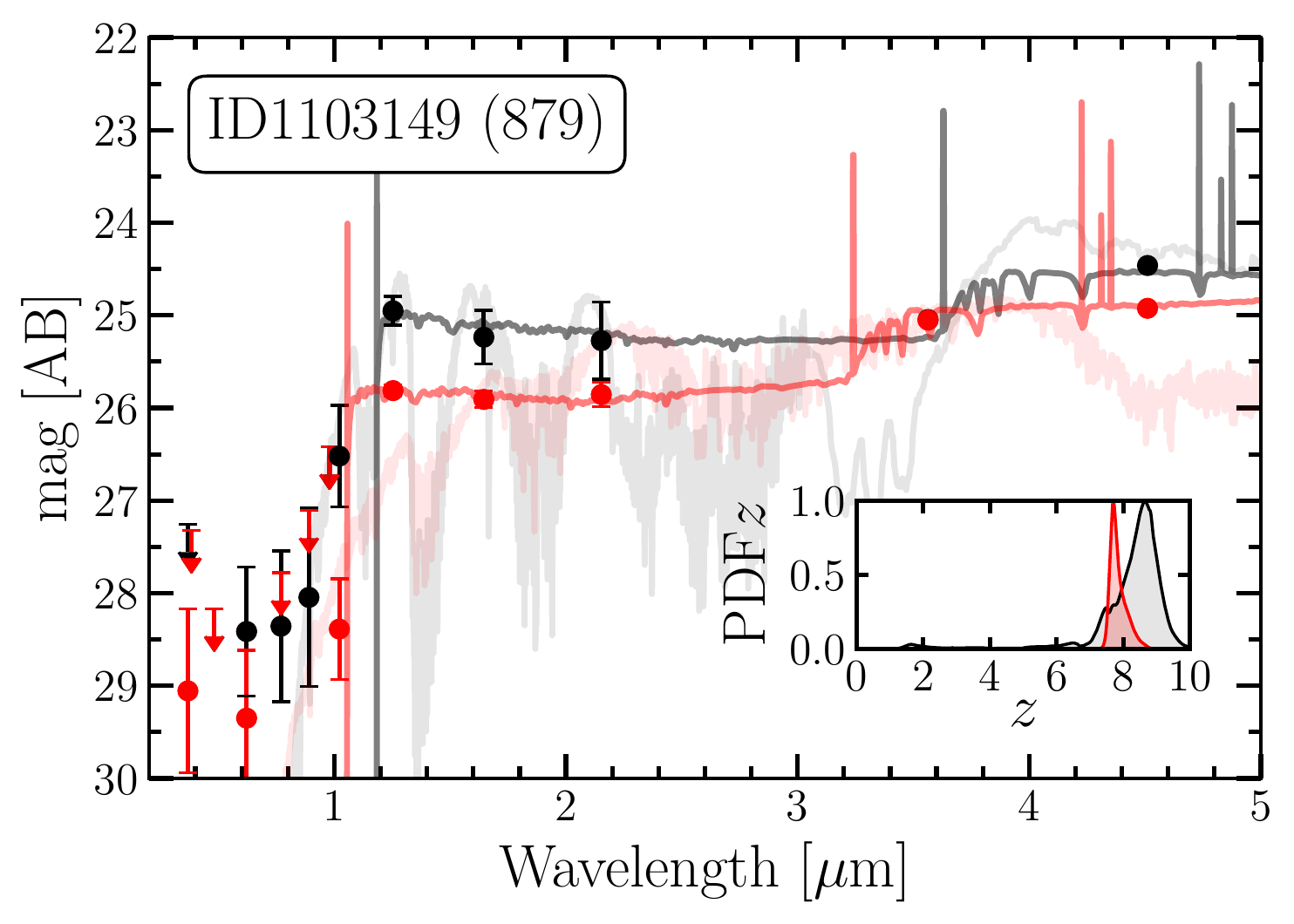}
  \includegraphics[width=0.33\textwidth]{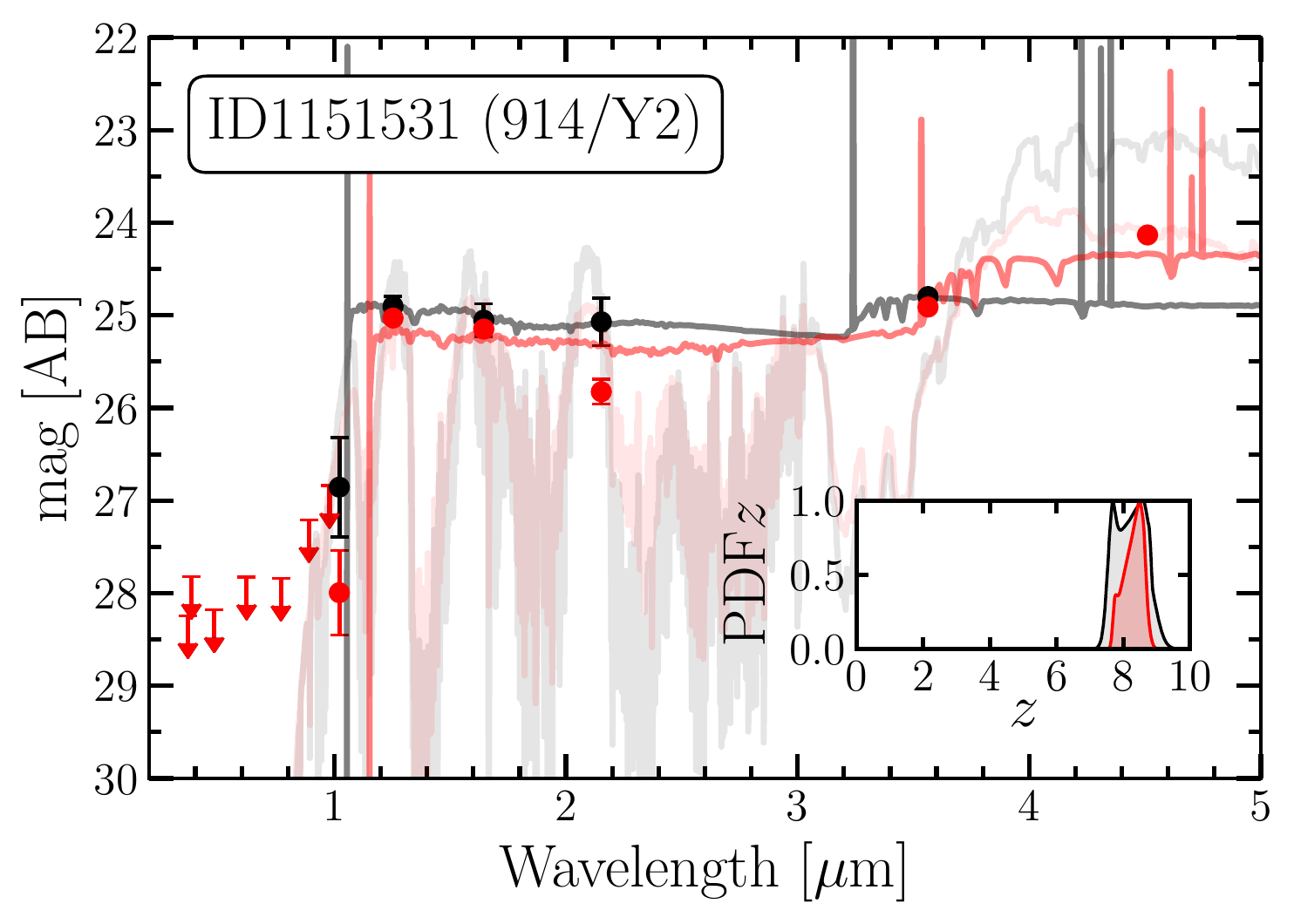}
  \includegraphics[width=0.33\textwidth]{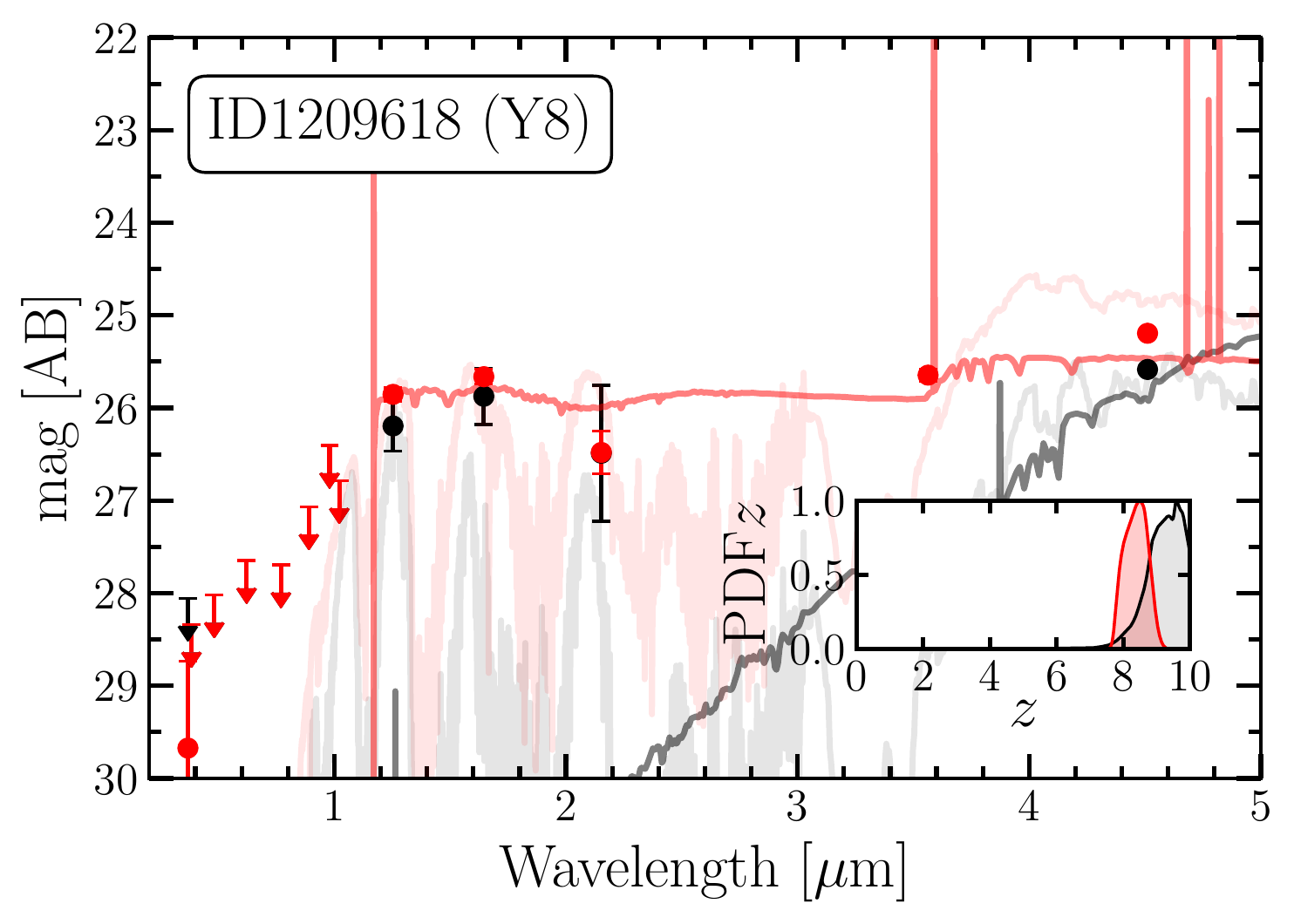}
  \includegraphics[width=0.33\textwidth]{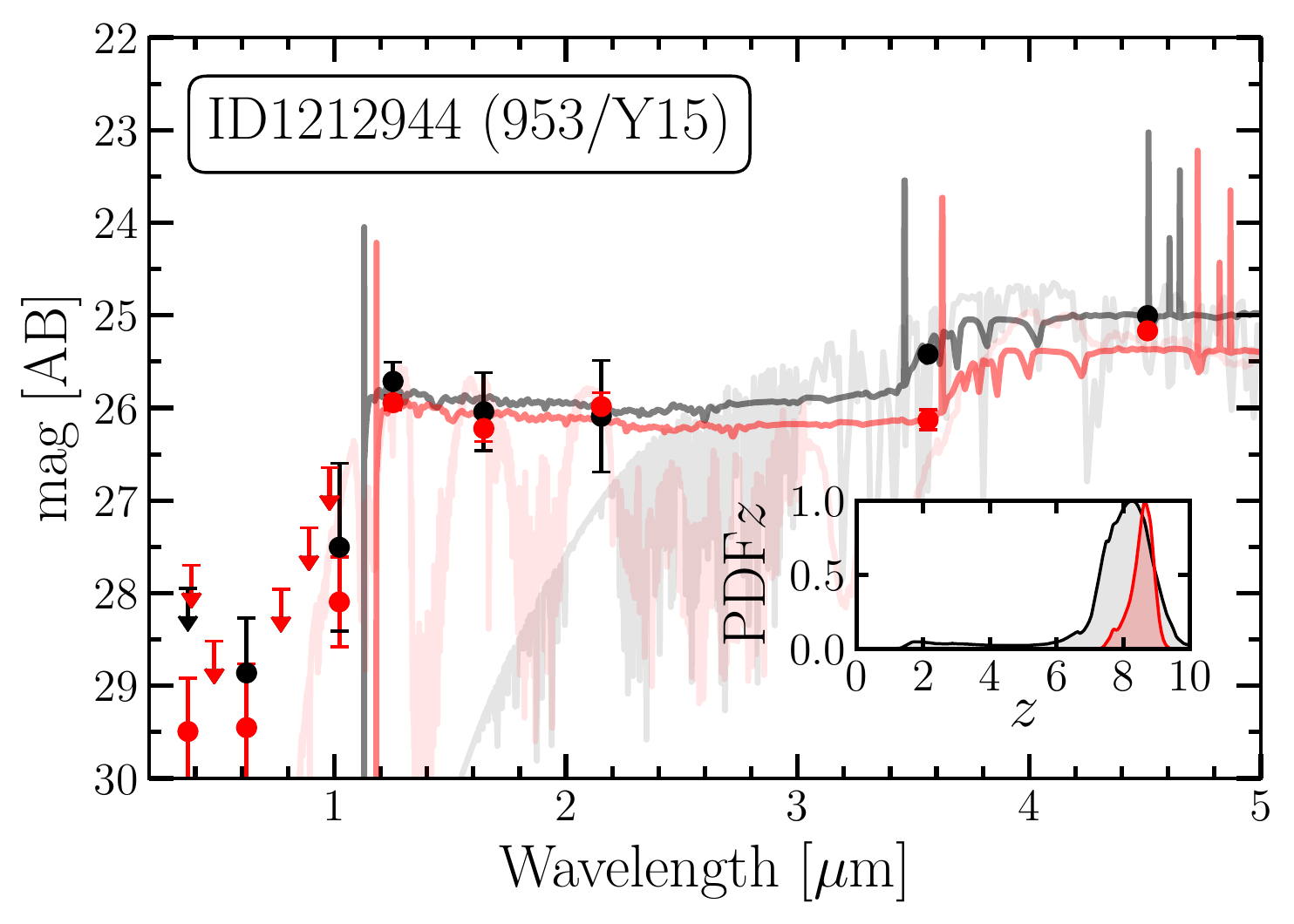}
  \includegraphics[width=0.33\textwidth]{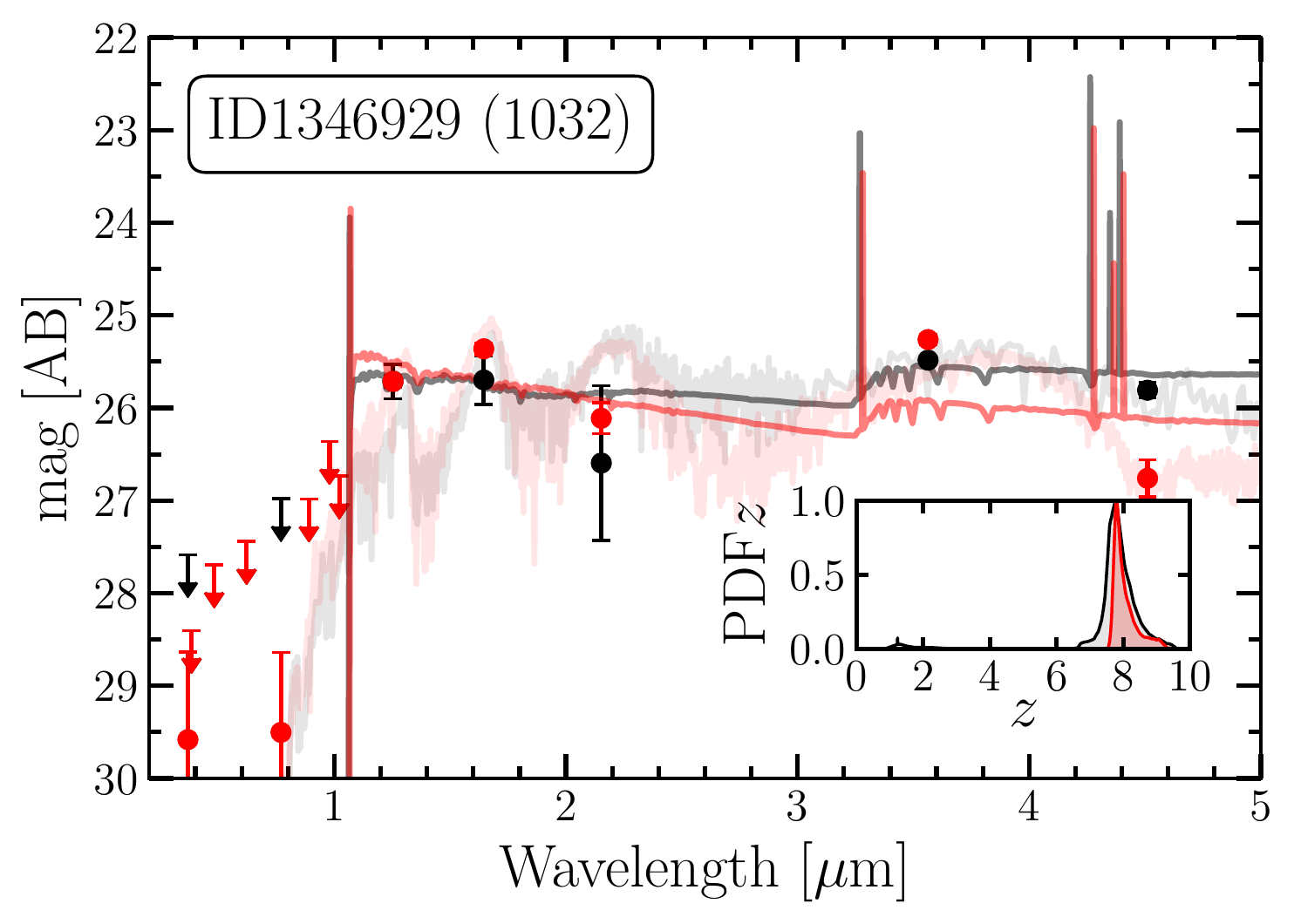}
  \includegraphics[width=0.33\textwidth]{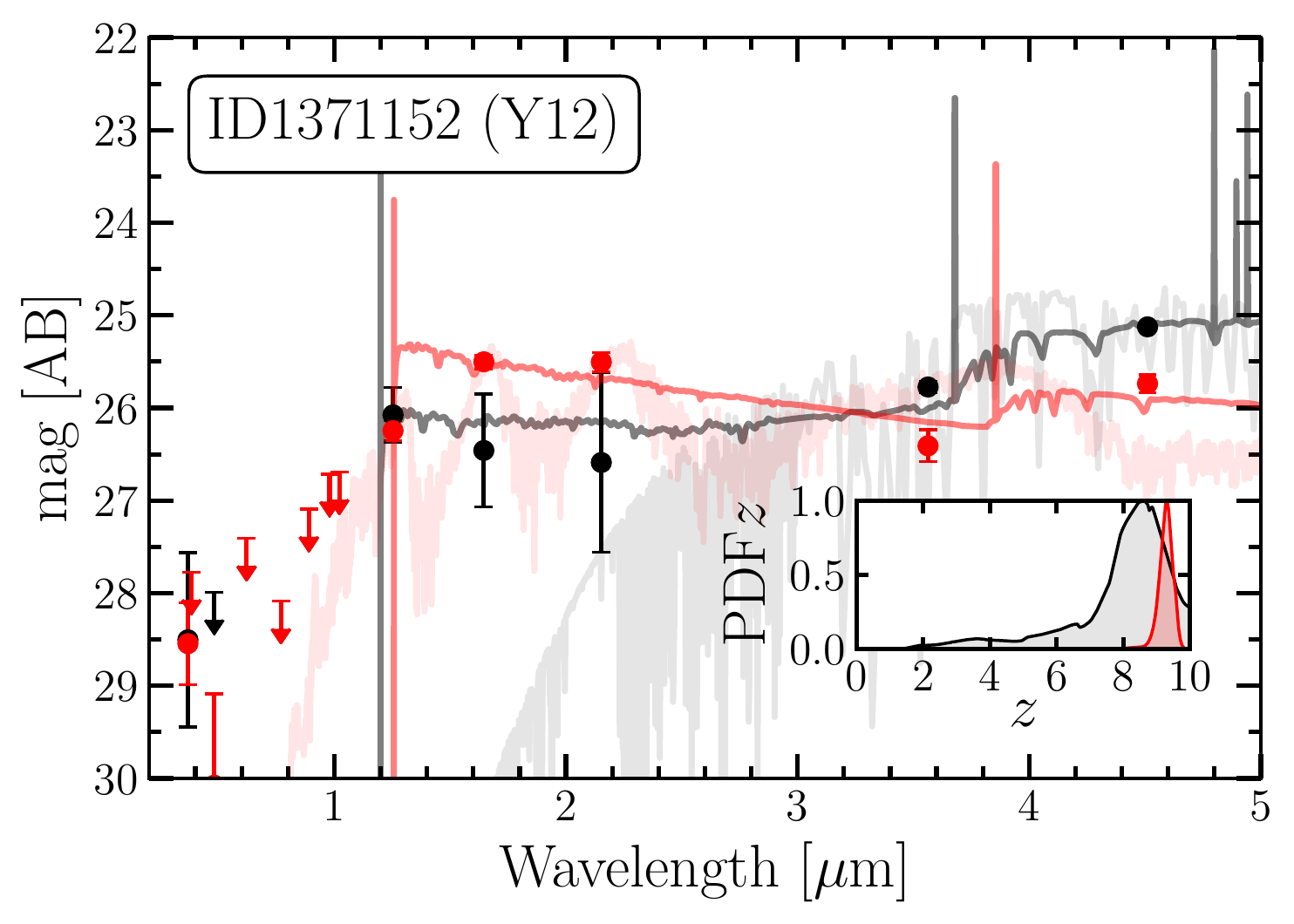}
  \caption{Same as Fig.~\ref{fig:seds_unblended_1} for the $z\geq7.5$ candidates from \citetalias{stefanon_brightest_2019} and \citetalias{bowler_lack_2020} which are recovered in the COSMOS2020 catalogue. The identifiers from those papers are indicated in parentheses.}
  \label{fig:seds_ancillary}
\end{figure}

The photometry of the $15$ recovered, previously identified candidates are shown in Fig.~\ref{fig:seds_ancillary}, and the corresponding stamp images are in Fig.~\ref{fig:snaps_ancillary_1} and \ref{fig:snaps_ancillary_2}. 

Among the $25$ previously identified candidates we find $15$ of them to be at high redshift. This represents $12$ of the $16$ candidates selected by \citetalias{bowler_lack_2020} in COSMOS, and $10$ of the $16$ candidates from \citetalias{stefanon_brightest_2019} (including $3$ selected in \citetalias{stefanon_brightest_2019} but not in \citetalias{bowler_lack_2020}). The identifiers from \citetalias{stefanon_brightest_2019} and \citetalias{bowler_lack_2020} are given in Table~\ref{tab:candidates_photoz}. 

For 9 of them, all photometric redshift estimates indicate a robust $z>7$ sources, and 13 of them belong to our gold sample. The candidates 356 (ID485056) and 1032 (ID1346929) present a significant fraction of the $\Delta \chi^2$ distribution falling in the star classification (Sect.~\ref{sec:brown_dwarfs}).

The candidates 301 (ID428351) and Y11 (ID859061) are blended with nearby sources and present significant optical fluxes. This was the reason for the candidate Y11 to be rejected from the \citetalias{bowler_lack_2020} selection. With the \classic{} catalogue, the photometric redshifts are $z\sim2$ for both sources. In contrast, these candidates are undetected in the optical with \farmer{} catalogue, leading to $z\sim8$. Similarly to the new blended candidates, only the photometry with \farmer{} is considered for these sources.
For the candidate 301, we note that the smaller $1.8\arcsec$ diameter apertures used in \citetalias{bowler_lack_2020} may have limited the impact of the nearby source on the optical photometry. For the candidate Y11, the $H$ and $K_s$-band flux is clearly concentrated in one location, while the $g$-band emission comes from a distinct nearby source, which also contributes to the $J$-band aperture flux. In this case, a smaller aperture may also limit the contamination.

The estimated redshifts and absolute UV magnitudes are in excellent agreement with those from \citetalias{stefanon_brightest_2019} and \citetalias{bowler_lack_2020}. We find that the absolute magnitudes from \farmer{} are systematically fainter for blended candidates, which is expected since the profile-fitting photometry separates the fluxes from nearby sources. 
In contrast, we find brighter magnitudes with \farmer{} for the two candidates 356 (ID485056) and 879 (ID1103149), which have no obvious neighbours. Nevertheless, resolved internal structures may be observed in the postage stamp images, and the \classic{} magnitudes are (in fact) in better agreement with \citetalias{bowler_lack_2020}. This situation may be due to the simple symmetric profile used to estimate the photometry in \farmer{}, or to the background subtraction. 

Candidate ID978062 is particularly bright, with a $H$-band magnitude of $24.5$. Its relatively broad \zPDF{}, with a full-width at half maximum (FWHM) larger than $1$, results from the Lyman break located between the $Y$ and the $J$ band. The resulting photometric redshift is higher ($z=8.86$ instead of $8.19$) than in \citetalias{bowler_lack_2020}, and the absolute magnitude brighter ($M_\text{UV}=-22.56$ instead of $-22.36$). This source has a spectroscopic redshift at 7.675 from the REBELS ALMA survey \citet{Bouwens21_REBELS}.

The candidate Y8 (ID1209618) has problematic \texttt{IRACLEAN} photometry with an extremely low [3.6] flux, which is not the case in \farmer{} catalogue. In addition, the \texttt{IRACLEAN} photometric uncertainties in both [3.6] and [4.5] are much smaller than in the near-infrared bands for the \classic{} catalogue, so that the main constraint on the SED is an unexpectedly red IRAC colour. All but one best-fit galaxy templates produce a redshift at $z>8$, except \lephare{} with the \classic{} catalogue.

\onecolumn

\begin{figure}
  \centering
  \includegraphics[width=\textwidth]{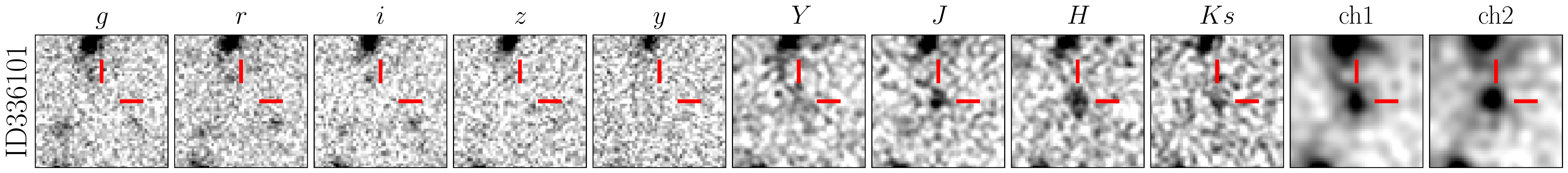}
  \includegraphics[width=\textwidth]{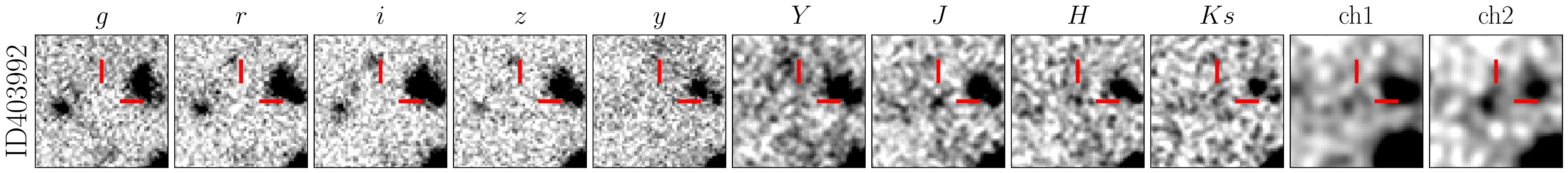}
  \includegraphics[width=\textwidth]{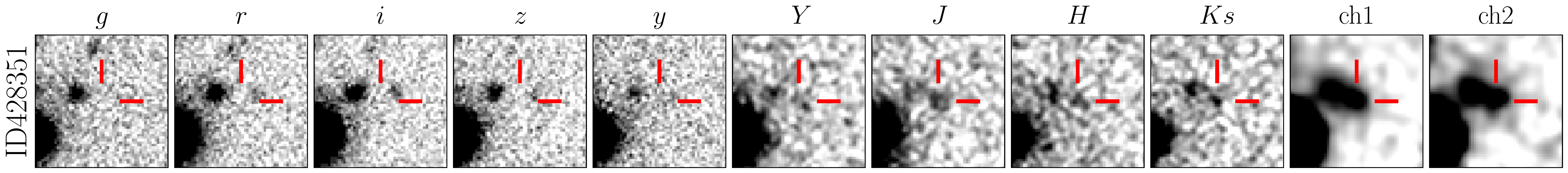}
  \includegraphics[width=\textwidth]{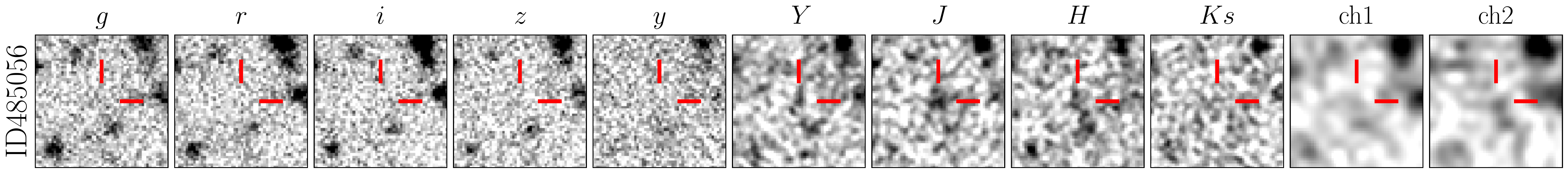}
  \includegraphics[width=\textwidth]{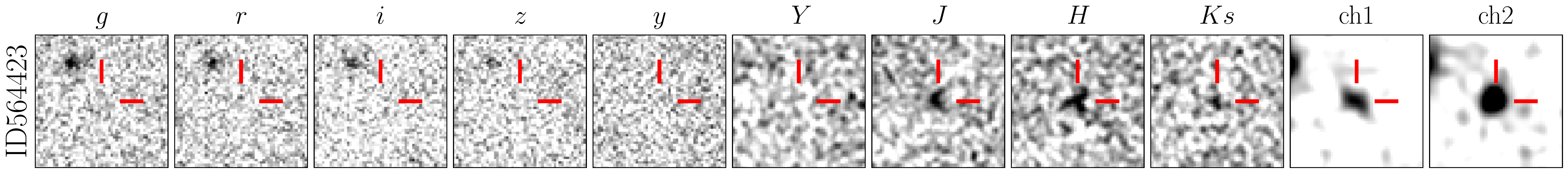}
  \includegraphics[width=\textwidth]{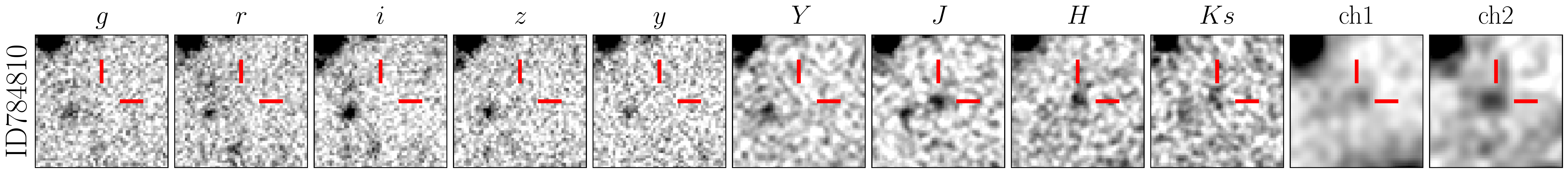}
  \includegraphics[width=\textwidth]{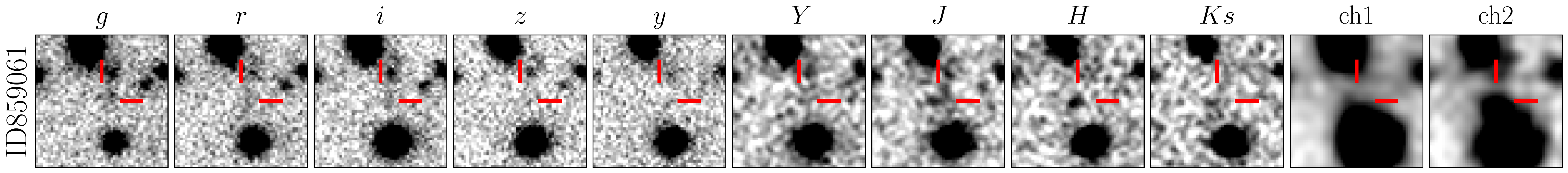}
  \includegraphics[width=\textwidth]{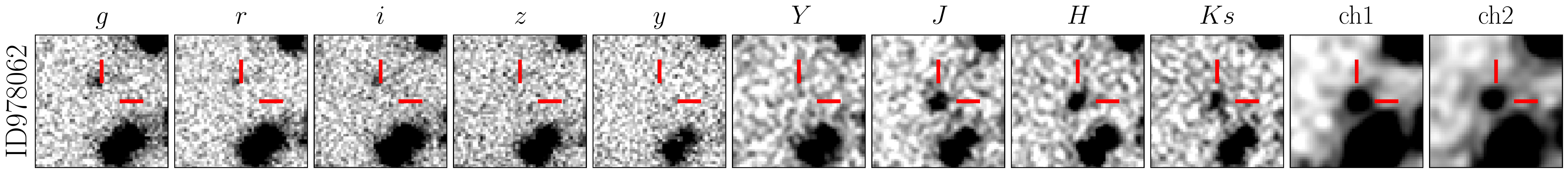}
  \includegraphics[width=\textwidth]{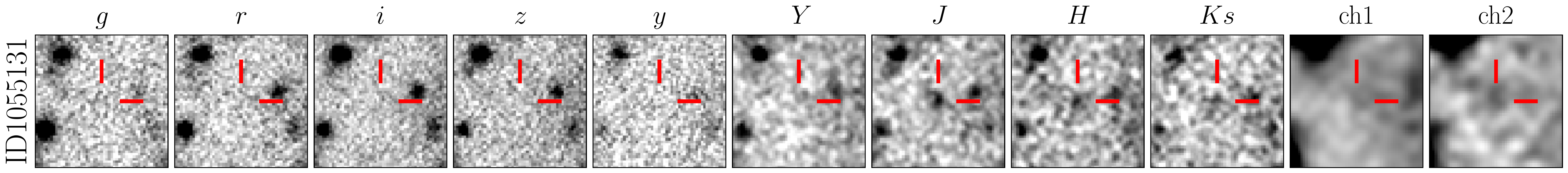}
  \includegraphics[width=\textwidth]{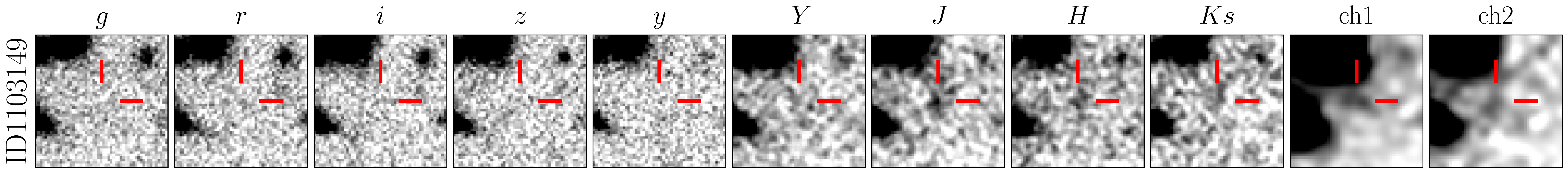}
  \includegraphics[width=\textwidth]{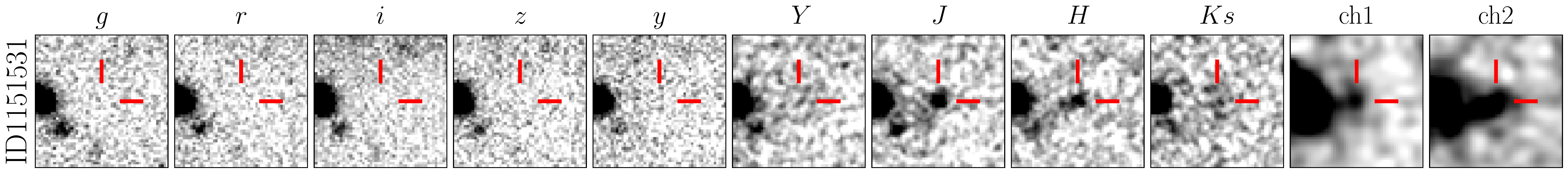}
  \caption{Same as Fig.~\ref{fig:snaps_unblended_1} for the $z\geq7.5$ candidates from \citetalias{stefanon_brightest_2019} and \citetalias{bowler_lack_2020} which are recovered in the COSMOS2020 catalogue.}
  \label{fig:snaps_ancillary_1}
\end{figure}

\begin{figure}
  \centering
  \includegraphics[width=\textwidth]{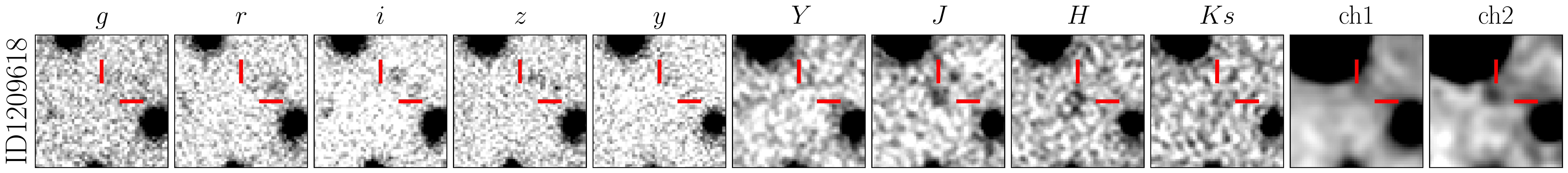}
  \includegraphics[width=\textwidth]{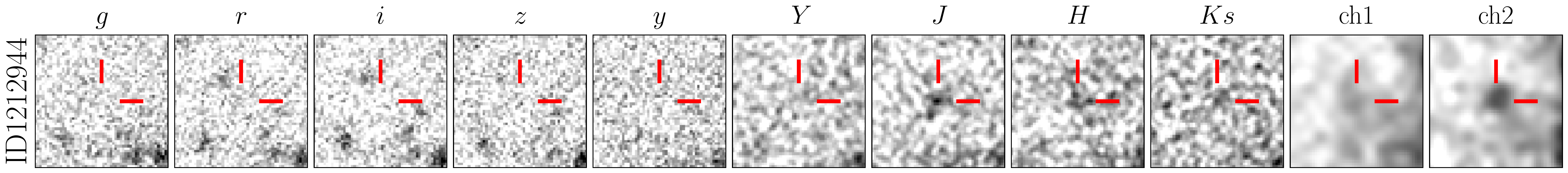}
  \includegraphics[width=\textwidth]{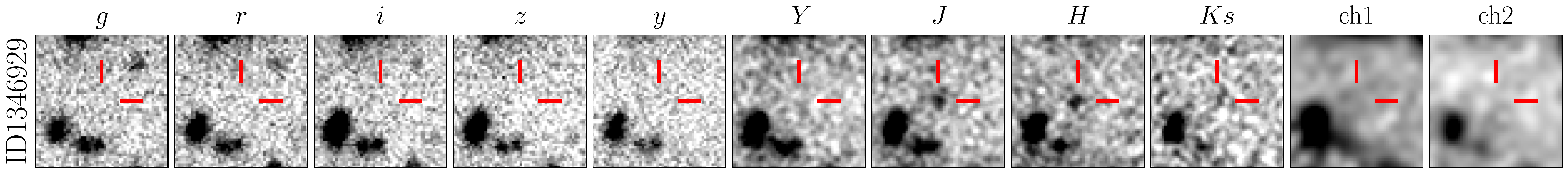}
  \includegraphics[width=\textwidth]{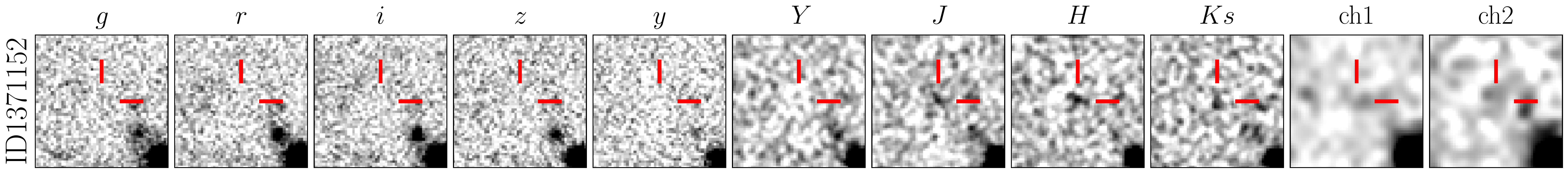}
  \caption{continued.}
  \label{fig:snaps_ancillary_2}
\end{figure}

\twocolumn

\subsection{Rejected candidates from the literature}
\label{sec:appendix_rejected_candidates}

We now describe the $10$ candidates from \citetalias{stefanon_brightest_2019} and \citetalias{bowler_lack_2020} which were not selected in this work. For four of these candidates (Y6, Y13, 919, 1212), the estimated \zPDF{} is double-peaked with strong low-redshift solutions. In addition, candidates Y6 and Y13 are detected at more than $2\sigma$ in the deep HSC $y$ and $z$ bands, respectively, and were already rejected in the \citetalias{bowler_lack_2020} galaxy sample. None of these sources were listed as robust by \citetalias{stefanon_brightest_2019} based on their \zPDF{}, so it is perhaps less of a surprise that we do not confirm them. For these reasons, these candidates are not included. Candidate 1212 is the brightest high-redshift galaxy identified in the COSMOS field in \citetalias{bowler_lack_2020}, with a photometric redshift of $z=9.1$ and an absolute UV magnitude of $M_\text{UV}=-23$. For this candidate, both the \classic{} and \farmer{} catalogues have a $3\sigma$ detection in the $r$ band, although this is not clear from the associated postage stamp. As a consequence, it is not kept, even though it remains an interesting candidate.

Our combined detection image fails to recover six candidates (Y7, Y9, Y14, Y15, 634, 1043) from \citetalias{stefanon_brightest_2019} and \citetalias{bowler_lack_2020}. The candidate Y7 is clearly visible in the $izYJHK$ detection image. While it is listed as robust by \citetalias{stefanon_brightest_2019} based on the \zPDF{}, it is not identified as a distinct object because of two large, bright nearby sources (\citetalias{bowler_lack_2020} arrived at the same conclusion), with a source at $J=20.7$. We matched these six sources with the official release catalogues from UltraVISTA DR4 \citep[][]{mccracken_ultravista_2012}. The detection is performed in each VIRCAM band. These sources do not have any counterpart in the $K_s$ catalogue. The source Y7 discussed previously is detected only in the $H$-band selected catalogue. The source Y14 from \citetalias{stefanon_brightest_2019} is detected in the $J$-band selected catalogue with a magnitude of $J=26.3\pm0.1$ mag, but not in the other bands.

\end{appendix}

\end{document}